\documentclass[aip,jcp,reprint,longbibliography,floatfix]{revtex4-2}
\usepackage{graphicx}
\usepackage{amsmath,amssymb,bm}
\usepackage{booktabs}
\usepackage{placeins}
\usepackage{microtype}
\usepackage{url}
\newcommand{\wavenum}{\ensuremath{\mathrm{cm}^{-1}}}

\newcommand{\mel}[3]{\langle #1\lvert #2\rvert #3\rangle}
\begin{document}
\title{Cartesian and Curvilinear Anharmonic Frequencies and Intensities:
Perturb-then-Diagonalize Polyad Transport and Adiabatic
Discrete-Variable Representation}
\author{Vincenzo Barone}
\affiliation{National Interuniversity Consortium for Materials Science and
Technology (INSTM), via G. Giusti 9, 50121 Firenze, Italy}
\email{vincebarone52@gmail.com}

\begin{abstract}
Anharmonic intensities reveal vibrational coupling that band positions
alone cannot determine. We derive explicit second-order infrared
transition moments in curvilinear normal coordinates, including kinetic
metric derivatives and transport through variational resonance polyads.
The vibrational perturbation-to-second-order-then-diagonalization
(VP2TD) treatment covers fundamentals, overtones, three-quantum
transitions, and hot bands. A general computational platform combines
Extended Symmetry-Adapted Local Coordinates (ESALCs), using Cartesian
(ESALC-C) or internal (ESALC-I) primitives, with interchangeable
quantum-chemical programs and parallel finite differences. Independent
oscillator matrices verify the intensity equations; a molecular water
check reproduces second-order amplitudes within
$8.4\times10^{-8}$ atomic units. Comparisons with established tools
reproduce a formaldehyde Fermi polyad within 0.163~cm$^{-1}$ and
0.061~km\,mol$^{-1}$ and water Raman activities within
$10^{-7}$~\AA$^6$. For 29 matched formaldehyde rotational line pairs,
the predicted hot-to-fundamental intensity ratio, using the measured
lower-state energy, is consistent with the nuclear-spin-corrected
median: the differences are 3.9\% (ESALC-I) and 3.8\% (ESALC-C), compared
with a 15\% uncertainty estimate for individual measured line
intensities. Common analytic reference derivatives
ensure identical harmonic observables, while internal coordinates
reduce and localize intermode anharmonic couplings. Rotation--vibration
constants extend the same tensor treatment. An integrated adiabatic discrete
variable representation treats large-amplitude motions, and
finite-level partition functions provide anharmonic thermochemistry.
\end{abstract}


\maketitle
\raggedbottom

\section{Introduction}
Vibrational spectroscopy connects an electronic-structure model to
observable molecular motion. Fundamental frequencies identify the
energetic scale of a vibration, while infrared (IR) and Raman intensities
report how electric properties change along that motion. Overtones,
combination bands, and hot bands carry additional information about
anharmonicity and coupling beyond the harmonic normal-mode picture. A useful theoretical method must
predict both band positions and transition strengths, including the
redistribution of intensity when nearly degenerate states mix
\cite{barone2005,willetts1990,vazquez2006,bloino2015}.

Two choices shape a vibrational calculation: the primitive coordinates
and the harmonic modes formed from them. Cartesian displacements have a
simple kinetic energy before molecular-frame reduction, while internal
stretches, bends, and torsions follow chemical motions and introduce a
geometry-dependent kinetic metric. Normal modes diagonalize the
quadratic forms in either representation. At higher order, curvilinear
normal modes retain the nonlinear internal-coordinate dependence and can
concentrate anharmonicity among fewer modes. The resulting reduction of
intermode couplings gives a direct chemical interpretation of the field.
A common equilibrium Hessian, property Jacobian, and molecular frame
ensure the same harmonic frequencies and intensities in both
representations.

Two practical obstacles have limited the routine use of internal
coordinates in general anharmonic calculations. The first is the automatic
construction of a complete, nonredundant, symmetry-adapted internal basis
for arbitrary molecular topologies, including rings and fused systems. The
second is the evaluation of the coordinate-dependent kinetic metric $G$ and
the derivatives required by the perturbative Hamiltonian. The present
implementation removes both obstacles: ORACLE and SMITH generate and certify
the ESALC-I basis automatically from a redundant pool of local primitives,
molecular geometry, and connectivity. The construction is democratic: it
treats symmetry-equivalent atoms and bonds on equal footing and does not
impose the sequential reference tree of a Z-matrix, which is unsuitable for
ring closure and becomes ambiguous for fused, bridged, or otherwise
nonstandard topologies. Rank selection then yields a complete nonredundant set,
using local symmetries to organize equivalent primitives within fragments and
chemical families, followed by the total molecular symmetry to assign the
global irreducible representations and selection rules,
and analytic $B$ and $B'$ provide $G$ and $G'$, while $G''$ is obtained by
finite differences of analytic $G'$ at the same displaced geometries used
for the potential derivatives. No molecule-specific internal-coordinate
definition or kinetic-energy derivation is required.

Second-order vibrational perturbation theory (VPT2) makes the weak
part of the anharmonic problem economical: once cubic, semidiagonal
quartic, and kinetic-metric derivatives are available, most energy
and transition-moment corrections are algebraic contractions.
Near-degenerate Fermi and Darling--Dennison interactions require a
different step. In vibrational perturbation to second order, then
diagonalization (VP2TD), selected resonant terms are removed from
perturbative denominators and restored in finite variational
polyads~\cite{mendolicchio2021,mendolicchio2022}. The same polyad
eigenvectors must transform transition moments, including an excited
initial state for a hot band, to predict intensity borrowing. We use an operational polyad defined by
the connected component of tested resonance couplings, rather than
assuming a conserved polyad quantum number exists in every case
\cite{krasnoshchekov2013polyad,franke2021vpt2k}. Explicit target bands or a spectral
window determine which observables must converge. Additional states
can enter as partners without becoming independent convergence
requirements.

The perturb-then-diagonalize ordering is particularly efficient for
vibrations because the nonresonant second-order corrections are
algebraic. In electronic structure, a correction outside a diagonalized
multiconfigurational reference generally requires a resolvent or an
equivalent state summation.\cite{andersson1992caspt2} The potential
representation is equally important: for a Morse oscillator, VPT2 from
its quadratic, cubic, and quartic derivatives gives the exact bound-state
energies. A variational calculation instead requires the higher-order
continuation of the potential.\cite{burcl2003morse,fortenberry2013quartic}
The program also integrates a discrete variable representation (DVR)
for large-amplitude motions (LAMs). Its adiabatic treatment retains a
periodic or multiwell potential and the response of the remaining
vibrations.

The distinction between general and transition-specific intensity
expressions is important. Reference~\onlinecite{willetts1990} established general perturbative
formulas for anharmonic IR intensities in rectilinear normal
coordinates. Reference~\onlinecite{vazquez2006} subsequently
gave a simpler algebraic expression for fundamental transition
moments, and Ref.~\onlinecite{bloino2015} developed explicit formulas
for different classes of transitions, including those involving
three quanta, all originating in the vibrational ground state.
Contact-transformation implementations can instead evaluate property
matrix elements between eigenstates of effective Hamiltonians.
AnharmoniCaOs\cite{mulas2018anharmonicaos} follows this route in rectilinear normal coordinates with
a dipole surface through second derivatives
and has been applied to thermally excited pyrene.\cite{chakraborty2021pyrene}
That work evaluates the transformed-property matrices numerically; it
does not report transition-specific algebraic reductions in force
constants and property derivatives.
Specialization removes algebraically vanishing terms before numerical
evaluation, making each coefficient easier to inspect and test. Our
contribution is a set of closed, transition-specific second-order amplitudes
for excited initial states, with both-state normalization and explicit
curvilinear metric terms. Existing methods already calculate hot-band
intensities using internal coordinates; the distinction here is their
algebraically reduced form. We derive those hot-band formulas together with explicit
curvilinear expressions containing the kinetic-metric derivatives and
transport through resonance polyads on both sides of the transition.
Their common algebra covers electric and magnetic properties, and their
reduction makes each physical contribution directly accessible.

VPT2 frequencies and intensities have a substantial literature and
several active implementations. Automated rectilinear calculations
and interfaces to electronic-structure programs are widespread
\cite{barone2005,barone2010,barnes2017,nelson2025}; higher-order
Van Vleck and numerical--analytic treatments are described in
Refs.~\onlinecite{sibert1988,krasnoshchekov2012,krasnoshchekov2014,krasnoshchekov2015uracil},
while AnharmoniCaOs combines a second-order Van Vleck Hamiltonian,
iterative resonance polyads, and finite-temperature state
sampling.\cite{mulas2018anharmonicaos,chakraborty2021pyrene}
Psience/PyVibPTn\cite{psience} also supports internal
coordinates, flexible perturbative orders, polyads, and anharmonic
transition intensities, including transitions from excited initial
states~\cite{boyer2022sparse,mccoy2022,boyer2026internals}.
Reference~\onlinecite{yangbloino2022} describes a general state-index construction for
transition moments between arbitrary initial and final states, while
reporting reduced algebraic formulas for transitions originating in
the vibrational ground state. That work did not give
the transition-specific reduced coefficients for hot bands. Variational alternatives remain essential where perturbative or low-order potential
models fail~\cite{bowman2003,christiansen2007,erfort2020}.
Earlier work established the curvilinear energy and frequency
equations~\cite{mendolicchio2022,mendolicchio2026}. The scientific
development here is the explicit second-order transition-property
theory in those coordinates: metric-dependent IR and Raman
intensities, property transport across resonance polyads, and hot
bands between anharmonic excited states. It also supplies overtone
and three-quantum amplitudes. The same tensor framework also describes chiroptical properties.

We construct Extended Symmetry-Adapted Local Coordinates (ESALCs) from
Cartesian or internal primitives after removing redundancies. ESALC-C uses Cartesian displacements
with translations and rotations removed in an Eckart frame; ESALC-I uses
internal coordinates localized on chemical groups. Harmonic normal modes
are constructed from either set. An autonomous calculation procedure
generates displaced Cartesian geometries and transforms the returned
energies, derivatives, and properties, including the kinetic metric.
Independent displaced points can be evaluated in parallel by different
quantum-chemical programs. LINK\cite{nanomatrix} supplies geometry optimization before the reference Hessian and
displaced derivatives are acquired. This general implementation is the second
contribution of the work: it connects the spectroscopic equations to a
complete coordinate and derivative calculation.

The following sections establish the common Hamiltonian and derive the
transition-property equations. Independent algebraic and molecular
validations precede the resonance and hot-band applications. Molecular examples compare rectilinear and
curvilinear representations, test hot-band intensities against
experiment, and quantify the localization of intermode couplings.
Further applications include rotation--vibration constants, integrated
DVR for LAMs, and anharmonic thermochemistry. The Supporting Information (SI) provide an equation
guide and a consolidated set of notation and unit conventions.

\section{Common rovibrational Hamiltonian}
\subsection{ESALC coordinates and the vibrational tangent space}
Analytic electronic derivatives are naturally expressed in Cartesian
coordinates, and many quantum-chemical programs already exploit molecular
symmetry in their evaluation. ESALCs organize the independent
finite-difference directions and the subsequent vibrational treatment.
For $N_{\mathrm{at}}$ atoms, let $T$ denote the dimension of the tangent
space generated by rigid translations and rotations: $T=6$ for a
nonlinear molecule and $T=5$ for a linear molecule. Removing that space
leaves $3N_{\mathrm{at}}-T$ vibrational directions. In ESALC-C, the
Cartesian primitives are projected onto this complement in the mass
metric and combined according to symmetry. Finite differences therefore
sample vibrational directions only, and tensor acquisition and
transformation can be organized into the symmetry-allowed blocks.

The construction also exploits chemical locality. A molecule is
partitioned into chemically defined fragments, whose primitive
coordinates are adapted to local symmetry before their assembly into
the molecular coordinate basis. This organization applies to both
Cartesian and internal primitives. ESALC-I additionally separates
internal coordinates into chemical families, such as bond stretches,
valence bends, out-of-plane motions, and torsions. Linear combinations
are formed only within the same family, retaining the interpretation
of each coordinate while obtaining a nonredundant basis. Fragment-local
adaptation provides a structured representation; exact molecular
selection rules follow from the symmetry operations of the whole
geometry, and allowed interfragment couplings are retained.

The internal primitives are explicit geometric functions. For bond
vectors $\mathbf r_{ab}=\mathbf x_a-\mathbf x_b$,
\begin{align}
 r_{ab}&=|\mathbf r_{ab}|,\qquad
 \theta_{abc}=\arccos\!\frac{\mathbf r_{ab}\cdot\mathbf r_{cb}}
 {r_{ab}r_{cb}},\nonumber\\
 \phi_{abcd}&=\operatorname{atan2}
 [\widehat{\mathbf b}_2\cdot(\mathbf n_1\times\mathbf n_2),
  \mathbf n_1\cdot\mathbf n_2],
 \label{eq:esalc-primitive-form}
\end{align}
where $\mathbf b_1=\mathbf x_b-\mathbf x_a$,
$\mathbf b_2=\mathbf x_c-\mathbf x_b$,
$\mathbf b_3=\mathbf x_d-\mathbf x_c$,
$\mathbf n_1=\mathbf b_1\times\mathbf b_2$, and
$\mathbf n_2=\mathbf b_2\times\mathbf b_3$; the dihedral is continued
smoothly around its reference value. For two symmetry-equivalent bonds,
the elementary combinations are
$s_\pm=(\Delta r_1\pm\Delta r_2)/\sqrt2$. Equivalent angles or
signed out-of-plane motions are combined within their own families.
These are local geometric coordinates; the subsequent GF eigenvectors
form the curvilinear normal modes from them. Thus symmetry adaptation
fixes the coefficients of the geometric functions, while their nonlinear
dependence on Cartesian positions supplies $B$, $B'$, and the metric.

\begin{figure*}[t]
\centering
\includegraphics[width=0.94\textwidth]{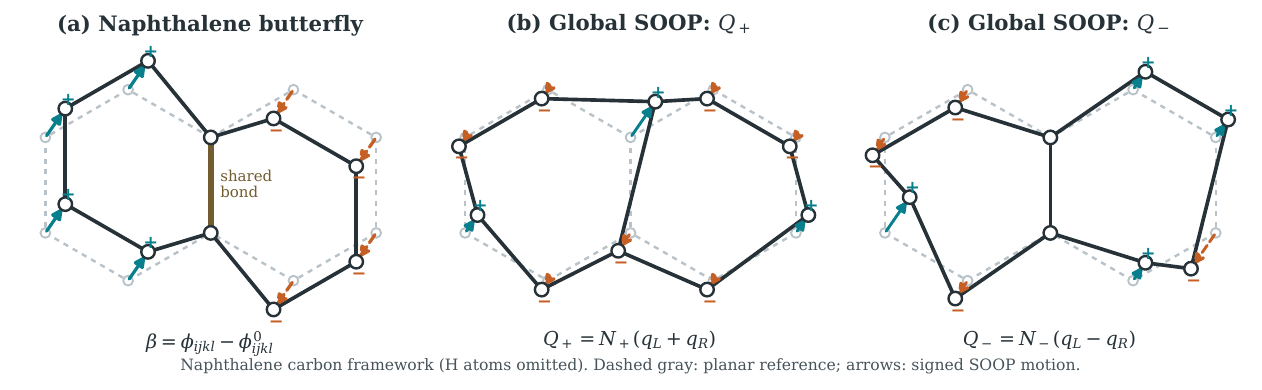}
\caption{Ring ESALC-I patterns for naphthalene (hydrogen atoms omitted).
(a) Butterfly folding about the shared bond. (b,c) Symmetric and
antisymmetric combinations of overlapping SOOP candidates. The shared-atom
overlap enters Eq.~\eqref{eq:fused-ring-soop}; all candidates are reduced and
symmetrized together. Arrow color and line style distinguish displacement
signs.}
\label{fig:esalci-ring-coordinates}
\end{figure*}

Ring ESALC-I coordinates follow the same construction. Let
$\bar\chi_j$ be the normalized, consistently oriented combination of the
two endocyclic selective out-of-plane (SOOP) primitives at vertex $j$ of
an ordered $n$-membered ring. Its independent puckering coordinates are
the $n-3$ cyclic combinations
\begin{align}
 q_{kc}&=\sqrt{\frac{2}{n}}\sum_{j=0}^{n-1}
       \cos\!\left(\frac{2\pi k j}{n}\right)\bar\chi_j,
       \nonumber\\
 q_{ks}&=\sqrt{\frac{2}{n}}\sum_{j=0}^{n-1}
       \sin\!\left(\frac{2\pi k j}{n}\right)\bar\chi_j,
 \label{eq:ring-soop-esalc}
\end{align}
with $k=2,\ldots,\lfloor(n-1)/2\rfloor$; an even ring also has
$q_{n/2}=n^{-1/2}\sum_j(-1)^j\bar\chi_j$. The omitted $k=0$ and
$k=1$ combinations are the out-of-plane translation and rigid tilts. These
are ring-local candidates, not independent coordinates assigned separately
to every cycle. For fused and polycyclic systems, all candidate displacement
columns are collected in one pool, dependent directions are removed by a
global rank-revealing reduction, and the surviving space is projected with
the total molecular symmetry. For two symmetry-related overlapping rings,
the resulting combinations have the form
\begin{equation}
 Q_\pm=\frac{q_L\pm q_R}{\sqrt{2(1\pm S_{LR})}},\qquad
 S_{LR}=\langle q_L\mid q_R\rangle,
 \label{eq:fused-ring-soop}
\end{equation}
where the overlap includes their shared atoms. A butterfly coordinate of two
fused rings is the signed dihedral
$\beta=\phi_{ijkl}-\phi^0_{ijkl}$ about their shared bond $(j,k)$,
where $i$ and $l$ belong to opposite ring-only sets. These two ring
families and representative displacement patterns are shown in
Fig.~\ref{fig:esalci-ring-coordinates}. Their detection, orientation,
cyclic combination, symmetry adaptation, and rank selection are performed
automatically from the molecular topology and reference geometry. Starting
from the full naphthalene pool, for example, preserves equivalence among both
rings and their shared sites before the nonredundant subspace is selected; no
atom ordering or Z-matrix traversal selects a privileged ring bond. Local
ring or fragment symmetries first form chemically compact candidates, and
the symmetry operations of the complete molecule then close and project the
final ESALC-I space. This final molecular-symmetry projection combines the
overlapping ring candidates into the actual molecular ESALCs; individual-ring
labels are not retained as independent vibrational coordinates.

More explicitly, the initial internal-coordinate set contains $m$
primitive functions $p_\mu(\mathbf x)$, usually with
$m>3N_{\mathrm{at}}-T$. Ring closures and relations among angles or
torsions make some of their differential displacements dependent.
The analytic primitive Wilson matrix,
$B^p_{\mu a}=\partial p_\mu/\partial x_a$, identifies these
dependencies at the reference geometry. SMITH, a module of nano-MATRIX,\cite{nanomatrix} first selects independent
coordinates from its normalized rows, retaining chemically defined
families and any designated special coordinates. Numerical
orthogonalization tests whether a candidate adds a new direction; the
retained coordinate keeps its primitive definition. A global rank
check confirms that the selected rows span all
$3N_{\mathrm{at}}-T$ vibrational directions. Removing a redundant row
thus removes a duplicate description of motion, rather than a
vibrational degree of freedom.

Molecular symmetry adaptation follows this nonredundant selection.
Within each homogeneous family, the source combinations must close
under the molecular symmetry operations. Symmetry projection and
orthonormalization of the coefficient vectors then produce independent
symmetry-adapted linear combinations, with a final check of their rank
and vibrational symmetry content. Writing the combined selection and
symmetry transformation as $C$, the ESALC-I coordinates are
\begin{align}
 s_i(\mathbf x)&=\sum_{\mu=1}^{m}C_{i\mu}
       [p_\mu(\mathbf x)-p_\mu(\mathbf x_0)],\nonumber\\
 B^s&=CB^p,\qquad
 G^s=CB^p\mathsf M^{-1}(B^p)^{\mathsf T}C^{\mathsf T},
 \label{eq:primitive-esalc-reduction}
\end{align}
where $\mathsf M$ is the diagonal Cartesian mass matrix. The
coefficients $C$ are fixed at the reference geometry; the primitive
functions and their derivatives retain their nonlinear geometry
dependence. Coefficient orthonormality does not make the physical
metric $G^s$ diagonal. The resulting ESALCs form an independent,
chemically organized coordinate basis, with the kinetic and potential
couplings retained for the subsequent Wilson kinetic-metric/force-matrix (GF) analysis.

The ORACLE and SMITH modules of nano-MATRIX\cite{nanomatrix} supply
molecular connectivity, fragments, symmetry, and the ESALC transformations;
the TRINITY module\cite{nanomatrix} performs the spectroscopic calculation.
The Wilson matrix $B$ and its first Cartesian derivative $B'$ are
analytic, as are the corresponding metric $G$ and its first derivative
$G'$. TRINITY obtains $G''$ by finite differences of analytic $G'$ at
the same single-mode displaced geometries used for the potential
field. Each displacement supplies an entire derivative slice, so
mixed components of $G''$ require no pair-mode displacements;
Eq.~\eqref{eq:analytic-metric-stencil} gives the explicit construction. The subsequent Cartesian
normal-mode or GF analysis may mix symmetry-compatible
coordinates across families: the family restriction concerns the
construction of the internal primitives' combinations, not a
truncation of the physical Hessian. The resulting normal modes define
the finite-difference stencil, while the independent electronic-structure program
continues to receive Cartesian geometries and return Cartesian
derivatives. This separation permits the same coordinate and
spectroscopic machinery to use different electronic-structure codes.

\begin{figure*}[t]
\centering
\includegraphics[width=\textwidth]{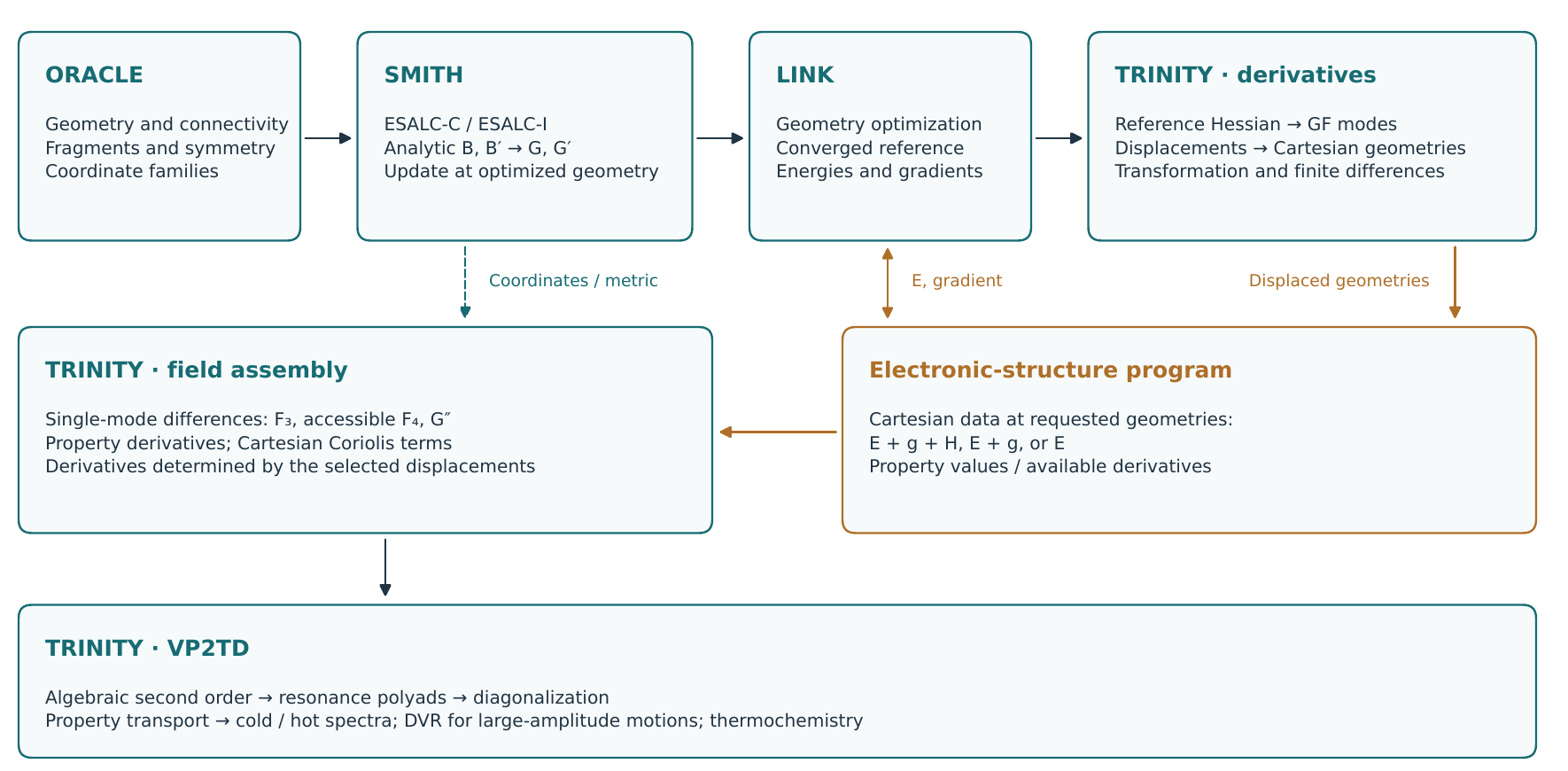}
\caption{Coordinate construction and spectroscopy with interchangeable
quantum-chemical programs. ORACLE identifies fragments and symmetry;
SMITH supplies ESALCs and analytic coordinate derivatives; LINK optimizes
the reference geometry; TRINITY forms
the anharmonic field and VP2TD spectra. Independent electronic
calculations return Cartesian energies and available derivatives.
The kinetic metric is determined from molecular geometry.}
\label{fig:workflow}
\end{figure*}

The division of responsibilities is summarized in
Fig.~\ref{fig:workflow}. The electronic interface can supply analytic
Hessians, gradients, or only energies. TRINITY selects a compatible
differentiation route and records which force-constant and property
components its stencil determines. In particular, one-mode differences
of gradients do not supply the same tensor coverage as one-mode
differences of Hessians; missing components are never identified with
calculated zeros. The coordinate construction, symmetry relations,
and geometric metric do not depend on that electronic choice.

\subsection{Metric and electronic derivatives}
Let $s=(s_1,\ldots,s_N)$ be a nonredundant coordinate representation and
$\varphi$ the three body-frame rotation parameters for a nonlinear
molecule. For either representation $X\in\{C,I\}$, the body-fixed
Cartesian geometry $\mathbf{x}^{X}(s)$ defines the vibrational and
rotational tangent vectors $t_{\alpha a}=\partial\mathbf{x}_a/
\partial\xi_\alpha$, $\xi=(s,\varphi)$. After center-of-mass
separation, form the complete covariant and contravariant metrics
\begin{equation}
 g_{\alpha\beta}(s)=\sum_a m_a t_{\alpha a}(s)\cdot
 t_{\beta a}(s),\qquad G(s)=g(s)^{-1}.
 \label{eq:metric}
\end{equation}
The vibrational block $G_v^{ij}$ is selected \emph{after} inverting
the full rovibrational matrix. At $J=0$, the flat-measure vibrational
operator takes the form used by the curvilinear generalized VPT2 (GVPT2) formalism,
\begin{equation}
 H_{J=0}=\frac12 p_iG_v^{ij}(s)p_j+V(s)+V_g(s),
 \label{eq:ham}
\end{equation}
where $V_g$ is the scalar term fixed by the determinant of the same
full metric.\cite{mendolicchio2026} The rotational tangent vectors
retain rotation--vibration coupling in this vibrational operator.

LINK supplies the stationary reference by optimizing the geometry with
electronic energies and Cartesian gradients in the coordinates supplied
by ORACLE and SMITH. After convergence, the molecular symmetry and
ESALC representation are reconstructed at the optimized geometry.
TRINITY then acquires the reference Hessian, performs harmonic analysis,
and freezes the resulting normal-mode transformation for displaced-point
calculations. A previously optimized geometry can enter at this stage
with its electronic method and convergence information retained.

At every geometry, the Cartesian Hessian is first symmetrized as
$H^x\leftarrow(H^x+H^{xT})/2$ before it enters the harmonic analysis
or any finite difference. The molecular point group further specifies
the stabilizer of each displaced geometry: the operations that leave
that geometry fixed are used to project its Hessian and properties at
the electronic source. Operations mapping it to a distinct displaced
geometry relate the corresponding stencil points covariantly. The
resulting Cartesian Hessian $H_{ab}^{x}$ and gradient $g_a^x$ give
the transformed force matrix
\begin{align}
 F_{ij}(s)&=J_{ai}(s)H_{ab}^{x}(s)J_{bj}(s)
  +g_a^x(s)\frac{\partial^2x_a}{\partial s_i\partial s_j},\\
 J_{ai}&=\frac{\partial x_a}{\partial s_i}.
 \label{eq:hessian}
\end{align}
For each requested ESALC displacement $s$, the calculation procedure follows the
electronic-structure program-neutral map
\begin{equation}
\begin{aligned}
 s&\longmapsto\mathbf x(s)
 \longmapsto\bigl(E,\mathbf g^x,H^x,\boldsymbol\mu,D^x\bigr)\\
 &\longmapsto\bigl(E(s),\mathbf g^s,F^s,\boldsymbol\mu(s),D^s\bigr),
\end{aligned}
 \label{eq:backend-map}
\end{equation}
where $\mathbf g^s=J^T\mathbf g^x$ and
$D^s_{\alpha i}=D^x_{\alpha a}J_{ai}$ when the electronic-structure program supplies
analytic Cartesian dipole derivatives. Property values may instead
be differentiated by the calculation procedure on the same ESALC stencil. The
electronic-structure program receives only Cartesian coordinates and returns
the available Cartesian energy, derivatives, and properties; it does
not construct ESALCs, the body-fixed metric, or its derivatives.
These are obtained independently from $\mathbf x(s)$, masses, and
analytic metric differentiation and geometry-only differences, as
specified below. The complete rovibrational metric in
Eq.~\eqref{eq:metric} also fixes the geometric scalar.
The gradient term is retained at every displaced point even when
the reference geometry is stationary. At equilibrium, Wilson GF
determines the shared harmonic reference,
\begin{equation}
 G_v(0)F(0)L=L\operatorname{diag}(\omega_1^2,\ldots,\omega_N^2).
 \label{eq:gf}
\end{equation}
The normal-mode transform $L$ is then frozen during the finite-
difference acquisition. Single-mode differences of Cartesian Hessians
give the complete cubic field and the quartic field with at most three
distinct mode indices. Pair-mode Hessian differences additionally give
the complete quartic field. The full metric gives $G^{ij}{}_{,k}$ and
the accessible $G^{ij}{}_{,kl}$ in those same modes. The resulting cubic, quartic,
and ordered kinetic operators are the $H_1$ and $H_2$ of the
curvilinear GVPT2 expansion~\cite{mendolicchio2026}.

For the reduced acquisition studied below, let $h$ be a displacement
in one dimensionless normal mode $q_k$. Only the reference and
$q_k=\pm h$ are used for differences of electronic Hessians and the
metric:
\begin{align}
 F_{ij,k}&\simeq\frac{F_{ij}(+h e_k)-F_{ij}(-h e_k)}{2h},\\
 F_{ij,kk}&\simeq\frac{F_{ij}(+h e_k)-2F_{ij}(0)+F_{ij}(-h e_k)}{h^2},
 \label{eq:singlemodeqff}
\end{align}
These Hessian differences are accompanied by evaluations of analytic
$G$ and $G'$ at the same geometries. The standard metric stencil
differentiates $G'$ to obtain all $G_{ij,kl}$, as specified in
Eq.~\eqref{eq:analytic-metric-stencil}.
The acquisition averages the distinct displaced-mode estimates of the
same force constant, following the established procedure.\cite{barone2005}  When the optional
energy--gradient estimator is selected, the gradients returned
together with each displaced Hessian provide, without additional
electronic-structure points,
\begin{equation}
 C^{(g,k)}_{ikk}=\frac{g_i(+h e_k)+g_i(-h e_k)-2g_i(0)}{h^2}.
 \label{eq:gradient-cubic-mean}
\end{equation}
For $i\ne k$, the two distinct Hessian-direction estimates
$C^{(H,k)}_{ikk}$ and $C^{(H,i)}_{ikk}$ and the optional gradient
estimate in Eq.~\eqref{eq:gradient-cubic-mean} can be combined with
independent Hessian and gradient weights, $w_H$ and $w_g$:
\begin{align}
 f_{ikk}&=\frac{w_H\bigl(C^{(H,k)}_{ikk}+C^{(H,i)}_{ikk}\bigr)
                 +w_g C^{(g,k)}_{ikk}}{2w_H+w_g},\\
 f_{iii}&=\frac{w_H C^{(H,i)}_{iii}+w_g C^{(g,i)}_{iii}}{w_H+w_g}.
 \label{eq:final-cubic-mean}
\end{align}
The weights are nonnegative and not both zero; $w_H$ is the weight
of each distinct Hessian-direction estimate. Different weights allow
for different numerical uncertainties in the two derivative estimators.
The implemented optional mean uses $w_H=w_g=1$, whereas the
Hessian-only mean corresponds to $w_g=0$.
For three distinct indices, the three Hessian displacement estimates
are averaged.  The two Hessian estimates of $F_{iijj}$ are likewise
averaged; no extrapolation across displacement steps is applied.
The common-field Gaussian\cite{gaussian} comparison below uses only the Hessian-direction
mean, after applying the Gaussian threshold to each directional
estimate; the gradient estimator remains an independent diagnostic.
The stored gradients and energies also permit independent checks of
diagonal fourth derivatives.  They are not substituted into the
quartic tensor, because the energy--gradient fourth derivative is
especially sensitive to cancellation in soft modes, as quantified
below.  Thus the quartic field retains the Hessian single-mode
stencil of Eq.~\eqref{eq:singlemodeqff}.
Thus the cubic field is complete, whereas the quartic potential
contains only the three-index semidiagonal elements $F_{ijkk}$ and
the value-only metric stencil supplies only $G_{ij,kk}$. The
standard analytic-metric route instead recovers all mixed $G_{ij,kl}$
from single-mode differences of $G_{ij,k}$, as specified below.
Four-distinct-index potential derivatives remain unavailable and
are recorded as unacquired rather than inferred from symmetry.
With analytic
dipole Jacobians $D_{ai}=\partial\mu_a/\partial q_i$, the same
single-mode differences recover $D_{ai,j}$ and the semidiagonal cubic
dipole derivatives $D_{ai,jj}$; with
dipole values alone they recover diagonal derivatives through
$D_{a,iii}$ using $q_i=\pm2h$, but no mixed dipole derivatives.
The reduced calculation retains Fermi polyads and omits quartic
Darling--Dennison couplings that this stencil cannot determine.
The available derivative set therefore defines the retained resonance
interactions.

Selected completeness tests use simultaneous displacements of two
normal modes. For $k\ne l$, the additional Hessian difference is
\begin{equation}
 F_{ij,kl}\simeq
 \frac{F_{ij}^{++}-F_{ij}^{+-}-F_{ij}^{-+}+F_{ij}^{--}}{4h^2},
 \label{eq:pairmodeqff}
\end{equation}
with $F_{ij}^{\sigma\tau}=F_{ij}(\sigma h e_k+\tau h e_l)$ for
$\sigma,\tau\in\{-1,+1\}$.
Together with the single-mode data, this supplies the complete
quartic energy field. Applying the same cross difference to analytic
dipole Jacobians supplies the complete cubic dipole field.
For the value-only metric controls, the same geometry-only pair
stencil completes $G_{ij,kl}$. The standard acquisition instead uses
SMITH's analytic Wilson matrix $B_{ia}=\partial s_i/\partial x_a$
and its Cartesian derivative $B'_{iac}$ to obtain, before modal scaling,
\begin{align}
 G_v^{ij}&=\sum_a B_{ia}m_a^{-1}B_{ja},\nonumber\\
 G_v^{ij}{}_{,k}&=\sum_{ac}m_a^{-1}
 (B'_{iac}B_{ja}+B_{ia}B'_{jac})J_{ck},\nonumber\\
 G_{ij,kl}&\simeq\frac{G_{ij,k}(+he_l)-G_{ij,k}(-he_l)}{2h}.
 \label{eq:analytic-metric-stencil}
\end{align}
Here each Cartesian component carries its atom's mass; the last
line is in the frozen dimensionless modal coordinate representation. The two derivative
orders $k,l$ are averaged. For the complete body-frame metric the
same derivative can be obtained from $G_{,k}=-Gg_{,k}G$, using the
analytic coordinate tangents and curvatures. Both routes retain the
rotational reduction; the Wilson expression supplies the internal
vibrational block, not a constant-metric Cartesian replacement.
Thus complete $G''$ requires no pair-mode electronic calculation
and no additional Hessian. The SI compare this acquisition with
value-only metric differences.

An electronic-structure program returning energies and Cartesian
gradients can supply the force field by
fitting, in the frozen normal-coordinate representation,
\begin{equation}
 E(q)=E_0+g_iq_i+\tfrac12F_{ij}q_iq_j
 +\tfrac16f_{ijk}q_iq_jq_k
 +\tfrac1{24}f_{ijkl}q_iq_jq_kq_l,
 \label{eq:eg-fit}
\end{equation}
to both $E(q)$ and $\partial E/\partial q_i$ at displaced geometries.
The geometry-dependent metric remains an independent calculation.
An independent energy--gradient-only acquisition example is reported
below, using a separate electronic field.

\section{Algebraic vibrational second order and resonance polyads}
Write the dimensionless vibrational Hamiltonian as
$H=H_0+\lambda H_1+\lambda^2H_2$, where $H_0$ contains the GF
harmonic oscillators, $H_1$ the complete cubic potential and
first metric-derivative operators, and $H_2$ the quartic potential,
second metric-derivative, and geometric scalar contributions. The
operators, in the ordered convention of Ref.~\onlinecite{mendolicchio2026},
are explicitly
\begin{align}
 H_0&=\tfrac12\sum_i\omega_i(p_i^2+q_i^2),\nonumber\\
 H_1&=\tfrac16 f_{ijk}q_iq_jq_k
       +\tfrac12 g_{ij,k}p_iq_kp_j,\nonumber\\
 H_2&=\tfrac1{24}f_{ijkl}q_iq_jq_kq_l
       +\tfrac14 g_{ij,kl}p_iq_kq_lp_j+V_g^{\rm eq},
 \label{eq:operators}
\end{align}
where $H$ denotes energy divided by $hc$, so its coefficients are
in wavenumbers; $q_i,p_i$ are dimensionless
canonical coordinates with $[q_i,p_j]=i\delta_{ij}$, and
$g_{ij,k}=\partial G_v^{ij}/\partial q_k$ and
$g_{ij,kl}=\partial^2G_v^{ij}/\partial q_k\partial q_l$ at the
reference. The force tensors $f$ are derivatives of $V$ with respect
to these same dimensionless modes. Equations~\eqref{eq:operators}--
\eqref{eq:chi-offdiag} below are written for ESALC-I. The rectilinear
ESALC-C reduction sets the internal-coordinate $g_{ij,k}$ and
$g_{ij,kl}$ to zero and retains the body-fixed Coriolis contribution
given explicitly below; removing $g$ without that contribution would
describe a different Hamiltonian.

For compact closed algebraic expressions, define the effective tensors
\begin{align}
 \eta_{ijkl}&=f_{ijkl}+g_{ij,kl}+g_{kl,ij},\nonumber\\
 b_{ijkl}&=f_{ijkl}+2g_{ij,kl}+4g_{ik,jl},\nonumber\\
 \sigma_{ijk}&=f_{ijk}-g_{ij,k}-g_{ik,j}-g_{jk,i},\nonumber\\
 \rho_{ijk}&=f_{ijk}-g_{ij,k}+g_{ik,j}+g_{jk,i}.
 \label{eq:effective-tensors}
\end{align}
Here $g$ denotes the frequency-scaled metric derivatives of the
dimensionless Hamiltonian, as in Ref.~\onlinecite{mendolicchio2026};
no unscaled internal-coordinate metric derivative is substituted in
Eq.~\eqref{eq:effective-tensors}. For a nonresonant level with
occupations $v_i$, the second-order energy is
\begin{equation}
 E_{\mathbf v}=E_0+\sum_i\omega_i v_i+
 \sum_{i\leq j}\chi_{ij}
 \left[v_iv_j+\tfrac12(v_i+v_j)\right].
 \label{eq:energy-algebraic}
\end{equation}
The zero-point term and the diagonal elements of the anharmonicity
matrix are
\begin{align}
 E_0={}&\tfrac12\sum_i\omega_i
 +\tfrac1{32}\sum_{ij}b_{iijj}\nonumber\\
 &-\sum_{ijk}\left[
 \frac{\sigma_{ijk}^2}{48(\omega_i+\omega_j+\omega_k)}
 +\frac{\rho_{kii}\rho_{kjj}}{32\omega_k}\right]
 +V_g^{\rm eq},
 \label{eq:zpe}\\
 16\chi_{ii}={}&\eta_{iiii}
 -\frac{\sigma_{iii}^2+9\rho_{iii}^2}{6\omega_i}\nonumber\\
 &-\frac12\sum_{j\ne i}\left[
 \frac{4\rho_{jii}^2}{\omega_j}
 +\frac{\sigma_{iij}^2}{2\omega_i+\omega_j}
 -\frac{\rho_{iij}^2}{2\omega_i-\omega_j}\right].
 \label{eq:chi-diag}
\end{align}
For $i\ne j$, the off-diagonal element is
\begin{widetext}
\begin{align}
4\chi_{ij}={}&\eta_{iijj}
-\frac12\left[
 \frac{\sigma_{iij}^2}{2\omega_i+\omega_j}
 +\frac{\rho_{iij}^2}{2\omega_i-\omega_j}
 +\frac{\sigma_{jji}^2}{2\omega_j+\omega_i}
 +\frac{\rho_{jji}^2}{2\omega_j-\omega_i}\right]
-\frac{\rho_{iii}\rho_{ijj}}{\omega_i}
-\frac{\rho_{jjj}\rho_{jii}}{\omega_j}\nonumber\\
&-\frac12\sum_{k\ne i,j}\left[
 \frac{\sigma_{ijk}^2}{\omega_i+\omega_j+\omega_k}
 -\frac{\rho_{ijk}^2}{\omega_i+\omega_j-\omega_k}
 +\frac{\rho_{ikj}^2}{\omega_i-\omega_j+\omega_k}
 -\frac{\rho_{jki}^2}{\omega_i-\omega_j-\omega_k}
 +\frac{2\rho_{kii}\rho_{kjj}}{\omega_k}\right].
\label{eq:chi-offdiag}
\end{align}
\end{widetext}
Equations~\eqref{eq:zpe}--\eqref{eq:chi-offdiag} are the curvilinear
algebraic VPT2 expressions of Ref.~\onlinecite{mendolicchio2026},
and are evaluated by TRINITY without a virtual-state sum. Terms with
small Fermi denominators are removed from these expressions and placed
in the corresponding variational polyad. The fundamental is then
$\nu_i=E_{1_i}-E_0=\omega_i+2\chi_{ii}
+\tfrac12\sum_{j\ne i}\chi_{ij}$ before polyad mixing.

The
curvilinear Fermi and Darling--Dennison matrix elements follow the
same contact-transformation framework as the reference
formulation~\cite{mendolicchio2026}. TRINITY forms accepted
1--2 Fermi and 1--1, 2--2, and 1--3 Darling--Dennison blocks from
one complete field. The full quartic tensor is retained where
the 2--2 and 1--3 classes require it. Each selected block is
diagonalized only after external nonresonant corrections are added.

Two definitions of a polyad are useful. A conserved polyad quantum
number $P(\mathbf v)=\sum_i c_i v_i$ exists only if its coefficient
vector satisfies $\mathbf c\cdot(\mathbf v_S-\mathbf v_R)=0$ for
every retained resonance; the integer null space of the resonance
vectors determines whether any nontrivial such label exists
\cite{krasnoshchekov2013polyad}. In the GVPT2 algorithm, a polyad
instead denotes a connected component of the graph whose vertices
are zero-order vibrational states and whose edges are \emph{selected}
resonant couplings\cite{martintaylor1997,rosnikpolik2014,yangbloino2022}.
Transitive membership does not imply a direct nonzero matrix element
between every pair, nor does a proximity window by itself create an
edge. We attach a conserved quantum number only when the resonance
vectors admit one; otherwise the connected component is the complete
operational definition.

For a complete-spectrum request, the target set contains every
fundamental and two-quantum state in the declared polyad-mode domain;
three-quantum targets are optional. A spectral-window or explicit-state
request instead converges only those specified bands. Other states may
enter their variational polyads as resonance partners without becoming
separate convergence targets. Large-amplitude torsions assigned to DVR
are excluded from this domain, so neither their overtones nor
combinations containing them enter a VP2TD polyad. Every state within
the declared total and per-mode occupation limits whose harmonic energy
falls inside a requested spectral window is a target, irrespective of
its zero-order intensity, because polyad
mixing can make an initially dark band observable. Each target seeds
an iteratively grown polyad; two quanta are therefore a minimum target
coverage, not a ceiling on interacting states. The default
variational limits are four quanta in total and four on an individual
mode; they are recorded as model truncations, and states reaching either
boundary are reported. Shell convergence at such a boundary applies
only within the capped variational model. These limits do not truncate the external perturbative corrections,
which are reduced analytically over the complete tensor support.
A $1:2$
Fermi pair first passes a harmonic-energy window and is then tested
using the so-called Martin energy criterion (Ref.~\onlinecite{martintaylor1997})
or the first-order wavefunction
mixing ratio $|V_{RS}|/|E_R^{(0)}-E_S^{(0)}|$; the coefficient test is
retained because energy screening alone can miss intensity borrowing
\cite{yangbloino2022,boyer2022}. The first accepted states seed a polyad. At
each expansion all states reachable through the allowed $1:2$,
$1:1$, $2:2$, and, where acquired, $1:3$ interactions are generated
as candidates. Only evaluated effective couplings that pass their
declared resonance tests become edges. The connected component is
diagonalized after each expansion, and the requested energies and
transported transition intensities of requested or newly bright bands
in that window are compared with the preceding
expansion. Expansion stops when the declared convergence
tolerances are met for the requested observables, or when no further
accepted neighbor exists; a computational limit marks the result as unconverged. The completed
component is then saved and the next requested polyad is treated.
This construction includes states reached only through a chain of
Darling--Dennison couplings. In the default settings, the harmonic windows are 200~cm$^{-1}$ for
Fermi and 100~cm$^{-1}$ for Darling--Dennison candidates. The Fermi
test accepts the energy index $|V_{RS}|^4/|\Delta_{RS}|^3\geq
1$~cm$^{-1}$ or a first-order mixing ratio
$|V_{RS}|/|\Delta_{RS}|\geq0.1$; the Darling--Dennison test uses
the effective contact-transformed coupling and its declared
normalization, with a 10~cm$^{-1}$ coupling cutoff or, for $1:1$
pairs, a mixing ratio of at least 0.3. These are reproducible
adjustable selection parameters
\cite{yangbloino2022}. The gap, coupling, and selection criterion are retained for each
accepted or rejected pair. The
starting states and convergence tolerances are stated separately from
computational resource limits. An optional intensity-aware selection uses
the actual occupation-dependent couplings, the exact
isolated-pair energy displacement, and both directed second-order
wavefunction coefficients for Darling--Dennison partners. The latter
retain contributions that cannot be inferred from the ratio of a
Hermitian effective coupling to the energy gap. Its equations, final
thresholds, direct-pair exclusion rules, and independent operator-matrix
checks are given in the SI. The reference comparisons use the
thresholds specified above. The following state-index form makes the
excluded perturbative channels explicit; transition
properties are evaluated by the reduced coefficient equations in the SI. For harmonic states $R,S$
in a selected polyad $P$, the Hermitian second-order matrix is
\begin{widetext}
\begin{equation}
 (H_{\rm VP2TD})_{RS}=
 E_R^{(0)}\delta_{RS}+(H_1)_{RS}+(H_2)_{RS}
 +\frac12\sum_{T\in\mathcal E_{RS}}(H_1)_{RT}(H_1)_{TS}
 \left[\frac1{E_R^{(0)}-E_T^{(0)}}
      +\frac1{E_S^{(0)}-E_T^{(0)}}\right].
 \label{eq:pt2d}
\end{equation}
\end{widetext}
Here $\mathcal E_{RS}$ contains the intermediate states whose
$R\leftrightarrow T$ and $S\leftrightarrow T$ channels are both
nonresonant. Membership of $T$ in the same connected polyad does not
remove its contribution when neither direct channel is resonant.
The first step excludes accepted Fermi and Darling--Dennison
channels from the perturbative denominators and supplies
the deperturbed diagonal energies and external corrections. The
second step restores their off-diagonal couplings inside $P$ and
diagonalizes $H_{\rm VP2TD}$,
$U^\dagger H_{\rm VP2TD}U=\operatorname{diag}(E_k)$.
The same eigenvectors $U$ mix the dressed transition moments below.
The algebraic $\chi$ matrix in Eqs.~\eqref{eq:chi-diag} and
\eqref{eq:chi-offdiag} supplies the nonresonant part of the first
step; Eq.~\eqref{eq:pt2d} specifies the restored polyad interactions.

For a comparison between two curvilinear coordinate representations with the same
harmonic reference, let their quadratic coordinate connection be
$q_i^{C}=q_i^{I}+\frac12\sum_{jk}A_{ijk}q_j^{I}q_k^{I}$ after linear
alignment of degenerate partners. Its anti-Hermitian oscillator
generator and the effective-Hamiltonian transport are
\begin{align}
 S&=-\frac{i}{4}\sum_{ijk}A_{ijk}\{q_jq_k,p_i\},\nonumber\\
 H_1^{C}&=H_1^{I}+[S,H_0],\nonumber\\
 H_2^{C}&=H_2^{I}+[S,H_1^{I}]
 +\tfrac12[S,[S,H_0]]+[T,H_0].
 \label{eq:polyad-gauge-transport}
\end{align}
Here $T$ is the next coordinate representation generator. For exactly degenerate
Darling--Dennison partners $R,S$,
$[T,H_0]_{RS}=(E_S^{(0)}-E_R^{(0)})T_{RS}=0$; therefore their
second-order matrix element follows from the quadratic connection.
For nondegenerate second-order couplings, the quadratic connection
alone is insufficient; the cubic coordinate transformation is retained below.
For the complete second-order property comparison we determine
the third-order coordinate transformation from the two body-fixed geometry maps,
\begin{align}
 q_i^C&=q_i^I+\tfrac12 A_{ijk}q_j^Iq_k^I\nonumber\\
 &\quad+\tfrac16 B_{ijkl}q_j^Iq_k^Iq_l^I+O(q^4),\nonumber\\
 q_i^I&=q_i^C+\tfrac12 U_{i,jk}q_j^Cq_k^C\nonumber\\
 &\quad+\tfrac16 U_{i,jkl}q_j^Cq_k^Cq_l^C+O(q^4),\nonumber\\
 U_{i,jk}&=-A_{ijk},\nonumber\\
 U_{i,jkl}&=-B_{ijkl}+A_{ijm}A_{mkl}\nonumber\\
 &\quad+A_{ikm}A_{mjl}+A_{ilm}A_{mjk}.
 \label{eq:chart-cubic-inverse}
\end{align}
Repeated mode indices are summed. Before the final orthogonal
alignment of degenerate partners, the dipole derivatives become
\begin{align}
 \bar D_{\alpha i}&=D^I_{\alpha i},\nonumber\\
 \bar D_{\alpha ij}&=D^I_{\alpha ij}
                         +D^I_{\alpha m}U_{m,ij},\nonumber\\
 \bar D_{\alpha ijk}&=D^I_{\alpha ijk}
 +D^I_{\alpha mi}U_{m,jk}\nonumber\\
 &\quad+D^I_{\alpha mj}U_{m,ik}
 +D^I_{\alpha mk}U_{m,ij}\nonumber\\
 &\quad+D^I_{\alpha m}U_{m,ijk}.
 \label{eq:chart-property-second}
\end{align}
The potential is composed with the same inverse map through quartic
order. For the contravariant metric we use
$G^C_{ij}(q^C)=J_{ia}G^I_{ab}(q^I(q^C))J_{jb}$, where
$J_{ia}=\partial q_i^C/\partial q_a^I$; its expansion supplies
$G'$ and $G''$ through second order. The dipole tensors, potential,
and metric are then rotated by the same harmonic partner alignment
and supplied to VP2TD. The cubic jet uses geometry evaluations only,
without extra electronic-structure points.
In practice, writing $X^I(q^I)=X^C(q^C(q^I))$ for the two aligned
Cartesian geometry maps and $J^C_{ai}=\partial X_a^C/\partial q_i^C$,
the cubic tensor is obtained by projection of
\begin{align}
 J^C_{ai}B_{ijkl}&=X^I_{a,jkl}-X^C_{a,jkl}\nonumber\\
 &\quad-X^C_{a,jm}A_{mkl}
 -X^C_{a,km}A_{mjl}\nonumber\\
 &\quad-X^C_{a,lm}A_{mjk},
 \label{eq:chart-cubic-geometry}
\end{align}
with the repeated Cartesian and mode indices contracted as usual.
The selected Fermi and degenerate Darling--Dennison polyad is
completed under every operation of the supplied finite real point
group before Eq.~\eqref{eq:pt2d} is assembled. A mass-weighted
Cartesian projector fixes a reproducible real frame for each
degenerate normal-mode subspace, and the full group average of
Eq.~\eqref{eq:nh3-real-projection} enforces symmetry on the resulting
Hamiltonian. This construction depends on the group representation,
not on a formula specialized to $C_{3v}$.

\subsection{Non-Abelian groups in a real canonical basis}
The canonical construction of Ref.~\onlinecite{mendolicchio2021} also applies to linear, symmetric,
and spherical tops. We retain real Cartesian components for each
two- or threefold degenerate vibration and use the same real
oscillator basis and VP2TD contractions as for Abelian groups.
Degenerate partners must be acquired in one consistent orientation,
and the variational basis must contain each complete degenerate
manifold. The couplings responsible for $S$-type splitting are
retained in that manifold even without an accidental Fermi
resonance; an energy-window or coupling-strength test intended for
accidental resonances cannot discard them.

Only after diagonalization do we convert to the conventional polar
or spherical labels. If $P$ is the block-unitary map from the real
canonical basis to the standard degenerate-oscillator basis, then
\begin{align}
 H_{\rm std}&=P^\dagger H_{\rm can}P,\nonumber\\
 M^{[r]}_{\rm std}&=P_f^\dagger M^{[r]}_{\rm can}P_i,
 \qquad r=0,1,2.
 \label{eq:nonabelian-postrotation}
\end{align}
The second relation applies to both ground-origin transitions and
hot bands, for which the initial manifold may itself be degenerate.
It also applies componentwise to the polarizability tensor. The
standard-basis transition moments may be complex, but no complex
force constants, metric tensors, or special non-Abelian perturbative
formulas enter the canonical calculation. Unitarity preserves the
polyad eigenvalues and the summed IR or Raman strengths over a
complete degenerate manifold; individual standard-basis lines are
obtained from the rotated moments before taking their squared
moduli. This is the representation strategy for both ESALC-C and
ESALC-I. Its numerical validation on non-Abelian molecules requires
complete degeneracy blocks and is separate from the Abelian
examples reported below.

\subsection{Explicit rectilinear ESALC-C limit}
Let $\mathbf e_{ai}$ be orthonormal mass-weighted Cartesian mode
vectors, $\zeta_{ij}^{\alpha}=\sum_a(\mathbf e_{ai}\times
\mathbf e_{aj})_\alpha=-\zeta_{ji}^{\alpha}$ their Coriolis
coefficients, and $\mu_{\alpha\beta}^{(0)}$ the equilibrium inverse
effective inertia tensor. In atomic units, with the dimensionless
$q_i,p_i$ of Eq.~\eqref{eq:operators}, the vibrational angular momentum
and the $J=0$ Coriolis operator are
\begin{align}
 \pi_\alpha&=\sum_{ij}\zeta_{ij}^{\alpha}
       \sqrt{\frac{\omega_j}{\omega_i}}\,q_i p_j,\nonumber\\
 H_{\rm Cor}^{(2)}&=\tfrac12\sum_{\alpha\beta}
       \mu_{\alpha\beta}^{(0)}\pi_\alpha\pi_\beta .
 \label{eq:coriolis-explicit}
\end{align}
The product ordering in Eq.~\eqref{eq:coriolis-explicit} is retained;
in particular, its scalar commutators must not be discarded. Thus the
rectilinear operators are $H_1^{C}=f_{ijk}q_iq_jq_k/6$ and
$H_2^{C}=f_{ijkl}q_iq_jq_kq_l/24+H_{\rm Cor}^{(2)}+U_W^{(0)}$,
where $U_W^{(0)}=-\operatorname{tr}\mu^{(0)}/8$ is the equilibrium
Watson scalar in this convention~\cite{watson1968}. Accordingly
$\sigma^C_{ijk}=\rho^C_{ijk}=f_{ijk}$ and
$\eta^C_{ijkl}=b^C_{ijkl}=f_{ijkl}$, but the Coriolis term must
still be added to the quartic operator and to polyad matrices.

For clarity, define $\kappa_{ij}=\sum_{\alpha\beta}
\mu_{\alpha\beta}^{(0)}\zeta_{ij}^{\alpha}\zeta_{ij}^{\beta}$.
The diagonal contractions of Eq.~\eqref{eq:coriolis-explicit} give
the explicit additions to the nonresonant energy parameters,
\begin{align}
 \Delta\chi_{ij}^{\rm Cor}
  &=\frac{\kappa_{ij}}2
    \left(\frac{\omega_i}{\omega_j}
         +\frac{\omega_j}{\omega_i}\right),\quad i<j,\nonumber\\
 \Delta\chi_{ii}^{\rm Cor}&=0,\nonumber\\
 \Delta E_0^{\rm Cor}
  &=\sum_{i<j}\frac{\kappa_{ij}}2
    \left[\frac14\left(\frac{\omega_i}{\omega_j}
                       +\frac{\omega_j}{\omega_i}\right)
          -\frac12\right]+U_W^{(0)}.
 \label{eq:coriolis-chi}
\end{align}
The off-diagonal Coriolis matrix elements are retained in
Eq.~\eqref{eq:pt2d} for accepted resonance polyads. In a numerical
implementation that already uses the complete Eckart rovibrational
metric of Eq.~\eqref{eq:metric}, its rotational reduction contains
these same terms; Eq.~\eqref{eq:coriolis-explicit} is the equivalent
rectilinear representation, not an additional correction to be
counted twice.
The displaced-geometry I/C comparisons use the complete Eckart
metric assembled from vibrational and rotational tangents. Its
derivatives supply the rotational reduction without adding
Eq.~\eqref{eq:coriolis-explicit} a second time. The common-field
Cartesian reconstructions instead use the analytic Coriolis tensor
built from the signed normal modes and rotational constants. The numerical
contraction and data are given in the SI.

\section{Unified anharmonic transition-property formalism}
We use one second-order transition-property construction for IR,
Raman, vibrational circular dichroism (VCD), and Raman optical
activity (ROA). Building on the general rectilinear formulas of
Ref.~\onlinecite{willetts1990}, their algebraic reduction for
fundamentals in Ref.~\onlinecite{vazquez2006}, and the
transition-specific state/property separation of
Ref.~\onlinecite{bloino2015}, let $P_\gamma$ denote one Cartesian component of
an electric dipole, polarizability, magnetic dipole, optical-activity,
or dipole--quadrupole response tensor.  With the same nonresonant
wave operators and resonant polyad eigenvectors for every component,
the normalized transition amplitude is
\begin{align}
 {\cal M}^{P}_{\gamma,KL}
 &={\sum}_{r=0}^{2}{\cal M}^{P,[r]}_{\gamma,KL}\nonumber\\
 &={\sum}_{fa}U^*_{fK}\left[\sum_{r=0}^{2}M^{P,[r]}_{\gamma,fa}\right]U_{aL}
 +O(\lambda^3).
 \label{eq:unified-property-transport}
\end{align}
Here $K$ and $L$ label final and initial polyad roots; the
perturbation parameter $\lambda$ counts one cubic force or one
additional order of a property surface.  The same formula includes
cold bands, hot bands, overtones, and three-quantum transitions.
For coordinate-valued properties the three surface orders are
$P_{\gamma,i}q_i$, $P_{\gamma,ij}q_iq_j/2$, and
$P_{\gamma,ijk}q_iq_jq_k/6$.  The magnetic dipole is time odd and
contains momenta: its coefficient surface is contracted with
$\{B_{\gamma,i}(q),p_i\}/2$, preserving the indicated ordering.
The SI provide the complete multimode reduction through second order
as explicit rational coefficient equations in the force constants,
metric derivatives, and property derivatives. These include ESALC-I
kinetic terms and both-state polyad transport, with specialized formulas
for fundamentals, two- and three-quantum transitions, hot bands, and
diagonal expectations. The coefficient indices are finite
tensor multi-indices, not intermediate vibrational states. Field
coefficients are reused across properties and transitions. Selected
resonant pairs use occupation-dependent reduced coefficient equations
with direct-pair selection factors, folded Fermi terms, and both-state normalization
before polyad mixing. Thus the operational path is algebraically reduced
also for resonances. We verify the reduced formulas against the separate
full contraction
on the same tensors, including random two- and three-mode fields,
excited initial occupations, and both coordinate and momentum
properties. The SI also give separate hot-band equations
for ESALC-I and ESALC-C, the explicit Coriolis correction to the
second-order property coefficients, and their direct reconstruction
from the same $\rho,\sigma,\eta,b$ tensors used for the energies. Independent dense two-mode and sparse three-mode
oscillator matrices additionally test the reduced amplitudes and
the ordering of the kinetic and magnetic terms.

For IR specifically, in dimensionless GF normal coordinates, expand
the electric dipole
through the cubic surface,
\begin{equation}
 \boldsymbol\mu(q)=\boldsymbol\mu_0+D_iq_i
 +\frac12D_{ij}q_iq_j+\frac16D_{ijk}q_iq_jq_k+\cdots.
 \label{eq:dipole}
\end{equation}
The same ESALC-C or ESALC-I coordinate representation and Eckart convention must be used
for $D_i$, $D_{ij}$, and $D_{ijk}$. For an isolated harmonic
fundamental, $M^{(0)}_i=D_i/\sqrt2$ and
\begin{equation}
 I_i^{\rm harm}=C_{\rm IR}\,\omega_i
       \lVert D_i\rVert^2/2.
 \label{eq:harmir}
\end{equation}
The leading direct electrical moments are
$D_{\alpha ii}/(2\sqrt2)$ for an overtone and $D_{\alpha ij}/2$
for a binary combination with $i\ne j$. Mechanical mixing adds to
these moments. Writing $\mathbf{ij}=\mathbf e_i+\mathbf e_j$,
the explicit first-order combination-band amplitude in the same
finite algebraic contractions used below is
\begin{align}
 {\cal M}^{\rm comb,[1]}_{ij,\alpha}
 ={}&\frac{D_{\alpha ij}}2
 +\sum_m {\cal R}_{m,\mathbf{ij}}
       {\cal C}_{m,\mathbf{ij}}^*
       {\cal D}^{1}_{\alpha,m0}\nonumber\\
 &+\sum_m {\cal D}^{1}_{\alpha,\mathbf{ij}m}
       {\cal R}_{m0}{\cal C}_{m0},\qquad i\ne j,
 \label{eq:combination-leading}\\
 I^{\rm comb,[2]}_{ij,0}
 ={}& C_{\rm IR}\nu_{ij,0}
       \sum_\alpha|{\cal M}^{\rm comb,[1]}_{ij,\alpha}|^2.
 \label{eq:combination-intensity}
\end{align}
Here $\mathcal C$, $\mathcal D^1$, and the nonresonant scalar
$\mathcal R$ are the finite Kronecker-delta and square-root
expressions in Eqs.~\eqref{eq:intensity-cubic-components},
\eqref{eq:intensity-dipole-components}, and
\eqref{eq:intensity-denominators}; if $\mathbf{ij}$ is in a
resonance polyad, its dressed amplitude is subsequently mixed by
Eq.~\eqref{eq:polyadir}. The quartic force tensor is equivalently expressed
through the effective curvilinear tensors by
\begin{align}
 f_{ijkl}&=\tfrac12(3\mathcal S\eta_{ijkl}
                         -\mathcal S b_{ijkl}),\nonumber\\
 \mathcal S T_{ijkl}&=\frac1{24}\sum_{\pi\in S_4}
 T_{\pi(i)\pi(j)\pi(k)\pi(l)}.
 \label{eq:quartic-eta-b}
\end{align}
This identity retains the complete second metric derivative in
the energy and intensity coefficients below. Accepted Fermi and
Darling--Dennison channels are excluded from perturbative
denominators and restored in a variational polyad. If $U_{fk}$
diagonalizes that polyad, the observable intensity is
\begin{equation}
 I_{k0}=C_{\rm IR}\,\nu_{k0}
 \sum_{\alpha=x,y,z}\left|
 \sum_{f\in{\cal P}_k}U_{fk}^{*}{\cal M}_{f0,\alpha}
 \right|^2 .
 \label{eq:polyadir}
\end{equation}
The state-index derivation of the complete fundamental and overtone
amplitudes follows. Its coefficient-reduced evaluation, including
resonant pairs, is given explicitly in the SI; the state-index
contraction is retained as a verification reference.
Let $\mathbf m,\mathbf n$ be nonnegative occupation vectors. For an
ordered list of mode indices $\boldsymbol\iota=(i_1,\ldots,i_r)$ and
signs $\boldsymbol\epsilon=(\epsilon_1,\ldots,\epsilon_r)$, define
the numerical coefficient
\begin{widetext}
\begin{align}
 {\cal T}_{\mathbf m\mathbf n}
       (\boldsymbol\iota;\boldsymbol\epsilon)
 &=
 \delta_{\mathbf m,\,\mathbf n+\sum_{t=1}^r
                       \epsilon_t\mathbf e_{i_t}}
 \prod_{t=1}^{r}
 \sqrt{n_{i_t}^{(t)}+\frac{1+\epsilon_t}{2}},
 \qquad
 \mathbf n^{(t)}=\mathbf n+
          \sum_{u=t+1}^{r}\epsilon_u\mathbf e_{i_u},
 \nonumber\\
 {\cal L}_{\mathbf m\mathbf n}
       (\boldsymbol\iota;\mathbf h)
 &=
 2^{-r/2}
 \sum_{\epsilon_1,\ldots,\epsilon_r=\pm1}
 \left[\prod_{t=1}^{r}(i\epsilon_t)^{h_t}\right]
 {\cal T}_{\mathbf m\mathbf n}
       (\boldsymbol\iota;\boldsymbol\epsilon),
 \qquad h_t\in\{0,1\}.
 \label{eq:ladder-explicit}
\end{align}
\end{widetext}
Here $\mathbf h$ is a binary index specifying the kinetic factors;
$\mathbf h=\mathbf0$ gives purely coordinate contractions. Terms
with negative intermediate occupations are zero. Equations below
contain only these finite Kronecker-delta and square-root
coefficients, force and metric tensors, and dipole derivatives.
The full cubic and quartic coefficients are
\begin{widetext}
\begin{align}
 {\cal C}_{mn}
 ={}&\frac{\sigma_{ijk}}{6\sqrt8}
   \bigl[{\cal T}_{mn}((i,j,k);(+,+,+))
        +{\cal T}_{mn}((i,j,k);(-,-,-))\bigr]\nonumber\\
 &+\frac{\rho_{ijk}}{2\sqrt8}
   \bigl[{\cal T}_{mn}((i,j,k);(+,+,-))
        +{\cal T}_{mn}((k,j,i);(+,-,-))\bigr]
  +\frac{\rho_{ijj}}{4\sqrt2}
   \bigl[{\cal T}_{mn}((i);(+))
        +{\cal T}_{mn}((i);(-))\bigr],
 \label{eq:intensity-cubic-components}\\
 {\cal B}^{I}_{mn}
 ={}&\frac{3{\cal S}\eta_{ijkl}-{\cal S}b_{ijkl}}{48}
         {\cal L}_{mn}((i,j,k,l);(0,0,0,0))
   +\frac{g_{ij,kl}}4
         {\cal L}_{mn}((i,k,l,j);(1,0,0,1))
   +V_g^{\rm eq}\delta_{mn},\nonumber\\
 {\cal B}^{C}_{mn}
 ={}&\frac{f_{ijkl}}{24}
         {\cal L}_{mn}((i,j,k,l);(0,0,0,0))
   +\frac{\mu_{\beta\gamma}^{(0)}
          \zeta_{ij}^{\beta}\zeta_{kl}^{\gamma}}2
       \sqrt{\frac{\omega_j\omega_l}{\omega_i\omega_k}}\,
         {\cal L}_{mn}((i,j,k,l);(0,1,0,1))
   +U_W^{(0)}\delta_{mn},
 \label{eq:intensity-quartic-components}\\
 {\cal D}^{1}_{\alpha,mn}
 ={}&D_{\alpha i}{\cal L}_{mn}((i);(0)),\qquad
 {\cal D}^{2}_{\alpha,mn}
  =\tfrac12D_{\alpha ij}{\cal L}_{mn}((i,j);(0,0)),\qquad
 {\cal D}^{3}_{\alpha,mn}
  =\tfrac16D_{\alpha ijk}
              {\cal L}_{mn}((i,j,k);(0,0,0)).
 \label{eq:intensity-dipole-components}
\end{align}
Repeated mode indices are summed. ESALC-I uses
${\cal B}={\cal B}^{I}$ and the curvilinear $\sigma,\rho$;
ESALC-C uses ${\cal B}={\cal B}^{C}$ and $\sigma=\rho=f_{ijk}$.
Write $m\sim n$ when $m=n$ or when $(m,n)$ is a directly
selected resonant pair, and define $\mathcal D_n=\{m:m\sim n\}$.
This neighborhood is distinct from the full diagonalization block
$\mathcal P_n$. The scalar denominators and retained internal
cubic coefficients are
\begin{align}
 {\cal R}_{mn}&=
 \begin{cases}
 (E_n^{(0)}-E_m^{(0)})^{-1},&m\not\sim n,\\
 0,&m\sim n,
 \end{cases}\nonumber\\
 {\cal C}^{d}_{mn}&=\mathbf1_{m\sim n}{\cal C}_{mn}.
 \label{eq:intensity-denominators}
\end{align}
For the fundamental of mode $i$, write $\mathbf i=\mathbf e_i$.
Parity removes the first-order cubic--linear-dipole contribution.
The entire nonresonant amplitude through second order is
\begin{align}
 {\cal M}^{\rm fund}_{i\alpha}
={}&\frac{D_{\alpha i}}{\sqrt2}
 +\frac{D_{\alpha iii}+\sum_{j\ne i}D_{\alpha ijj}}{4\sqrt2}
 \nonumber\\
&+\sum_m\left[
 {\cal R}_{m\mathbf i}{\cal C}_{m\mathbf i}^{*}
                         {\cal D}^{2}_{\alpha,m0}
 +{\cal D}^{2}_{\alpha,\mathbf i m}
                         {\cal R}_{m0}{\cal C}_{m0}
 \right]\nonumber\\
&+\sum_m{\cal R}_{m\mathbf i}
 \left[
 {\cal B}_{m\mathbf i}^{*}
 +\sum_t{\cal C}_{mt}^{*}{\cal R}_{t\mathbf i}
                           {\cal C}_{t\mathbf i}^{*}
 -\sum_t{\cal R}_{mt}{\cal C}_{mt}^{*}
                           ({\cal C}^{d}_{t\mathbf i})^{*}
 \right]{\cal D}^{1}_{\alpha,m0}
 \nonumber\\
&+\sum_m{\cal D}^{1}_{\alpha,\mathbf i m}{\cal R}_{m0}
 \left[
 {\cal B}_{m0}
 +\sum_t{\cal C}_{mt}{\cal R}_{t0}{\cal C}_{t0}
 -\sum_t{\cal R}_{mt}{\cal C}_{mt}{\cal C}^{d}_{t0}
 \right]\nonumber\\
&+\sum_{mn}{\cal R}_{m\mathbf i}{\cal C}_{m\mathbf i}^{*}
      {\cal D}^{1}_{\alpha,mn}
      {\cal R}_{n0}{\cal C}_{n0}
 \nonumber\\
&-\frac12\sum_{g\in{\cal D}_{\mathbf i}}\sum_t
 {\cal R}_{t\mathbf i}{\cal C}_{t\mathbf i}^{*}
 {\cal R}_{tg}{\cal C}_{tg}
 {\cal D}^{1}_{\alpha,g0}
 -\frac12\sum_{g\in{\cal D}_0}\sum_t
 {\cal D}^{1}_{\alpha,\mathbf i g}
 {\cal R}_{tg}{\cal C}_{tg}^{*}
 {\cal R}_{t0}{\cal C}_{t0}.
 \label{eq:fundamental-full-amplitude}\\
 I_{i0}^{\rm fund}
 ={}&C_{\rm IR}\,\nu_{i0}
       \sum_{\alpha=x,y,z}
       \left|{\cal M}^{\rm fund}_{i\alpha}\right|^2 .
 \label{eq:fundamental-full-intensity}
\end{align}
\end{widetext}
Every coefficient in Eq.~\eqref{eq:fundamental-full-amplitude}
is an explicit tensor contraction from
Eqs.~\eqref{eq:ladder-explicit}--\eqref{eq:intensity-denominators};
there is no unspecified transition operator. The fourth and fifth
lines include the quartic potential, the second metric derivative
for ESALC-I, or the Coriolis term for ESALC-C. The last line is
the full second-order normalization contribution.
For the overtone $0\rightarrow 2\mathbf e_i$, excitation parity
eliminates the second-order amplitude: the complete amplitude at
the same perturbative order consists of the direct quadratic-dipole
term and the two first-order cubic mixing terms. Performing their
algebraic contractions and the finite occupation sums gives
\begin{widetext}
\begin{align}
 {\cal M}^{\rm over}_{i\alpha}
 ={}&\frac{D_{\alpha ii}}{2\sqrt2}
 +\frac{D_{\alpha i}}{\sqrt2\,\omega_i}
       \left(\frac{\rho_{iii}}4-\frac{\sigma_{iii}}{12}\right)
 \nonumber\\
 &+\sum_{j\ne i}\frac{D_{\alpha j}}{4\sqrt2}
   \left[
    \rho_{iij}\,
       \frac{1-\mathbf1_{\mathbf e_j\sim2\mathbf e_i}}
            {2\omega_i-\omega_j}
    -\frac{\sigma_{iij}}{2\omega_i+\omega_j}
   \right],\label{eq:overtone-full-amplitude}\\
 I_{2_i,0}^{\rm over}
 ={}&C_{\rm IR}\,\nu_{2_i,0}
       \sum_{\alpha=x,y,z}
       \left|{\cal M}^{\rm over}_{i\alpha}\right|^2 .
 \label{eq:overtone-full-intensity}
\end{align}
\end{widetext}
The fraction with $2\omega_i-\omega_j$ is defined to be zero
when $\mathbf e_j$ and $2\mathbf e_i$ are a directly accepted
Fermi pair; that coupling is then restored by its variational
eigenvectors. Equation~\eqref{eq:overtone-full-amplitude} includes
both the internal-coordinate $\rho,\sigma$ kinetic corrections and,
by setting $\rho=\sigma=f$, the ESALC-C limit. A quartic or Coriolis
term cannot change the even-parity overtone amplitude at this
order, although it changes its transition frequency and the
variational mixing. For a mixed polyad, the observable intensity is
obtained by inserting the explicit fundamental and overtone
amplitudes from Eqs.~\eqref{eq:fundamental-full-amplitude} and
\eqref{eq:overtone-full-amplitude} into Eq.~\eqref{eq:polyadir}
before taking the squared modulus.
The intensity expression may be truncated consistently at second
perturbative order; the implementation also reports the squared
intensity of the summed second-order moment. Both conventions use
the same $U$ as the energy calculation.
More explicitly, for a transition with the GVPT2 energy difference
$\nu_{k0}$, the order-resolved intensity is
\begin{align}
 I_{k0}^{[0]}&=C_{\rm IR}\nu_{k0}|M^{(0)}_{k0}|^2,\\
 I_{k0}^{[1]}&=2C_{\rm IR}\nu_{k0}
 \operatorname{Re}(M^{(0)*}_{k0}\!\cdot M^{(1)}_{k0}),\\
 I_{k0}^{[2]}&=C_{\rm IR}\nu_{k0}
 \bigl(|M^{(1)}_{k0}|^2+
 2\operatorname{Re}(M^{(0)*}_{k0}\!\cdot M^{(2)}_{k0})\bigr).
 \label{eq:intensityorders}
\end{align}
\subsection{Raman property extension}
The same second-order property transport applies componentwise to
the six independent entries of the electronic polarizability
tensor, acquired from Cartesian electronic-structure program derivatives and transformed
to either ESALC normal-mode coordinate representation. For any initial and final VP2TD
roots, including a hot band, let
${\cal A}_{if,\alpha\beta}=
\langle\Psi_f|\hat\alpha_{\alpha\beta}|\Psi_i\rangle$
after dressing both states and applying their polyad eigenvectors.
For a symmetric transition tensor, real in the canonical basis but
possibly complex after conversion of degenerate partners, the Raman
activity is
\begin{align}
 \bar\alpha_{if}&=\tfrac13({\cal A}_{xx}+{\cal A}_{yy}
                         +{\cal A}_{zz}),\nonumber\\
 \gamma_{if}^{2}&=\tfrac12\bigl[|{\cal A}_{xx}-{\cal A}_{yy}|^2
 +|{\cal A}_{yy}-{\cal A}_{zz}|^2
 +|{\cal A}_{zz}-{\cal A}_{xx}|^2\nonumber\\
 &\hspace{22mm}+6(|{\cal A}_{xy}|^2+|{\cal A}_{xz}|^2
 +|{\cal A}_{yz}|^2)\bigr],\nonumber\\
 S_{if}^{\rm R}&=(45|\bar\alpha_{if}|^2+7\gamma_{if}^{2})
 a_0^6.
 \label{eq:raman-activity}
\end{align}
Here $a_0$ denotes the Bohr radius; the code reports
$S_{if}^{\rm R}$ as an activity in \AA$^6$.
Laser-frequency and thermal population factors needed
for an experimental Raman intensity are separate. The present
common-field Gaussian comparison uses the static Raman activity
defined in Eq.~\eqref{eq:raman-activity}.

\subsection{Chiroptical transition tensors}
For a chiral molecule, the same second-order state and polyad
transport can act on property tensors beyond the electric dipole.
Writing the electric and magnetic transition moments as
$\boldsymbol{\mu}_{if}$ and $\boldsymbol{m}_{if}$ in one common gauge,
the VCD rotational strength is
\begin{equation}
 R_{if}=\operatorname{Im}\!\left[
 \sum_{a}\mu_{if,a}m_{if,a}^{*}\right].
 \label{eq:vcd-rotatory}
\end{equation}
The magnetic transition operator is time-odd and contains nuclear
momenta; its coordinate-dependent coefficient must therefore be
differentiated separately from the electric property surface.
For off-resonant Raman optical activity with scattered circular
polarization in backscattering [SCP(180)], let
$\mathcal A$, $\mathcal G'$ and $\mathcal Q$ denote the transported
transition tensors for the dynamic electric polarizability,
electric--magnetic optical activity, and electric dipole--quadrupole
polarizability, respectively, at laser frequency $\omega_L$.
With Cartesian indices $a,b,c,d$ and the Levi--Civita tensor
$\epsilon_{abc}$, define
\begin{align}
 \bar\alpha&=\tfrac13\sum_a\mathcal A_{aa},\nonumber\\
 \beta^2(\mathcal A)&=\tfrac12\operatorname{Re}\!\left[
 3\sum_{ab}\mathcal A_{ab}\mathcal A_{ab}^{*}
 -\bigl|\sum_a\mathcal A_{aa}\bigr|^2\right],\nonumber\\
 \beta(\mathcal G')&=\tfrac12\operatorname{Re}\!\left[
 3\sum_{ab}\mathcal A_{ab}\mathcal G_{ab}^{\prime *}
 -\sum_a\mathcal A_{aa}\sum_b\mathcal G_{bb}^{\prime *}\right],\nonumber\\
 \beta(\mathcal Q)&=\tfrac{\omega_L}{2}\operatorname{Re}\!\left[
 \sum_{abcd}\mathcal A_{ad}\epsilon_{abc}
                  \mathcal Q_{bcd}^{*}\right].
 \label{eq:roa-invariants}
\end{align}
In the tensor and unit convention used here, the common-prefactor-free
Raman and ROA activities and their circular intensity difference (CID) are
\begin{align}
 S_{if}^{\rm R}&=45|\bar\alpha|^2+7\beta^2(\mathcal A),\nonumber\\
 S_{if}^{\rm ROA}&=\frac{2}{c}\left[12\beta(\mathcal G')
                                   +4\beta(\mathcal Q)\right],\nonumber\\
 {\rm CID}_{if}&=S_{if}^{\rm ROA}/S_{if}^{\rm R}.
 \label{eq:roa-scp180}
\end{align}
All three response tensors must use the same electronic method,
origin and gauge; Gaussian's stored frequency-normalized optical
and quadrupole tensors are converted before contraction.  The
anharmonic transition-amplitude program accepts these tensor surfaces
and transports their elements through the same polyad eigenvectors
used for the energies.  The molecular chiroptical control below
checks the harmonic response tensors and invariant contractions.
\subsection{Three-quantum transitions and hot bands}
Matrix and contact-transformation approaches can evaluate transitions
from excited vibrational states without reducing their matrix elements
to transition-specific coefficient formulas.  The equations below
provide that algebraic specialization as closed second-order amplitudes,
separated by perturbative order, and apply the polyad transformations of both the
initial and final states.  The curvilinear kinetic terms have no
rectilinear analogue.
The finite occupation basis also contains the three distinct classes
$3\mathbf e_i$, $2\mathbf e_i+\mathbf e_j$ ($i\ne j$), and
$\mathbf e_i+\mathbf e_j+\mathbf e_k$ (all indices distinct).
For a transition from the ground state to one of these states, the
zeroth- and first-order dipole moments vanish by harmonic selection rules and perturbative parity.
The direct electric part of the second-order amplitude is explicitly
\begin{equation}
\begin{aligned}
 {\cal D}^{3}_{\alpha,3_i,0}
   &=\frac{\sqrt3}{12}D_{\alpha iii},\\
 {\cal D}^{3}_{\alpha,2_i1_j,0}
   &=\frac14D_{\alpha iij},\qquad i\ne j,\\
 {\cal D}^{3}_{\alpha,1_i1_j1_k,0}
   &=\frac{D_{\alpha ijk}}{2\sqrt2},\qquad i,j,k\ \mathrm{distinct}.
\end{aligned}
\label{eq:three-quantum-direct}
\end{equation}
Mechanical, mixed, and variational contributions must be added; the
direct cubic dipole alone does not define the three-quantum intensity.
The ESALC-C limit agrees with the state classes and 1--3 resonance
channels described by the established rectilinear treatment.\cite{bloino2015}
The same state
construction applies to ESALC-I, with its metric terms retained.

Hot bands require a more general initial occupation $a$ instead of
the ground state. We take $a=\mathbf e_i$ here and let $f$ be any
higher-energy state represented in the VP2TD basis. Equations
\eqref{eq:ladder-explicit}--\eqref{eq:intensity-denominators} already
specify every numerical coefficient, so the following expressions
contain no unresolved creation or annihilation operators. Define the
external first- and second-order amplitudes and their overlap by
\begin{widetext}
\begin{align}
 A_{ma}&={\cal R}_{ma}{\cal C}_{ma},\nonumber\\
 B_{ma}&={\cal R}_{ma}\left[\mathcal B_{ma}
       +\sum_t {\cal C}_{mt}A_{ta}
       -\sum_{g\in\mathcal D_a}A_{mg}{\cal C}^{d}_{ga}\right],\nonumber\\
 N_{ab}&=\sum_t A^*_{ta}A_{tb}.
 \label{eq:hot-external-amplitudes}
\end{align}
\end{widetext}
The resonant denominators are zero by the definition of ${\cal R}$;
their couplings occur only in the finite VP2TD Hamiltonian. The
complete transition moment through second perturbative order is
\begin{widetext}
\begin{align}
 M^{[0]}_{fa,\alpha}
 ={}&{\cal D}^{1}_{\alpha,fa},\nonumber\\
 M^{[1]}_{fa,\alpha}
 ={}&{\cal D}^{2}_{\alpha,fa}
 +\sum_m A^*_{mf}{\cal D}^{1}_{\alpha,ma}
 +\sum_n {\cal D}^{1}_{\alpha,fn}A_{na},\nonumber\\
 M^{[2]}_{fa,\alpha}
 ={}&{\cal D}^{3}_{\alpha,fa}
 +\sum_m A^*_{mf}{\cal D}^{2}_{\alpha,ma}
 +\sum_n {\cal D}^{2}_{\alpha,fn}A_{na}
 +\sum_m B^*_{mf}{\cal D}^{1}_{\alpha,ma}
 +\sum_n {\cal D}^{1}_{\alpha,fn}B_{na}\nonumber\\
 &+\sum_{mn}A^*_{mf}{\cal D}^{1}_{\alpha,mn}A_{na}
 -\frac12\sum_{g\in\mathcal D_f}N_{fg}{\cal D}^{1}_{\alpha,ga}
 -\frac12\sum_{g\in\mathcal D_a}{\cal D}^{1}_{\alpha,fg}N_{ga}.
 \label{eq:hot-pair-moment}
\end{align}
\end{widetext}
For real force and property tensors the conjugations reduce to the
real contractions used in the implementation. In particular,
$M^{[0]}_{2_i,1_i,\alpha}=D_{\alpha i}$, whereas
$M^{[0]}_{3_i,1_i,\alpha}=0$ and the latter hot band starts with the
quadratic dipole and first-order mechanical terms.

The VP2TD eigenvectors act on \emph{both} states of a hot-band
transition, rather than only on the final state as in a cold band:
\begin{align}
 \widetilde M^{[r]}_{KL,\alpha}
 &=\sum_{f\in\mathcal P_K}\sum_{a\in\mathcal P_L}
       U^*_{fK}M^{[r]}_{fa,\alpha}U_{aL},\nonumber\\
 \nu_{KL}&=E_K-E_L>0,\nonumber\\
 I^{\rm intrinsic}_{KL}
 &=C_{\rm IR}\nu_{KL}\sum_\alpha
       \left|\sum_{r=0}^{2}\widetilde M^{[r]}_{KL,\alpha}\right|^2.
 \label{eq:hot-pt2d-intensity}
\end{align}
The order-truncated intensity follows Eq.~\eqref{eq:intensityorders}
with $0\mapsto L$ and $k\mapsto K$. The intrinsic intensity is
reported per occupied initial level. For a thermal absorption
spectrum its line strength is multiplied separately by
\begin{equation}
 \begin{gathered}
 p_L(T)\,[1-\exp(-hc\nu_{KL}/k_{\rm B}T)],\\
 p_L(T)=\frac{\exp[-hc(E_L-E_0)/k_{\rm B}T]}{Z_{\rm vib}(T)}.
 \end{gathered}
 \label{eq:hot-population}
\end{equation}
The same equations apply to ESALC-C after setting the metric
derivatives to zero and retaining the Cartesian Coriolis term in
${\cal B}^{C}$, and to ESALC-I with ${\cal B}^{I}$.
As an explicit scalar check on the tensor coefficients, consider one
uncoupled mode of frequency $\omega$ and suppress its repeated mode
index on $\eta$, $b$, $\sigma$, and $\rho$. Taking $D_{\alpha11}$
as first-order and $D_{\alpha111}$ as second-order property terms,
Eqs.~\eqref{eq:ladder-explicit}--\eqref{eq:fundamental-full-amplitude}
reduce to
\begin{widetext}
\begin{align}
 M_{10,\alpha}=\frac{1}{\sqrt2}\bigg\{D_{\alpha1}
 +\frac{D_{\alpha111}}4
 -\frac{D_{\alpha11}}{\omega}
       \left(\frac{3\rho}{8}+\frac{\sigma}{24}\right)
 +D_{\alpha1}\left[
       \frac{b-3\eta}{32\omega}
       +\frac{7\rho\sigma}{96\omega^2}
       +\frac{\sigma^2}{288\omega^2}\right]\bigg\}+O(\lambda^3),
 \label{eq:one-mode-fundamental-ir}\\
 M_{20,\alpha}^{[1]}=\frac{1}{\sqrt2}\left[
       \frac{D_{\alpha11}}2
       +\frac{D_{\alpha1}(3\rho-\sigma)}{12\omega}\right],
 \label{eq:one-mode-overtone-ir}
\end{align}
\end{widetext}
Hence $I_{10}=C_{\rm IR}\nu_{10}\sum_\alpha|M_{10,\alpha}|^2$
and the leading overtone intensity is
$I_{20}=C_{\rm IR}\nu_{20}\sum_\alpha|M_{20,\alpha}^{[1]}|^2$.
Equation~\eqref{eq:one-mode-fundamental-ir} shows explicitly where
the quartic $\eta$ and $b$ tensors enter an anharmonic fundamental
intensity; Eq.~\eqref{eq:one-mode-overtone-ir} separates direct
electrical intensity from the $\rho$/$\sigma$ mechanical borrowing.
For multiple modes and resonances, the implementation evaluates
the equivalent reduced coefficient equations in the SI, followed
by the polyad eigenvectors of Eq.~\eqref{eq:polyadir}.
For a dark harmonic transition, the leading nonzero contribution
may therefore be $I^{[2]}$, including electrical anharmonicity
and mechanical intensity borrowing.

\section{Adiabatic DVR for large-amplitude motion}
\label{sec:adiabatic-dvr}
A torsion or other large-amplitude coordinate can be treated explicitly
while the remaining vibrations form a transverse bath. We select a
chemically defined internal coordinate $\phi$ and use LINK to optimize
all unconstrained coordinates at fixed $\phi$. The resulting path
provides the relaxed electronic energy, molecular geometry, and required
properties. Cartesian gradients and Hessians are needed along the path
when its transverse vibrational response is included. The constrained
bath is obtained by projecting the Lagrangian Hessian, which includes
the curvature of the constraint, onto its tangent space after removal
of overall translations and rotations.

For bath occupations $\mathbf a$, the harmonic transverse energies give
\begin{align}
 \epsilon_{\mathbf a}^{\mathcal B}(\phi)
 &=\sum_{j\in\mathcal B}\omega_j^{\perp}(\phi)
        \left(a_j+\tfrac12\right),\nonumber\\
 U_{\mathbf a}^{\rm ad}(\phi)
 &=V_{\rm rel}(\phi)+\epsilon_{\mathbf a}^{\mathcal B}(\phi)
       -\epsilon_{\mathbf0}^{\mathcal B}(\phi_0).
 \label{eq:dvr-adiabatic-potential}
\end{align}
Energies in this section are in wavenumbers; the constant bath reference
at the minimum $\phi_0$ is restored for total zero-point energies and
conformer free energies. Following the transverse modes by overlap
connects the same bath state between successive geometries. The ground
bath defines the torsional ladder, while excited bath channels provide
the corresponding adiabatic vibrational transitions.

The path inertia $I_{\rm eff}(\phi)$ is computed from its mass-weighted
tangent after Eckart removal of overall translation and rotation.
Defining $\mathcal G(\phi)=\hbar^2/[hcI_{\rm eff}(\phi)]$, the
one-coordinate Hamiltonian in a flat integration measure is
\begin{align}
 \widehat H_{\mathbf a}^{\rm LAM}
 &=-\frac12\frac{d}{d\phi}\mathcal G(\phi)\frac{d}{d\phi}
       +U_{\mathbf a}^{\rm ad}(\phi)+V_{\rm geom}(\phi),\nonumber\\
 V_{\rm geom}(\phi)
 &=-\frac18\mathcal G''(\phi)
       +\frac{[\mathcal G'(\phi)]^2}{32\mathcal G(\phi)}.
 \label{eq:dvr-path-hamiltonian}
\end{align}
For a periodic rotor, the potential, metric, and property components are
represented by real Fourier functions consistent with equivalent-atom
permutations. The even part can be written as a polynomial in
$\cos[s(\phi-\phi_0)]$, where $s$ is the rotor symmetry number;
sine terms retain allowed odd property components. The signed angle
remains the integration coordinate. A Fourier DVR then diagonalizes
Eq.~\eqref{eq:dvr-path-hamiltonian}; confined conformer wells use a
finite interval and their bound levels. Fixed-geometry scans and a
constant effective inertia also provide reduced rotor models, as
specified in the individual applications. Selected transverse states
can be retained as coupled channels when their explicit response is
required.

Property matrix elements are quadratures of the DVR wavefunctions and
the property surface. For a fixed bath channel,
\begin{equation}
 P_{\mathbf a;nm}=\int\psi_{\mathbf a n}^{*}(\phi)
       P_{\mathbf a}^{\rm eff}(\phi)
       \psi_{\mathbf a m}(\phi)\,d\phi,
 \label{eq:dvr-property-average}
\end{equation}
where $P_{\mathbf a}^{\rm eff}$ contains the electronic property and
any retained perturbative bath correction. Diagonal elements give
state averages, off-diagonal elements give transition moments, and
Boltzmann weighting gives thermal averages. The explicit torsion is
excluded from the perturbative bath indices before adding its DVR
contribution. Potential-fit order, grid size, and the number of retained
levels are checked against the relevant transition or thermal average.
The SI give the detailed constrained-bath and coupled-channel
constructions used in the molecular examples.

\section{Computational strategy and validation design}
The scientific and computational developments are tested at different
levels. Finite oscillator matrices give a reference for the new
curvilinear transition coefficients, including metric derivatives
and excited-state initial vectors. Same-field ESALC-C/I calculations
test harmonic invariance and the effect of the coordinate representation on
anharmonic couplings. Gaussian \texttt{Freq=Anharm} checkpoints test
the rectilinear derivative acquisition and resonance treatment on
identical electronic data. These comparisons answer different
questions: agreement with Gaussian checks the Cartesian limit,
whereas oscillator-matrix and same-field tests exercise the new
internal-coordinate formulas.
Table~\ref{tab:validation-map} collects the principal checks and
distinguishes independent numerical or experimental comparisons from
same-field coordinate-consistency tests.

\begin{table*}[tp]
\caption{Principal validation results. References include independent
oscillator calculations, established spectroscopic tools, and
experiment. The last two rows compare coordinate representations.
Differences are maxima unless a ratio is specified.}
\label{tab:validation-map}
\begin{ruledtabular}
\begin{tabular}{p{0.20\textwidth}p{0.34\textwidth}p{0.37\textwidth}}
Quantity & Comparator & Result\\
Curvilinear IR & Independent oscillator matrices on a molecular H$_2$O field
 & Second-order transition-amplitude difference $8.4\times10^{-8}$~a.u.\\
Fermi IR transport & Gaussian H$_2$CO three-root Fermi polyad
 & 0.163~\wavenum{} in energy; 0.061~km\,mol$^{-1}$ in intensity\\
Hot-band intensity & 29 matched H$_2$CO rotational line pairs
 & With the measured $\nu_6$ population, the thermal ratio differs
 from the measured median by 3.9\% (ESALC-I) and 3.8\% (ESALC-C)\\
Static Raman & Gaussian H$_2$O nine-band common-field calculation
 & $8.1\times10^{-8}$~\AA$^6$ in activity\\
NH$_3$ symmetry gauge & Six-state $C_{3v}$ polyad, transported I/C fields
 & 0.033~\wavenum{} maximum root difference\\
Dimer coordinate representation transport & Cyclic formic-acid dimer, transported I/C fields
 & 0.322~\wavenum{} in fundamentals; 0.601~km\,mol$^{-1}$ in IR\\
\end{tabular}
\end{ruledtabular}
\end{table*}

The coordinate procedure transforms the Cartesian data from each
displaced geometry into force, metric, and property derivatives.
Independent electronic calculations can run in parallel. The principal
coordinate comparisons use one electronic-structure surface and all
vibrational modes. Multilevel fields and reduced-dimensional treatments
are also available; the separate stretching controls in the SI specify
their selected subspaces.
An energy--gradient acquisition example for H$_2$CO determines the full
cubic and quartic potential from 68 fitting geometries, with 16
independent validation points. With second-order M{\o}ller--Plesset perturbation theory (MP2)
and the def2-TZVP basis, its energy residual is
$1.02\times10^{-8}$~hartree; the SI give the fitting and validation
details.
The algebraic property stage distributes independent final states
and populated initial states across parallel processes while keeping each
variational polyad intact. A saved-field performance control is
reported in the Supporting Information.
For an experimental comparison, a higher-level harmonic field may
replace the lower-level quadratic term before any perturbative or
polyad calculation. In a common mass-normalized mode gauge with
diagonal $F_2$ and unit equilibrium metric, the frequency-only form is
\begin{equation}
 (F_2^{\rm comp})_{ii}=(F_2^{\rm low})_{ii}
 \left(\frac{\omega_i^{\rm high}}{\omega_i^{\rm low}}\right)^2,
 \qquad F_{n\geq3}^{\rm comp}=F_{n\geq3}^{\rm low}.
 \label{eq:composite-harmonic-replacement}
\end{equation}
Equation~\eqref{eq:composite-harmonic-replacement} is a
frequency-only diagonal replacement. A complete higher-level Hessian
expressed in the lower-level normal modes generally also contains
off-diagonal quadratic terms, because the two sets of normal modes
need not coincide. In a common mass-normalized vibrational space,
$F_2^{\rm comp}=R\,\mathrm{diag}[(\omega^{\rm high})^2]R^{\mathsf T}$,
where $R$ maps the higher-level modes into the lower-level basis.
These off-diagonal terms vanish when the higher-level normal modes
are used as the basis; the lower-level anharmonic force, metric, and
property tensors must then be transformed consistently into that basis.
If the lower-level modes are retained, the off-diagonal quadratic
terms must be included explicitly in the vibrational Hamiltonian.
Published frequencies alone determine only the diagonal replacement
in Eq.~\eqref{eq:composite-harmonic-replacement}, not the rotation $R$.
The VP2TD solver constructs its unperturbed energies, nonresonant
denominators, resonance tests, and variational polyads from the
chosen composite harmonic Hamiltonian. The higher-level information
therefore enters before the perturbative and polyad calculations.

\subsection{Molecular set and electronic calculations}
For a semi-rigid molecule, ESALC-C and ESALC-I describe the same
Hamiltonian. Agreement in a common harmonic basis tests consistency;
residual differences reflect finite perturbative order and numerical
differentiation, with greater sensitivity near resonances. The
interpretive advantage of ESALC-I is assessed through intermode coupling
patterns and the separation of large-amplitude coordinates.
This work tests the vibrational formalism and its implementation, so the
electronic level is held fixed within each comparison rather than
optimized separately for each molecule. Most calculations use the
Becke three-parameter Lee--Yang--Parr hybrid functional (B3LYP) with
the 6-31G* basis, equivalently denoted 6-31G(d). The formic-acid,
glycine, and acetic-acid calculations include third-generation
dispersion with Becke--Johnson damping (D3BJ).
Each coordinate representation and independent reference
implementation receives the same electronic data, isolating differences
in the vibrational treatment. Deviations from experiment can therefore
combine errors in the electronic surface, finite perturbative order,
and, for fitted line strengths, the rovibrational model. The transition-
moment equations are validated separately against independent oscillator
matrices and established implementations using identical electronic data.

\begin{figure*}[t]
\centering
\includegraphics[width=0.76\textwidth]{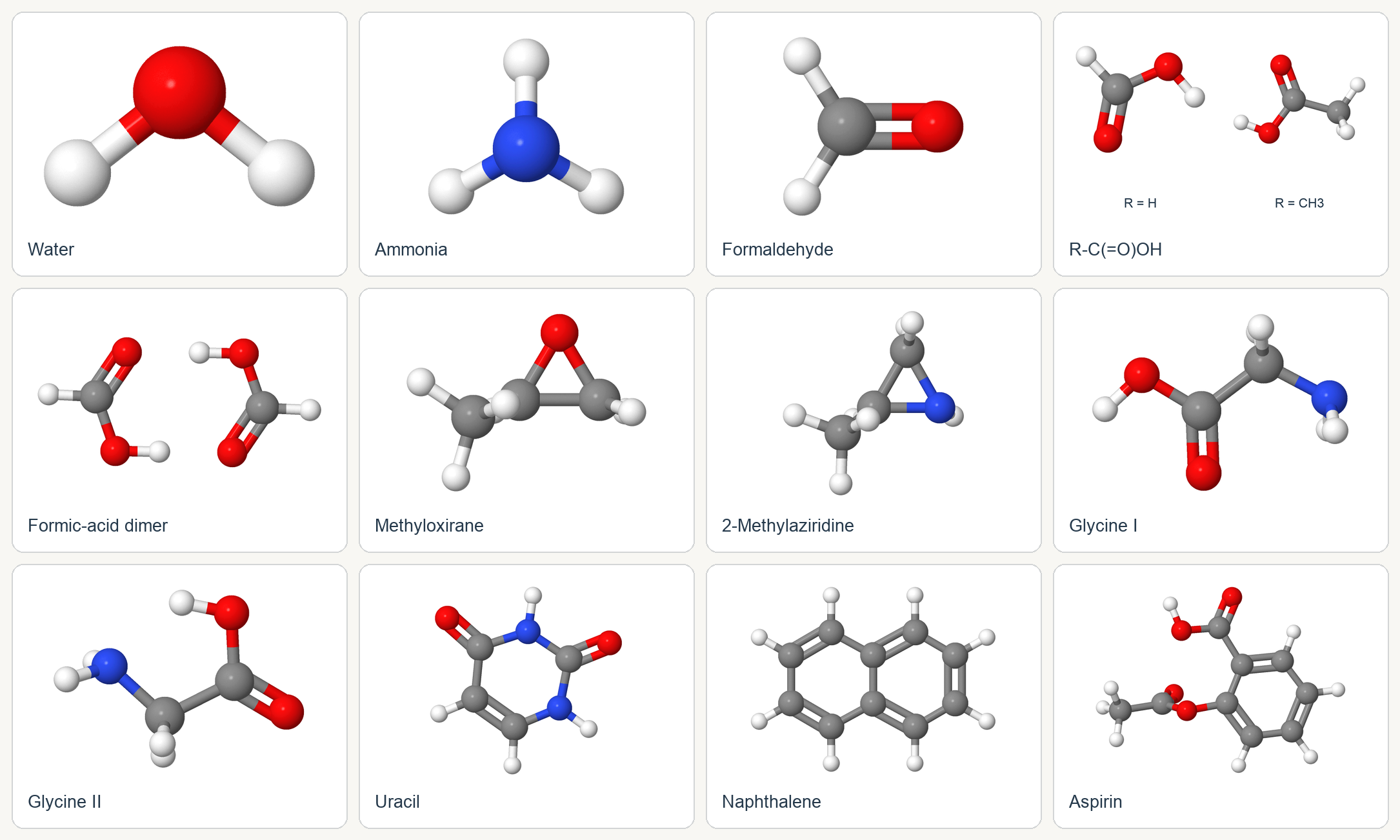}
\caption{Reference molecular structures: H$_2$O, NH$_3$, H$_2$CO,
the carboxylic acids R--C(=O)OH with R=H and CH$_3$, the cyclic
formic-acid dimer, the isoelectronic pair $(R)$-methyloxirane and
$(S)$-2-methylaziridine, glycine conformers I and
II, uracil, naphthalene, and aspirin conformer II.}
\label{fig:benchmark-structures}
\end{figure*}

Figure~\ref{fig:benchmark-structures} collects the reference
geometries used below. The cyclic formic-acid dimer is included
alongside the monomer to test whether the coordinate comparison
remains stable for a hydrogen-bonded complex. The global-minimum trans (C1)
structure of $(S)$-2-methylaziridine represents the expectation-value
and periodic-rotor application. It is valence-isoelectronic with
$R$-methyloxirane: replacement of O by NH preserves the electron count.
All structures were
rendered from the saved reference geometries with Jmol;\cite{jmol} the rigid
display rotations do not change the calculations.
The two electronic-structure programs tested here are Gaussian\cite{gaussian}
and the Python-based simulations of chemistry framework
(PySCF).\cite{pyscf} Gaussian supplies the molecular Hessian and property
derivative fields used for the main spectroscopic comparisons, as well
as the independent rectilinear reference. PySCF is tested in the water
dipole-value control and a separate energy--gradient acquisition;
reduced water and hydrogen-peroxide stretch controls using saved
PySCF fields are documented in the Supporting Information. The
energy--gradient example employs MP2 with frozen oxygen 1s orbitals and the
def2-TZVP basis. Each comparison below identifies its electronic level.
Table~\ref{tab:electronic-backends} summarizes the electronic-data
interfaces available for derivative acquisition. LINK
accepts energies and Cartesian gradients from each listed program;
the Hessian column distinguishes direct acquisition or import from
finite differences of gradients. Property derivatives can be obtained
from analytic output where the interface provides it or from property
values on the same displacement grid. The complete spectroscopic
comparisons reported here use Gaussian fields and the specified PySCF
controls; the other rows specify the available electronic-data interfaces.
\begin{table*}[tp]
\caption{Electronic-structure programs and derivative interfaces available to VP2TD.}
\label{tab:electronic-backends}
\begin{ruledtabular}
\begin{tabular}{llll}
Program & Point energies and derivatives & Cartesian Hessian source & Use in this work\\
Gaussian\cite{gaussian} & energy, gradient & analytic Hessian & molecular benchmarks\\
PySCF\cite{pyscf} & energy, gradient & direct output & water and derivative controls\\
ORCA\cite{orca-software} & energy, gradient & direct output/import & interface only\\
Molpro\cite{molpro-software} & energy, gradient & output import or gradient differences & interface only\\
TURBOMOLE\cite{turbomole-software} & energy, gradient & gradient differences & interface only\\
CFOUR\cite{cfour-software} & energy, gradient & analytic Hessian & interface only\\
xTB\cite{xtb-software} & energy, gradient & numerical Hessian & interface only\\
\end{tabular}
\end{ruledtabular}
\end{table*}
The electronic coordinate comparisons use single-mode displacements
$q_k=\pm h$ with $h=0.16$ in dimensionless normal coordinates.
A common dimensionless step samples the same fraction of the harmonic
oscillator length for every mode: the corresponding mass-weighted
physical displacement scales as $\omega_k^{-1/2}$. Thus $h=0.16$
is not a fixed Cartesian displacement in angstroms.
Its choice balances the $O(h^2)$ truncation error of the central
stencil against numerical noise in the electronic derivatives.
In particular, Hessian noise is amplified as $h^{-1}$ in the cubic
constants and as $h^{-2}$ in the quartic constants, so decreasing
$h$ does not necessarily improve the acquired field.
The same-convention water control at $h=0.12$ and 0.16 gives maximum
changes of 0.320~\wavenum{} in fundamentals and
0.084~km\,mol$^{-1}$ in anharmonic fundamental intensities.
These measured sensitivities support $h=0.16$ as the common working
step for the coordinate comparisons; the two-step check measures
finite-step sensitivity rather than an extrapolated zero-step error.
Full-precision derivatives and consistent electronic convergence,
integration grids, symmetry, and orientation are used with this stencil.

Direct Gaussian force-constant comparisons instead match its physical
step, $\Delta Q=0.01$~\AA$\sqrt{\mathrm{amu}}$, which corresponds to
frequency-dependent dimensionless steps $h_k$. The methyloxirane
polynomial re-expansion and geometry-only metric checks specify their
distinct steps separately. Water uses matched ESALC-I and
ESALC-C coordinate representations built from one analytic Cartesian Hessian; the
harmonic frequencies agree to $1.2\times10^{-9}$~\wavenum{}.
Acquisition conventions, grid and step controls, and the residual
reference-gradient treatment are specified in the Supporting
Information.

\section{Validation of transition-property formulas}
Tensor contractions with explicit sums over intermediate vibrational
states provide a general reference implementation for validating the
specialized algebraic expressions, using identical force, metric, and
property derivatives and the same perturbative order. The comparisons
cover ground and excited initial states, normalization of both states,
and the retained resonant contributions. This sum-over-states
implementation supports tests of individual contributions. Routine calculations use the
specialized algebraic formulas, which eliminate explicit intermediate-state
sums before numerical evaluation.

Independent oscillator-matrix calculations provide a further check of
the operator ordering and perturbative coefficients. Comparisons with
molecular properties from established tools then connect those coefficients
to observable transition strengths, including resonance mixing.

\subsection{Curvilinear two-mode operator control}
We constructed a reproducible two-mode Hamiltonian with nonzero
off-diagonal $g_{ij,k}$, $g_{ij,kl}$, cubic and quartic potential
derivatives, and dipole derivatives through third order. Direct
diagonalization of independently assembled oscillator matrices tests
the ordered $p_iq_kp_j$ operators against the ESALC-I algebra without
using the ESALC-C branch. The extrapolated exact coefficients
$\chi_{11}$ and $\chi_{12}$ differ from
Eqs.~\eqref{eq:chi-diag} and \eqref{eq:chi-offdiag} by at most
$1.3\times10^{-7}$~\wavenum{}. The transition moments test the
second-order normalization and mixed metric terms of
Eq.~\eqref{eq:fundamental-full-amplitude} and the overtone
expression, Eq.~\eqref{eq:overtone-full-amplitude}.
Appendix~\ref{app:synthetic} gives the independent construction,
scaling results, and finite-basis check.

\subsection{Independent molecular curvilinear intensity check}
The H$_2$O check uses a Gaussian B3LYP/6-31G(d)
ESALC-I field and its analytic dipole Jacobians to test the
anharmonic IR amplitude against an independent finite-oscillator
matrix calculation. The matrix representation includes the ordered
$p_iq_kp_j$ metric terms and the same dipole surface through cubic
order. With $H=H_0+\lambda H_1+\lambda^2 H_2$ and
$\mu=\mu_1+\lambda\mu_2+\lambda^2\mu_3$, polynomial fits to
finite harmonic-oscillator transition moments at $|\lambda|\leq0.02$ reproduce the
ESALC-I algebraic coefficients for all three fundamentals. The
maximum absolute differences across modes and Cartesian dipole
components are $3.9\times10^{-12}$, $4.2\times10^{-14}$, and
$8.4\times10^{-8}$~a.u. at orders zero, one, and two, respectively.
The ten-function-per-mode basis and all electronic points were read
from saved data. This provides an independent molecular test of the curvilinear coefficients. The tensor archive and a standalone
independent numerical verification script accompany the Supporting Information;
the separate frequency and metric controls are documented there.

\subsection{Independent oscillator and Morse controls}
A coordinate-space sinc DVR independently verifies the rectilinear
fundamental and overtone amplitudes. The reduced and tensor equations
agree within $10^{-12}$ atomic units; their difference from the exact
DVR increases with perturbation strength as expected from truncation.
On the molecular water field, independent oscillator matrices test the
curvilinear coefficients as described above. Local exponential
continuations of the stretching potentials reproduce the VPT2
fundamentals within 1.50~\wavenum{} for H$_2$O and within
1.45~\wavenum{} in the selected H$_2$CO and H$_2$O$_2$ subspaces.
The SI give the model potentials, transition moments, and convergence
tests, including the Morse consistency controls.

\subsection{Three-quantum and hot-band oscillator controls}
We checked the three-quantum and hot-band transitions against an independently
assembled 12-level oscillator matrix. The one-mode model has
$\omega=1000$, $f_3=120$, $f_4=35$~\wavenum{} and dipole derivatives
$(D_1,D_2,D_3)=(0.8,0.4,0.12)$ in atomic units. The ESALC-I control
adds $(g_1,g_2)=(18,12)$~\wavenum{}; the ESALC-C control sets both
metric derivatives to zero. The three perturbative moment
coefficients of Eq.~\eqref{eq:hot-pair-moment} agree with direct
oscillator-matrix Rayleigh--Schr\"odinger contractions within
$2\times10^{-11}$ atomic units for $0\rightarrow3_1$,
$1_1\rightarrow2_1$, and $1_1\rightarrow3_1$ in both coordinate
limits. The SI tabulate the nonzero components. The dipole scale makes the
higher-order coefficients clearly visible.

\begin{figure*}[t]
\centering
\includegraphics[width=0.82\textwidth]{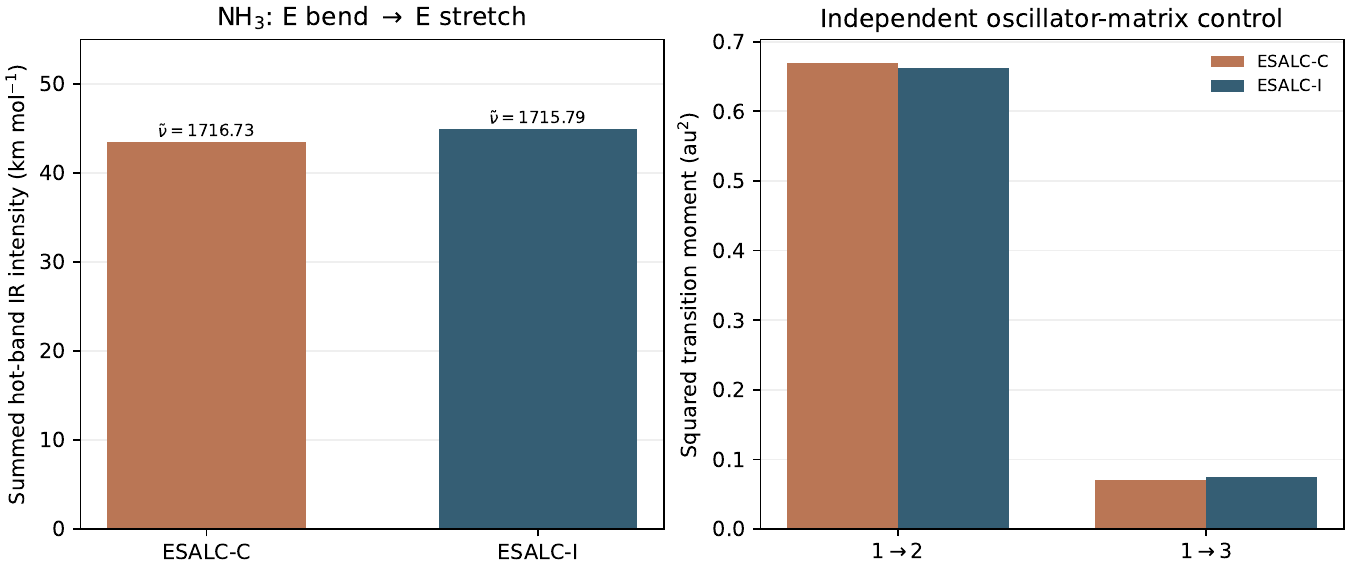}
\caption{Hot-band capabilities in both coordinate representations.
Left: intrinsic NH$_3$ IR intensity summed over the degenerate
$E$-bend-to-$E$-stretch channels, with the transition wavenumber
annotated above each bar. Right: squared transition moments from
the independent one-mode oscillator-matrix control; these are not
molecular IR intensities. The $1\rightarrow3$ band requires the
anharmonic transition operator.}
\label{fig:hot-band-controls}
\end{figure*}

As a separate VP2TD control, a two-mode complete quartic field with
$\omega=(1000,3010)$~\wavenum{} and a symmetrized
$f_{1112}=150$~\wavenum{} forms a near-resonant
$3_1/1_2$ polyad. The nominally dark three-quantum level borrows
ground-state intensity, and a hot transition originating in the
mixed fundamental carries its initial-state eigenvector coefficient.
This explicitly tests Eq.~\eqref{eq:hot-pt2d-intensity} on both sides
of a transition. The same equations apply to fundamental-origin
transitions in both ESALC-C and ESALC-I.
Figure~\ref{fig:hot-band-controls} makes the distinction visible:
the independent oscillator calculation checks individual
fundamental-origin transition moments, while the molecular NH$_3$
calculation sums the intrinsic intensities over the degenerate
$E$-bend-to-$E$-stretch hot-band partners. The latter are
43.446 and 44.924~km\,mol$^{-1}$ in ESALC-C and ESALC-I,
respectively; their centroid transitions are 1716.728 and
1715.787~\wavenum{}. The comparison uses the same saved electronic
field and the second-order transition-moment transport on both
coordinate representations.

\subsection{Common-field static Raman comparison}
We performed a Gaussian B3LYP/6-31G(d) SuperFine grid
\texttt{Freq=(Anharm,Raman)} calculation for H$_2$O. From its
formatted checkpoint we read the analytic Cartesian polarizability
Jacobian and its numerical directional derivatives; the cubic and
quartic normal-mode force constants come from the same job.
After matching mode order and signs, TRINITY applies the
second-order transition-property transport and evaluates
Eq.~\eqref{eq:raman-activity}. The harmonic activities coincide
with Gaussian in dimensionless normal-coordinate units. The nine
anharmonic fundamentals, overtones, and combinations agree within
$8.1\times10^{-8}$~\AA$^6$; their largest position difference is
$4.8\times10^{-4}$~\wavenum{}. Table~\ref{tab:h2o-raman}
shows the fundamentals and one overtone; all nine bands and the
reconstruction details are in the SI.
\begin{table}[t]
\caption{Static anharmonic Raman activities (\AA$^6$) for H$_2$O
from one Gaussian electronic job. Mode labels follow Gaussian's
printed normal-mode order.}
\label{tab:h2o-raman}
\begin{ruledtabular}
\begin{tabular}{lrr}
Band & Gaussian & TRINITY\\
$\nu_1$ & 0.38544915 & 0.38544915\\
$\nu_2$ & 0.08780914 & 0.08780906\\
$\nu_3$ & 0.19164488 & 0.19164489\\
$2\nu_2$ & 0.01192191 & 0.01192191\\
\end{tabular}
\end{ruledtabular}
\end{table}
Analytic Cartesian polarizability Jacobians are evaluated on
the matched ESALC-C and ESALC-I single-mode stencils, both with
$h=0.16$ in dimensionless normal coordinates. Both coordinate representations use the
same central B3LYP/6-31G(d) Hessian and electronic method; each
property Jacobian is transformed through the tangent of its own
Eckart-oriented coordinate representation. The harmonic Raman activities agree to
better than $10^{-12}$~\AA$^6$ across the two coordinate representations. The maximum
ESALC-C/I difference among the nine second-order Raman activities
is $5.6\times10^{-5}$~\AA$^6$; their maximum transition-energy
difference is 0.173~\wavenum{}. Table~\ref{tab:h2o-raman-charts}
lists representative bands. The complete nine-band comparison and
its reproducible acquisition are given in the SI.
\begin{table}[t]
\caption{Matched H$_2$O static anharmonic Raman activities
(\AA$^6$). Gaussian is the independent rectilinear reference;
ESALC-C and ESALC-I use matched $h=0.16$ single-mode stencils.}
\label{tab:h2o-raman-charts}
\begin{ruledtabular}
\begin{tabular}{lrrr}
Band & Gaussian & ESALC-C & ESALC-I\\
$\nu_2$ & 0.08780914 & 0.08780517 & 0.08786117\\
$\nu_1$ & 0.38544915 & 0.38546274 & 0.38544329\\
$\nu_3$ & 0.19164488 & 0.19164254 & 0.19164245\\
$2\nu_2$ & 0.01192191 & 0.01191466 & 0.01191110\\
\end{tabular}
\end{ruledtabular}
\end{table}

\subsection{Molecular coordinate consistency: water and hydrogen peroxide}
The analytic-dipole water comparison uses a common reference Jacobian,
ensuring harmonic intensity identity. The SI also report a separate
water control based on finite differences of dipole values.

An independent rectilinear comparison uses Gaussian B3LYP/6-31G(d), a
SuperFine integration grid, and a tightly optimized reference geometry.
The ESALC-I
and ESALC-C fields use the same Cartesian Hessian and dipole
Jacobian at the reference and the same $h=0.16$ single-mode
displacements.  Table~\ref{tab:water-gaussian} compares their
fundamentals directly with Gaussian \texttt{Freq=Anharm} on this
electronic surface.  All three harmonic frequencies agree to the
reported precision; the maximum ESALC-C/Gaussian fundamental
difference is 0.132~\wavenum{}, and the maximum intensity
difference is 0.011~km\,mol$^{-1}$.  The corresponding ESALC-I
differences are 0.104~\wavenum{} and 0.006~km\,mol$^{-1}$.
The harmonic IR intensities of the two ESALC coordinate representations also coincide
with Gaussian because this acquisition uses analytic Cartesian
dipole Jacobians, unlike the value-dipole PySCF control above.

\begin{table*}[tp]
\caption{H$_2$O fundamentals on one Gaussian B3LYP/6-31G(d)
surface.  Frequencies are in \wavenum{} and intensities in
km\,mol$^{-1}$.  C and I denote ESALC-C and ESALC-I.}
\label{tab:water-gaussian}
\begin{ruledtabular}
\begin{tabular}{lrrrrrr}
Mode & $\nu_{\rm Gaussian}$ & $\nu_C$ & $\nu_I$
 & $I_{\rm Gaussian}$ & $I_C$ & $I_I$\\
HOH bend & 1662.643 & 1662.676 & 1662.604 & 73.848 & 73.837 & 73.851\\
OH symmetric & 3555.551 & 3555.522 & 3555.491 & 0.448 & 0.448 & 0.442\\
OH antisymmetric & 3655.784 & 3655.916 & 3655.888 & 14.463 & 14.461 & 14.462\\
\end{tabular}
\end{ruledtabular}
\end{table*}

The H$_2$O$_2$ comparison uses Gaussian B3LYP/6-31G(d) analytic
Cartesian Hessians and dipole Jacobians at the reference and at
single-mode displacements $q_k=\pm0.16$. The reference geometry is
projected onto $C_2$ before any electronic point is acquired; the
reference Hessian and dipole Jacobian are symmetry averaged before
their modal transformation. The ESALC-I coordinate representation contains two OH
stretches, the OO stretch, two HOO angles, and the HOOH torsion;
ESALC-C uses Eckart-aligned Cartesian modes. Both use the same
frozen-reference-gradient convention. The semidiagonal quartic
field and the property derivatives are acquired from single-mode
points only. Table~\ref{tab:h2o2} gives the results without resonance
polyads.

\begin{table*}[!tp]
\centering
\caption{Matched-field H$_2$O$_2$ results from symmetry-consistent
Gaussian B3LYP/6-31G(d) single-mode stencils.
Frequencies are in $\wavenum$ and integrated IR intensities in
km\,mol$^{-1}$. The two harmonic columns coincide to the digits shown.}
\label{tab:h2o2}
\begin{tabular}{lrrr}
\toprule
Mode & $\omega_I/\omega_C$ & $\nu_I/\nu_C$
 & $I_I^{\rm anh}/I_C^{\rm anh}$\\
\midrule
HOOH torsion & 359.736/359.736 & 301.631/302.013 & 196.038/196.161\\
OO stretch & 956.990/956.990 & 932.881/932.891 & 1.517/1.410\\
HOO bend 1 & 1334.353/1334.353 & 1296.947/1296.993 & 107.885/107.868\\
HOO bend 2 & 1467.119/1467.119 & 1431.708/1431.677 & 0.687/0.683\\
OH stretch 1 & 3704.125/3704.125 & 3517.511/3517.216 & 2.385/2.392\\
OH stretch 2 & 3706.591/3706.591 & 3528.180/3528.227 & 20.694/20.685\\
\bottomrule
\end{tabular}
\end{table*}

The harmonic frequencies agree within $1.18\times10^{-9}$~\wavenum{}
and the harmonic IR intensities within
$3.91\times10^{-11}$~km\,mol$^{-1}$. The OO dipole derivative has no
component perpendicular to the molecular $C_2$ axis, as required by
symmetry; projecting the acquired $D_1,D_2,D_3$ tensors a second time
does not alter the displayed intensities. The largest fundamental
frequency difference is 0.382~\wavenum{} on the torsion, and the
largest IR difference is 0.123~km\,mol$^{-1}$. The OO IR difference
is 0.107~km\,mol$^{-1}$. These residuals refer to the reduced
anharmonic stencil, rather than to symmetry leakage.
The SI compare dipole-value and analytic-Jacobian
stencils, including the effect of mixed $G''$ terms. Analytic dipole
Jacobians are the preferred source of intensity derivatives: in the
H$_2$CO control, reducing the stencil changes intensities by at most
0.505~km\,mol$^{-1}$. The larger changes from a value-only H$_2$O$_2$
stencil establish the importance of retaining these mixed property
derivatives.

\section{Rectilinear comparison with established tools}
The ESALC-C acquisition was checked against single-calculation Gaussian
\texttt{Freq=Anharm} checkpoints, using their unrounded Cartesian
energy, gradient, Hessian, and dipole records. Reconstructing the
acetic-acid field from its reference and 36 displaced points gives
a first fundamental within $2.2\times10^{-7}$~\wavenum{} of the
checkpoint value. The analogous 30-mode uracil comparison gives
a lowest fundamental within $2.1\times10^{-7}$~\wavenum{}.
The directional averaging, cutoff order, Eckart transport, and
checkpoint precision controls are specified in the SI.
The checkpoint reconstruction retains the mixed quartic constants
in the Darling--Dennison polyad Hamiltonian, while the perturbative
fundamentals require only the semidiagonal quartic subset.

The acetic-acid UltraFine calculation provides an independent
molecular check of polyad intensity transport. Across its 18
fundamentals, the maximum frequency and IR-intensity differences
from Gaussian are $5.9\times10^{-5}$~\wavenum{} and
$8.2\times10^{-4}$~km\,mol$^{-1}$, respectively. The isolated
991.829~\wavenum{} band agrees within
$3.3\times10^{-6}$~km\,mol$^{-1}$. Its nine-state mid-IR polyad includes
the closed 2:2 combination--overtone Darling--Dennison coupling,
selected with the reduced-coupling criterion, and a weak 1:1 coupling;
transition moments are dressed through second order and then rotated by the
complete polyad eigenvectors. For resonant fundamentals, each selected 1:2
or 1:1 pole is removed before the finite mechanical terms and normalization
are collected. This pole-first form is algebraically identical to the
ordinary expression outside resonances and preserves the correct finite
remainder inside the deperturbed polyad. Table~\ref{tab:acetic-ir} shows
representative roots. The SI provide all nine polyad assignments and the
componentwise common-field property and Coriolis analysis.

\begin{table*}[tp]
\caption{Acetic-acid UltraFine same-electronic-job fundamental frequencies
(\wavenum{}) and IR intensities (km\,mol$^{-1}$).  VP2TD transition
strengths are assigned to the diagonalized root.}
\label{tab:acetic-ir}
\begin{ruledtabular}
\begin{tabular}{lrrrr}
Band & $\nu_{\rm Gaussian}$ & $\nu_{\rm VP2TD}$
 & $I_{\rm Gaussian}$ & $I_{\rm VP2TD}$\\
Soft torsion (isolated control) & 67.200 & 67.200 & 0.220 & 0.221\\
Isolated 991.829 band & 991.829 & 991.829 & 62.625 & 62.625\\
Mid-IR polyad root & 1184.696 & 1184.696 & 165.414 & 165.414\\
C=O stretch & 1824.936 & 1824.936 & 256.891 & 256.891\\
\end{tabular}
\end{ruledtabular}
\end{table*}

The soft-torsion row tests the common-field derivatives. This mode
is excluded from the mid-IR resonance polyad and assigned to the
separate DVR treatment.

\section{Resonances and hot-band intensities}
The following molecular examples test the transport of transition
moments through resonance polyads, including degenerate partners and
vibrationally excited initial states. Measured hot-band strengths then
connect the intrinsic moments to thermally populated spectra.

\subsection{H$_2$CO Fermi-polyad control}
The saved Gaussian B3LYP/6-31G(d) anharmonic calculation of
H$_2$CO supplies an independent VP2TD control on the same
electronic reference used for the ESALC-C acquisition. The
reference self-consistent-field energies differ by less than $10^{-8}$~hartree and
the six harmonic frequencies by at most 0.0086~\wavenum{}.
The corresponding harmonic IR intensities differ by at most
0.013~km\,mol$^{-1}$. ESALC-C is the Cartesian normal-mode
representation, so this comparison also checks its reduction to
the conventional rectilinear calculation. The reference calculation uses the established program's default
anharmonic displacement settings; TRINITY uses $h=0.16$ in
dimensionless normal coordinates.
Two accepted $1{:}2$ Fermi interactions mix the high-frequency
CH fundamental with two combination states. Table~\ref{tab:fermi}
compares all three \emph{polyad eigenvalues and mixed IR
intensities}; comparing a single assigned fundamental would be
misleading because its dominant character changes between
nearby roots. The largest
TRINITY--Gaussian difference is 0.163~\wavenum{} in energy and
0.061~km\,mol$^{-1}$ in intensity, with the same electronic model.
This directly checks application of the same variational
eigenvectors to the transition moments in
Eq.~\eqref{eq:polyadir}. The reference labels are ordered by
the increasing energy of the three polyad roots. Both intensities
use the transition energy times the squared norm of the final
mixed transition moment, with the same perturbative truncation.

\begin{table*}[tp]
\centering
\caption{H$_2$CO Fermi-polyad comparison at B3LYP/6-31G(d).
The two combination-band labels use the Gaussian mode ordering.
Frequencies are in \wavenum{} and integrated intensities in
km\,mol$^{-1}$.}
\label{tab:fermi}
\begin{tabular}{lrrrr}
\toprule
Polyad root & $\nu_{\rm Gaussian}$ & $\nu_{\rm TRINITY}$
 & $I_{\rm Gaussian}$ & $I_{\rm TRINITY}$\\
\midrule
$5_1$       & 2718.760 & 2718.632 & 94.481 & 94.446\\
$6_1+3_1$   & 2844.980 & 2844.817 & 85.990 & 85.929\\
$6_1+2_1$   & 3083.984 & 3083.878 & 4.004 & 3.999\\
\bottomrule
\end{tabular}
\end{table*}

\noindent The ESALC-I calculation on the same saved electronic points gives
untransformed Fermi roots at 2730.484, 2834.204, and
3082.611~\wavenum{}. Transporting its quadratic coordinate transformation
to the Cartesian harmonic gauge brings these eigenvalues to
2718.583, 2844.837, and 3083.880~\wavenum{}, within
0.050~\wavenum{} of ESALC-C root by root.
The complete intensity comparison transforms $D_1,D_2,D_3$ through the
cubic inverse coordinate transformation and includes the second-order
response and normalization of both vibrational states. Mixed
dipole-Jacobian points supply the complete $D_3$ for this control; the
mechanical field uses the single-mode Hessian stencil. Transporting both
the mechanical and dipole fields gives Fermi-root intensities of
94.4109, 85.8363, and 3.9945~km\,mol$^{-1}$. The largest difference
from direct ESALC-C is 0.093~km\,mol$^{-1}$ for these roots and
0.393~km\,mol$^{-1}$ over all retained states. The SI isolate the
property transformation from differences in the acquired mechanical fields.

\subsection{Ammonia: a non-Abelian real-basis control}
NH$_3$ tests the canonical procedure of
Ref.~\onlinecite{mendolicchio2021} on a molecule with $C_{3v}$
symmetry. We optimized the geometry at B3LYP/6-31G* with
Gaussian, projected the final coordinates onto exact $C_{3v}$,
and evaluated the Hessians and Cartesian dipole Jacobians at
that geometry. The ESALC-I coordinate representation contains three N--H lengths
and three H--N--H angles combined into one $A_1$ and one $E$
block for each family. ESALC-C uses the corresponding
Eckart-aligned Cartesian normal modes. Both routes use the
single-mode $\pm0.16$ stencil, the Gaussian SuperFine grid,
and the same 25 electronic points in total (one reference and
12 displaced points per coordinate representation). The residual Cartesian reference
gradient is $4.29\times10^{-7}$~hartree/bohr and is subtracted
consistently from the displaced gradients.

The six harmonic frequencies in the two coordinate representations coincide within
$1.1\times10^{-5}$~\wavenum; the two $E$ pairs have harmonic
frequencies 1726.7604 and 3568.1742~\wavenum. The canonical
VP2TD basis includes the complete near-degenerate manifolds
encountered in this calculation. We explicitly form the real
Cartesian representation of the $C_{3v}$ rotation and reflection
on each polyad and project its effective Hamiltonian,
\begin{equation}
 \overline H_P=\frac{1}{|C_{3v}|}
 \sum_{g\in C_{3v}}D_P(g)^{\mathsf T}H_P D_P(g).
 \label{eq:nh3-real-projection}
\end{equation}
This projection removes finite-grid violations of the exact
symmetry; it does not supply missing electronic derivatives or
alter the anharmonic order. The mode-action covariance residual
is below $6\times10^{-12}$ in both coordinate representations, the polyad basis is
closed under all six operations, and the projected matrix is
invariant to below $9\times10^{-12}$~\wavenum. Its eigenvectors
and transition moments remain real throughout VP2TD. Rotating
the six-state Fermi Hamiltonian into the standard circular basis
afterward preserves its eigenvalues within
$3.2\times10^{-12}$~\wavenum.

Table~\ref{tab:nh3-nonabelian} reports the symmetry-projected spectrum before coordinate transport.

\begin{table}[t]
\caption{NH$_3$ B3LYP/6-31G* projected VP2TD transition energies
(\wavenum) and ground-origin IR intensities (km\,mol$^{-1}$).
An $E$ entry denotes a degenerate pair and reports the sum of
its two intensities. The four rows marked $P$ belong to the
same six-state Fermi polyad.}
\label{tab:nh3-nonabelian}
\begin{ruledtabular}
\begin{tabular}{lrrrr}
Assignment & $\nu_I$ & $\nu_C$ & $I_I$ & $I_C$\\
$A_1$ umbrella mode & 1055.653 & 1055.992 & 158.806 & 158.684\\
$E$ bend & 1681.096 & 1681.107 & 29.259 & 29.257\\
$A_1$ stretch ($P$) & 3243.319 & 3238.298 & 1.579 & 1.536\\
$E$ mixed, lower ($P$) & 3340.756 & 3342.057 & 0.057 & 0.102\\
$A_1$ mixed, upper ($P$) & 3374.004 & 3374.500 & 0.668 & 0.665\\
$E$ stretch, upper ($P$) & 3396.882 & 3397.835 & 3.342 & 3.262\\
\end{tabular}
\end{ruledtabular}
\end{table}
The projected states in Table~\ref{tab:nh3-nonabelian} preserve complete
$E$ pairs. A reproducible Cartesian projector fixes their real
orientation. Comparing the effective Hamiltonians requires consistent
coordinate transport in addition to symmetry projection. The complete
six-state calculation transports Fermi and Darling--Dennison couplings
through second order. Table~\ref{tab:nh3-transport-summary} isolates the
effect of the retained terms; the SI give the step and symmetry controls.
No additional electronic Hessians are required for this transport.

\begin{table}[t]
\caption{NH$_3$ $A_1$ Fermi and six-state polyad comparison at
$h=0.16$. $\Delta\nu$ is the I/C difference for the lowest $A_1$
root in the first three rows and the largest I/C root difference
in the final six-state row.}
\label{tab:nh3-transport-summary}
\begin{ruledtabular}
\begin{tabular}{lr}
Treatment & $|\Delta\nu|$ (\wavenum{})\\
Projected, native coordinate representations & 5.021\\
Mixed $G''$ completed & 4.033\\
Three-state Fermi transport & 0.021\\
Six-state Fermi--Darling--Dennison transport & 0.033\\
\end{tabular}
\end{ruledtabular}
\end{table}

The SI list the transported energies and intensities of all six roots.

The automatic VP2TD route constructs the complete symmetry-closed
polyad and reproduces these six energies. For properties it rotates
the perturbatively dressed native-basis transition moments with the
transported eigenvectors, $\widetilde{\boldsymbol M}^{(r)}
=U^{\mathsf T}R\boldsymbol M^{(r)}$, where $R$ is the orthogonal
native-to-common-gauge occupation-space action. The second-order inverse-coordinate property transport reduces the
maximum I/C root-intensity difference from 0.106 to
0.001~km\,mol$^{-1}$ for the six-state block. The detailed
step-dependence and partner couplings are reported in the SI.
The separate three-quantum and fundamental-origin hot-band tests
use the same spectroscopy solver; this six-state comparison
concerns the $v\leq2$ manifold.
A Gaussian \texttt{Freq=Anharmonic} calculation at this same geometry,
method, and SuperFine grid agrees with the ESALC-C harmonic frequencies
within $5.2\times10^{-5}$~\wavenum{}.  Its anharmonic output
flags five inconsistent directional estimates of cubic or
quartic force constants and cannot apply its symmetry-operation
analysis to the non-Abelian group.  It splits the two components of
the $E$ bending fundamental by 0.515~\wavenum{}, whereas the
symmetry-projected ESALC-C pair is exactly degenerate.  The summed
$E$-bend intensities are 29.262 and 29.257~km\,mol$^{-1}$.
Gaussian therefore supplies the independent harmonic control for
NH$_3$; the symmetry-closed ESALC-I/C comparison tests its
anharmonic polyads.

For a further check, we take both components of the lower $E$
bending fundamental as hot-band origins and both components of
the upper $E$ stretching root as final states. Summing the four
intrinsic strengths gives 44.924 and 43.446~km\,mol$^{-1}$ in
ESALC-I and ESALC-C, respectively, at transition energies
1715.787 and 1716.728~\wavenum. Applying the complex circular
transformation only \emph{after} the real calculation, on both
ends of the $2\times2$ transition-moment matrix as in
Eq.~\eqref{eq:nonabelian-postrotation}, preserves its squared
norm within $9\times10^{-19}$ atomic units in either representation.
Temperature populations are excluded from these intrinsic
hot-band strengths.

\subsection{Measured water hot-band moments}
Reference~\onlinecite{toth1999h2ohot} fitted high-resolution H$_2$O line strengths in the
$(020)\leftarrow(010)$, $(100)\leftarrow(010)$, and
$(001)\leftarrow(010)$ hot bands with an interacting-state
rovibrational model and reported their leading effective transition
moments. We computed the three corresponding
intrinsic vibrational transition vectors from the same B3LYP/6-31G(d)
field in both ESALC coordinate representations, transporting the moment
through second order from the $(010)$ initial state. Table~\ref{tab:water-hot-moments}
compares their norms. The $2\nu_2\leftarrow\nu_2$ result lies within
the reported experimental uncertainty. On the scale of squared fitted
moments, the weaker $\nu_1\leftarrow\nu_2$ and
$\nu_3\leftarrow\nu_2$ bands carry only 4.01\% and 1.70\% of the
dominant band's strength. Their B3LYP-C absolute moment residuals
are 0.00764 and 0.01478~D, respectively. Relative to the fitted
moments, the calculated values are lower by 19.7\% and 58.6\%.
Their squared moments are smaller by factors of 1.55 and 5.83,
respectively; these are substantial deviations for the two weak bands.
The two coordinate representations agree closely for all three transitions. A separate
reconstruction from Gaussian's own single-calculation force and dipole
derivatives reproduces the ESALC-C moments within
$2.1\times10^{-5}$~D (SI), excluding the independent-point stencil
as the source of these gaps. The leading fitted parameters of Ref.~\onlinecite{toth1999h2ohot} and
the isolated vibrational matrix elements are related through the
rotational interaction model; the latter two discrepancies cannot
be assigned uniquely to the electronic dipole surface without
matching that model. The primary molecular comparisons use the same B3LYP field.
The MP2 check with an augmented correlation-consistent polarized
valence triple-zeta basis (aug-cc-pVTZ) below also reproduces the dominant band
and gives smaller moments for the two weak bands.
Small moments can be sensitive to partial
cancellation between direct electrical derivatives and mechanically
induced cubic mixing, and effective weak-band parameters can be
affected proportionally more by rovibrational interference, including
Herman--Wallis terms.
The MP2/aug-cc-pVTZ comparison in Table~\ref{tab:water-hot-moments}
uses the same transition-moment definition. Its dominant-band moment is
0.19550~D; the two weak moments are 0.01904 and 0.00695~D. Both electronic
methods reproduce the dominant transition on the scale of the measured
uncertainty.

\begin{table*}[tp]
\caption{H$_2$O hot-band transition moments (D). The values of Ref.~\onlinecite{toth1999h2ohot} are
leading effective parameters from fitted rovibrational line strengths;
calculated values are norms of intrinsic vibrational transition vectors.
The relative strength is the squared experimental moment normalized to the
$(020)\leftarrow(010)$ value; it excludes thermal populations and
rotational line-strength factors. MP2 uses the single-calculation rectilinear
reconstruction.}
\label{tab:water-hot-moments}
\begin{ruledtabular}
\begin{tabular}{lrrrrr}
Band & Ref.~\onlinecite{toth1999h2ohot} & Relative strength (\%) & B3LYP-C & B3LYP-I & MP2-C\\
$(020)\leftarrow(010)$ & 0.1936(97) & 100.00 & 0.19013 & 0.19017 & 0.19550\\
$(100)\leftarrow(010)$ & 0.03876(19) & 4.01 & 0.03112 & 0.03112 & 0.01904\\
$(001)\leftarrow(010)$ & 0.02523(75) & 1.70 & 0.01045 & 0.01045 & 0.00695\\
\end{tabular}
\end{ruledtabular}
\end{table*}

\subsection{Experimental formaldehyde hot-band origins}
Formaldehyde provides an external hot-band test on the same saved
curvilinear field. We replaced its six B3LYP/6-31G(d) harmonic
frequencies by the gas-phase coupled-cluster singles, doubles, and perturbative-triples
[CCSD(T)]/cc-pVQZ values
$(1190,1271,1537,1781,2933,3004)$~\wavenum{} reported in Ref.~\onlinecite{begue2006formaldehyde}, in the common
$\nu_4,\nu_6,\nu_3,\nu_2,\nu_1,\nu_5$ order. The saved cubic,
semidiagonal quartic, curvilinear metric, and dipole tensors were
retained. Equation~\eqref{eq:composite-harmonic-replacement} was
applied before solving VP2TD, including its resonance denominators
and final polyad. Table~\ref{tab:h2co-hot-experiment} compares the
two $\nu_4$-origin CH-stretch hot-band centers with the measured
band origins~\cite{perez2006h2cohot}. These are vibrational band
centers, so rotational line contours do not enter the comparison.

\begin{table}[t]
\caption{H$_2$CO $\nu_4$-origin hot bands (\wavenum{}). The
composite column uses high-level harmonics inside VP2TD and the
saved B3LYP higher derivatives. The last column is calculated minus
observed.}
\label{tab:h2co-hot-experiment}
\begin{ruledtabular}
\begin{tabular}{lrrr}
Band & Composite & Observed & Difference\\
$\nu_1+\nu_4-\nu_4$ & 2770.079 & 2774.271 & $-4.192$\\
$\nu_5+\nu_4-\nu_4$ & 2816.266 & 2829.262 & $-12.996$\\
\end{tabular}
\end{ruledtabular}
\end{table}

The second band belongs to a CH-stretch Fermi region, where the
quadratic substitution strongly changes the polyad mixing. For the
computed $\nu_4$ lower-state energy of 1170.768~\wavenum{}, its
Boltzmann population relative to the vibrational ground state is
0.00338 at 296~K and 0.0344 at 500~K. The intrinsic transition
moments are temperature independent in this model; thermal
populations activate the hot-band absorption. Room-temperature
spectra have independently resolved 46 H$_2$CO lines from four hot
bands~\cite{germann2024h2cohot}.

Those measured line intensities provide a direct control of the
$2\nu_6\leftarrow\nu_6$ moment. We parsed the published experimental
line lists~\cite{germann2024h2cohot} and paired 29 of its 37 hot-band
lines with $\nu_6\leftarrow0$ lines having identical upper and lower
$(J,K_a,K_c)$. At fixed lower-state $K_a$, the $B_2$ vibrational
symmetry of $\nu_6$ interchanges the ortho and para nuclear-spin
weights relative to the $A_1$ ground state~\cite{alrefaie2015h2co}.
The weight ratio $g_{\nu_6}/g_0$ is 3 for even $K_a$ and $1/3$ for
odd $K_a$. Dividing each measured hot/fundamental line-strength
ratio by that factor gives a median of 0.004545 (interquartile
range 0.004048--0.004815); the even- and odd-$K_a$ subsets yield
0.004583 and 0.004272 separately. At the approximately 294.95~K
measurement temperature, the VP2TD vibrational prediction is
\begin{equation}
 R_{6}^{\rm vib}(T)\simeq
 \exp[-hcE(\nu_6)/(k_{\rm B}T)]\,
 \frac{I_{2\nu_6\leftarrow\nu_6}^{\rm intrinsic}}
      {I_{\nu_6\leftarrow0}^{\rm intrinsic}}.
 \label{eq:h2co-hot-line-ratio}
\end{equation}
The ratio of stimulated-emission factors is taken as unity in
this mid-infrared comparison.
Reference~\onlinecite{alderzi2021h2comarvel} reports the empirical
$\nu_6$ term value
$E^{\rm exp}(\nu_6)=1249.09442(10)$~\wavenum{},
whereas the composite fields give 1250.22319 and
1249.94797~\wavenum{} in ESALC-I and ESALC-C. With their computed
populations, Eq.~\eqref{eq:h2co-hot-line-ratio} gives 0.004345 and
0.004354. Substituting the measured $E(\nu_6)$ in the Boltzmann
factor \emph{only} gives a common population factor of 0.00225833.
The predicted intrinsic intensity ratios remain 1.93454 and
1.93581, giving thermal ratios of 0.004369 and 0.004372, respectively
3.9\% and 3.8\% below the spin-corrected median. Thus the
population correction changes the two ratios by 0.55\% and
0.42\%; at this temperature a hypothetical 10~\wavenum{} error
would instead change a Boltzmann factor by about 4.8\%.
The unmodified B3LYP field with its computed populations gives
0.004153 and 0.004150. Matching the rotational quantum numbers cancels the
leading line-strength factors, although vibration--rotation changes
are not removed exactly. The source gives a 15\% uncertainty estimate
for individual absolute intensities. The paired-line ratio is thus
an experimental test of the hot-band intensity transport in both
coordinate representations, not a ranking of the coordinate representations. Their agreement is
the expected cross-representation consistency for this semi-rigid molecule.
The 3.9\%/3.8\% differences increase to approximately 4.4\%/4.2\%
when the computed lower-state energies are used. All are smaller than
the reported 15\% uncertainty estimate for individual absolute line
intensities and establish compatibility at this precision. The line-pair analysis and the remaining three measured hot-band
classes are documented in the SI.

\subsection{Trans-formic acid}
Trans-HCOOH was calculated at B3LYP-D3BJ/6-31G* from a
very-tight optimized minimum, using analytic Cartesian Hessians and
dipole Jacobians. The nonredundant ESALC-I coordinate representation comprises four
bond lengths, three bond angles, and two torsions, with its Wilson
$B$ and Cartesian derivative $B'$ supplied analytically by the
frozen SMITH definition. The same mass-weighted Eckart frame is
automatically imposed at every internal-coordinate realization.
The standard dimensionless normal-mode step is $h=0.16$.
There are 37 distinct electronic points: one common reference plus
one positive and one negative displacement along each of the nine
normal modes in each representation. No electronic Hessian or dipole
is evaluated at a displacement involving two modes. Mixed $G''$
components are instead obtained by differencing analytic $G'$ at
the single-mode points. The harmonic ESALC-I/ESALC-C maximum
frequency difference is $2.53\times10^{-9}$~\wavenum{}; the
maximum harmonic IR difference is
$5.39\times10^{-10}$~km\,mol$^{-1}$. Thus Eckart orientation and
harmonic equivalence are numerical protocol requirements, not an
assumption inferred from the anharmonic result.

The complete monomer fundamental spectrum is listed in Table~\ref{tab:formic}.

\begin{table*}[tp]
\centering
\caption{Trans-HCOOH at B3LYP-D3BJ/6-31G* with the
single-mode $h=0.16$ stencil. I/C denote ESALC-I/ESALC-C;
frequencies are in \wavenum{} and IR intensities in
km\,mol$^{-1}$. Modes are ordered by harmonic frequency.}
\label{tab:formic}
\scriptsize
\setlength{\tabcolsep}{4pt}
\begin{tabular}{lrrrr}
\toprule
Mode & $\omega_I/\omega_C$ & $\nu_I/\nu_C$
 & $I_I^{\rm harm}/I_C^{\rm harm}$
 & $I_I^{\rm anh}/I_C^{\rm anh}$\\
\midrule
1 & 626.779/626.779 & 620.026/620.000 & 49.583/49.583 & 46.980/47.154\\
2 & 708.605/708.605 & 668.865/668.861 & 153.627/153.627 & 150.407/150.441\\
3 & 1054.205/1054.205 & 1030.377/1030.135 & 0.916/0.916 & 0.674/0.640\\
4 & 1147.048/1147.048 & 1114.566/1114.552 & 238.248/238.248 & 233.945/233.706\\
5 & 1326.407/1326.407 & 1253.514/1256.314 & 8.992/8.992 & 12.522/12.908\\
6 & 1423.172/1423.172 & 1398.236/1397.363 & 4.353/4.353 & 2.605/2.317\\
7 & 1855.470/1855.470 & 1821.067/1821.068 & 315.142/315.142 & 297.316/297.306\\
8 & 3087.287/3087.287 & 2925.290/2925.363 & 53.819/53.819 & 53.171/53.144\\
9 & 3665.706/3665.706 & 3471.096/3471.172 & 40.469/40.469 & 30.653/30.648\\
\bottomrule
\end{tabular}
\end{table*}

The largest difference between the two anharmonic fundamentals is
2.800~\wavenum{} for mode~5; the CH and OH stretching fundamentals
(mode~8 and mode~9) differ by 0.073 and 0.076~\wavenum{},
respectively. The largest anharmonic intensity difference is
0.386~km\,mol$^{-1}$ for mode~5. Both coordinate representations select the same
Fermi-connected mode-5/mode-6 polyad, including the two-quantum
state~19; the corresponding mode-5 coupling magnitudes are
41.726 and 38.076~\wavenum{} in ESALC-I and ESALC-C. This
bare-coupling difference largely reflects the coordinate gauge:
the quadratic coordinate transformation maps the leading ESALC-I coupling
to 38.160~\wavenum{}. Transporting the cubic coordinate transformation and the available field through
second order reduces the largest Fermi-polyad root difference to
0.105~\wavenum{} (SI). Thus the
untransformed difference is not evidence of an inequivalent
semi-rigid description. The remaining residual is consistent with
the finite-order and single-mode-stencil approximations; this test
does not separate their individual contributions. Over all nine
fundamentals in the same evaluation, the maximum frequency and IR
residuals after full field and property transport are
0.242~\wavenum{} and 0.239~km\,mol$^{-1}$.

\subsection{Cyclic formic-acid dimer}
\begin{figure*}[t]
\centering
\includegraphics[width=0.87\textwidth]{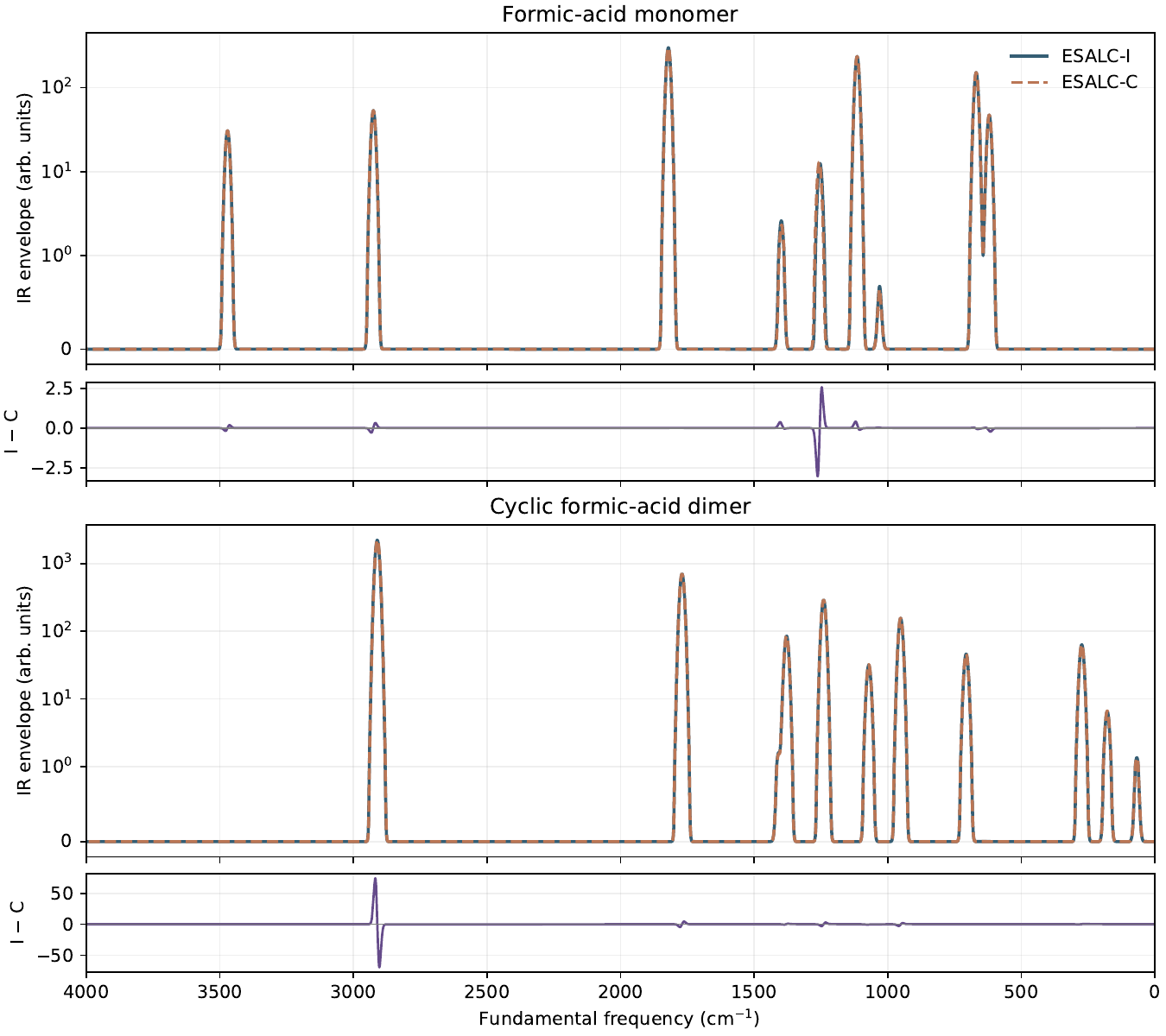}
\caption{VP2TD infrared spectra of trans-formic acid and its cyclic dimer
at B3LYP-D3BJ/6-31G*. ESALC-I and ESALC-C use the same electronic
surface. Lorentzian broadening is applied to the calculated lines;
the common-gauge comparisons are reported in the text and SI.}
\label{fig:formic-ir-envelopes}
\end{figure*}

The cyclic dimer was evaluated on the same B3LYP-D3BJ/6-31G*
surface with the same $h=0.16$ single-mode protocol. Its 24
vibrational modes retain harmonic I/C identity within
$9.6\times10^{-9}$~\wavenum{} and harmonic IR identity within
$6.8\times10^{-7}$~km\,mol$^{-1}$. The largest fundamental
difference is 5.965~\wavenum{} for the weak mode~21; the largest
intensity difference is 3.385~km\,mol$^{-1}$ for mode~24, compared
with an intensity near 940~km\,mol$^{-1}$. The strong mode~22
is 2910.426/2910.553~\wavenum{} and 1279.091/1278.682~km\,mol$^{-1}$
in ESALC-I/ESALC-C. The monomer and dimer band envelopes in
Fig.~\ref{fig:formic-ir-envelopes} visualize both the hydrogen-bond
shift and the small differences between coordinate representations.
These are untransformed differences. Transporting the complete
second-order field and dipole derivatives to the common Cartesian
harmonic gauge reduces the largest fundamental difference to
0.322~\wavenum{} and the largest IR difference to
0.601~km\,mol$^{-1}$. The largest transported polyad-root
residual is 0.392~\wavenum{} (SI). Thus the prominent native
mode-21 and mode-24 differences chiefly reflect the coordinate convention.
For the remaining mode-12 IR residual, the SI one-tensor evaluation
assigns $-0.594$~km\,mol$^{-1}$ to semidiagonal $F_{ijkk}$ terms
already acquired by the single-mode Hessian stencil; terms with
four distinct indices change this band's intensity by less than
$10^{-6}$~km\,mol$^{-1}$. This identifies the source of that
residual without assuming the four-distinct terms are negligible
for other bands.
The Gaussian broadening is for display only; no line-shape model
enters VP2TD.

\section{Localization of couplings and molecular applications}
The internal-coordinate representation exposes the chemical origin of
intermode couplings and separates soft torsions from the remaining
vibrations. Methyloxirane illustrates this localization together with
the common treatment of electric and magnetic properties.

\subsection{Chiral molecular control: methyloxirane}
Methyloxirane supplies a chiral test with measured and modeled
anharmonic VCD and ROA spectra~\cite{fuse2022vcd,yang2022roa}.
The latter study found weak methyl-rotation coupling to many other
modes and a significant contribution from three-quantum transitions
and a CH-stretching Fermi interaction to its ROA spectrum~\cite{yang2022roa}.  We optimized the $R$
enantiomer at B3LYP/6-31G* with the Gaussian electronic-structure program and acquired
harmonic VCD, dynamic Raman and ROA tensors at 532~nm.
Across all 24 modes, the largest absolute differences from the
Gaussian harmonic properties are
$5.0\times10^{-5}$~\wavenum{} in frequency,
$4.8\times10^{-5}$ in VCD rotational strength,
$4.7\times10^{-5}$ in Raman activity,
$4.7\times10^{-4}$ in ROA activity, and
$4.9\times10^{-4}$ in CID, each at or below its printed rounding.
All 24 modes are tabulated in the SI. This check is harmonic:
it validates units, signs, tensor ordering, and the optical-frequency
normalization. The molecular second-order VCD comparison appears
below; the anharmonic ROA calculation is treated separately.

\begin{figure*}[t]
\includegraphics[width=\textwidth]{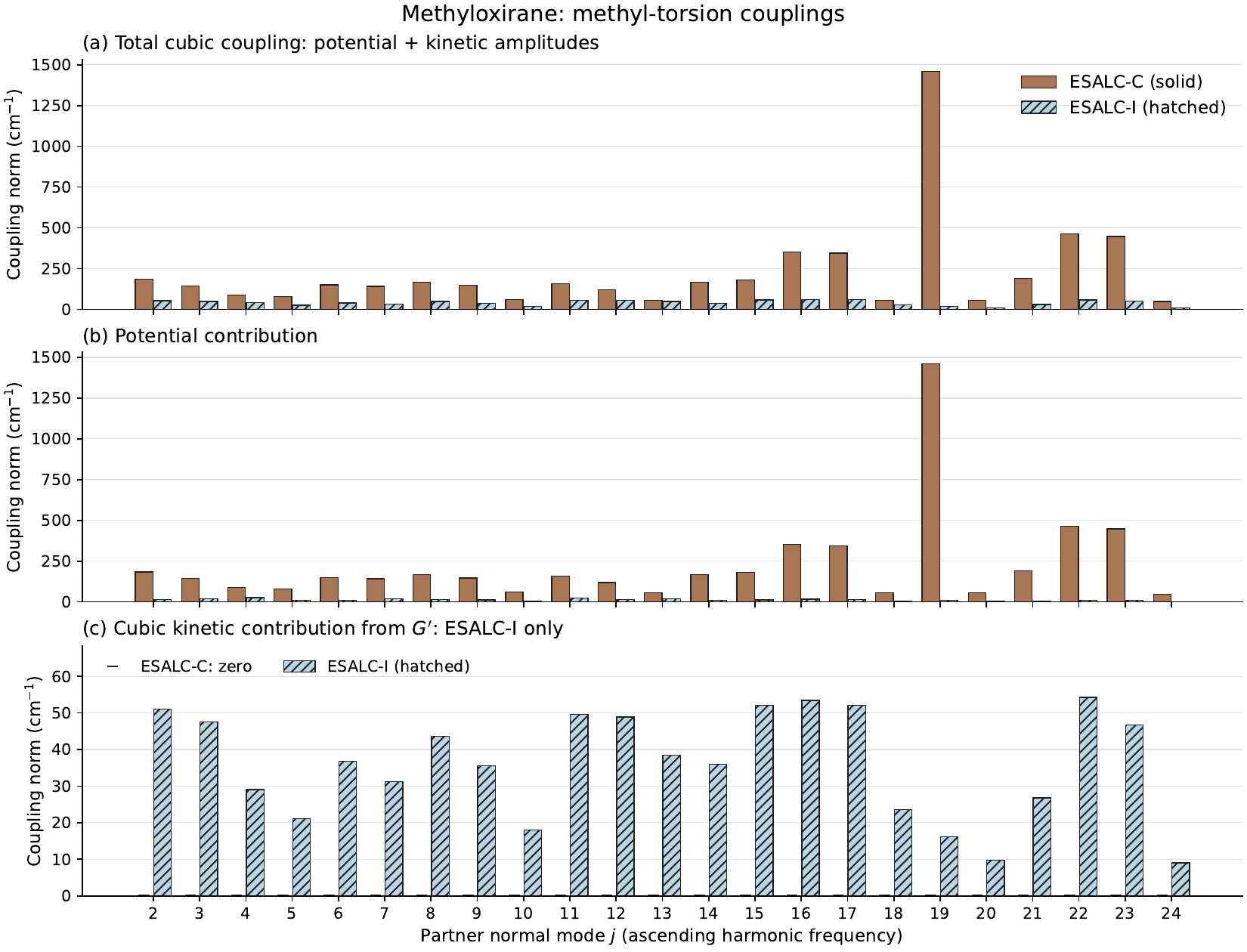}
\caption{Methyl-torsion cubic couplings in $R$-methyloxirane at B3LYP/6-31G*.
The upper panel shows $C_j$, defined in the text; lower panels separate
potential and kinetic amplitudes. Their coherent sum determines the
upper-panel norm. Solid brown bars: ESALC-C; hatched blue bars: ESALC-I.
The kinetic panel contains only ESALC-I bars; ESALC-C is identically
zero (baseline marks). Coriolis enters at second order.}
\label{fig:methyloxirane-torsion-coupling}
\end{figure*}

The complete common-field cubic field also permits a direct test of the
separation of the methyl torsion ($\tau$, mode~1) from the other
normal modes.  We constructed a full-rank ESALC-I coordinate representation with SMITH,
including methyl dihedrals, and transported the same electronic
force field through its analytic Wilson $B'$; ESALC-C and ESALC-I
have the same harmonic eigenvectors and frequencies to numerical
precision. We measure the coupling to partner $j$ by
$C_j=[\sum_{\nu:\nu_j>0}|\langle\nu|H_3^{(\tau)}|1_\tau\rangle|^2]^{1/2}$.
Figure~\ref{fig:methyloxirane-torsion-coupling} compares
the cubic coupling from the torsional fundamental to states
containing each partner mode, retaining only terms of $H_3$ that
contain $\tau$. The Euclidean norm over partners falls from
1761.6 to 206.2~\wavenum{} upon moving from ESALC-C to ESALC-I.
The largest ESALC-C coupling is to mode~19 (a CH stretch),
1460.7~\wavenum{}, compared with 17.8~\wavenum{} in ESALC-I.
At the transition-matrix level, the ESALC-I potential and kinetic
contributions have Euclidean partner norms of 67.0 and
186.4~\wavenum{}, respectively; their coherent sum is used
throughout VP2TD. The ESALC-C vibrational metric is constant, so
its $G'$ contribution vanishes; its separate Coriolis contribution
is handled outside this cubic vibrational comparison.

For ESALC-I VP2TD, the Cartesian
quartic field and dipole surface were re-expanded through the
nonlinear SMITH coordinate representation; 49 inexpensive evaluations of these saved
polynomials and analytic $B'$ supplied the ESALC-I single-mode
semidiagonal field and dipole derivatives. Thus both columns in
Table~\ref{tab:methyloxirane-esalci-vp2td} use identical electronic
information. The harmonic IR intensities coincide within
$3.8\times10^{-7}$~km\,mol$^{-1}$. Reducing the re-expansion step
from 0.02 to 0.01 dimensionless normal-coordinate units changes the
ESALC-I fundamentals by at most 0.047~\wavenum{}, the IR
intensities by at most 0.007~km\,mol$^{-1}$, and the VCD strengths
by at most 0.015 in $10^{-44}$~esu$^2$\,cm$^2$.
The transition intensities
and rotatory strengths are retained through second order in the
perturbation; the electric and axial moments are transported through
the same VP2TD polyad eigenvectors. Harmonic VCD strengths in ESALC-I
and ESALC-C agree with the checkpoint contraction to within
$8.3\times10^{-7}$ in $10^{-44}$~esu$^2$\,cm$^2$; the anharmonic $R$ values in
Table~\ref{tab:methyloxirane-esalci-vp2td} therefore test the
curvilinear property transport and the different mechanical fields.
The CH-stretching region contains strongly mixed polyad roots, whose
positions and intensities are more sensitive to the coordinate
representation than to the tested displacement step. In particular,
the large intensity differences for modes 21 and 23 describe different
mixing within the two truncated representations; without an independent
reference they do not rank their accuracy.

\begin{table*}[tp]
\caption{Selected methyloxirane VP2TD frequencies (\wavenum{}), IR intensities
(km\,mol$^{-1}$), and VCD strengths ($10^{-44}$~esu$^2$\,cm$^2$)
at B3LYP/6-31G*. Both representations use the same electronic field;
ESALC-I uses a semidiagonal stencil with $h=0.01$. Modes follow ascending
harmonic frequencies. The SI list all modes.}
\label{tab:methyloxirane-esalci-vp2td}
\begin{ruledtabular}
\begin{tabular}{rrrrrrrr}
Mode & $\omega$ & $\nu_{\rm C}$ & $\nu_{\rm I}$ & $I_{\rm C}$ & $I_{\rm I}$ & $R_{\rm C}$ & $R_{\rm I}$ \\
1 & 212.27 & 202.89 & 202.87 & 0.43 & 0.44 & 2.44 & 2.95 \\
2 & 364.11 & 364.36 & 364.36 & 3.99 & 4.01 & -12.12 & -12.55 \\
6 & 912.74 & 898.51 & 898.51 & 3.23 & 2.97 & 23.41 & 22.81 \\
16 & 1513.55 & 1472.48 & 1472.65 & 2.43 & 3.22 & 0.64 & 0.47 \\
19 & 3049.94 & 2916.19 & 2916.41 & 4.67 & 3.40 & -0.69 & -0.66 \\
21 & 3106.46 & 2999.46 & 2990.74 & 10.57 & 2.51 & 8.72 & 2.23 \\
23 & 3134.18 & 3004.08 & 3000.88 & 16.26 & 36.44 & -0.60 & -7.90 \\
24 & 3180.46 & 3026.63 & 3026.63 & 35.64 & 47.89 & -5.72 & -7.72 \\
\end{tabular}
\end{ruledtabular}
\end{table*}

The integrated DVR treatment of the hindered methyl rotation uses 24
relaxed geometries at $15^\circ$ intervals. A third-order, threefold
Fourier potential has a barrier of 890.46~\wavenum{}. Its effective
inertia matches the 212.27~\wavenum{} harmonic torsion, incorporating
the local response of the other modes. The first excited torsional
manifold has a centroid of 200.88~\wavenum{}. Grid and potential-fit
controls are given in the SI.

The Supporting Information collects the ESALC-I/C frequency and
intensity differences across the complete molecular set, including the
resonant roots before common-representation transport.

\section{Expectation values and rotation--vibration properties}
\subsection{Thermal expectation values}
The diagonal property expression describes vibrational averaging of an
electronic response, including the specific optical rotation at a fixed
light wavelength. The reduced diagonal formula in the SI gives both the property-curvature
term and the mechanical correction involving cubic force constants and
first metric derivatives. A thermal average replaces each
$n_j+1/2$ by $\tfrac12\coth[hc\omega_j/(2k_{\rm B}T)]$.
For a state-specific expectation value, we have also implemented the
next nonzero order beyond the standard second-order vibrational
average: the fourth-order correction, including Hamiltonian terms
through $H_5$. Here the subscript denotes the coordinate degree,
so $H_5$ contains the quintic potential and the third metric derivative.
The correction includes the fourth property derivative and the normalized
$P_3H_3$, $P_2H_4$, $P_2H_3^2$, $P_1H_3H_4$, $P_1H_3^3$, and
$P_1H_5$ contributions. The SI give the explicit reduced expressions.
For the one-dimensional torsional application below, the fifth potential
and third metric derivatives are obtained from the fitted periodic
surfaces. Including them gives practically exact agreement with the
DVR ground-state property average for conformer C1, as quantified below.
A diagonal Morse estimate of $P_1H_5$ is also available when only
lower derivatives are known, with a quartic consistency diagnostic.
Rotation--vibration property coefficients are
vibrationally averaged by the same equations and contracted with the
angular-momentum second moment of the selected rotational state.

\subsection{Optical rotation of $(S)$-2-methylaziridine}
Reference~\onlinecite{regan2026optical}
provides a molecular reference for $(S)$-2-methylaziridine. At 300~K
and 633~nm the reported B3LYP calculation changes the equilibrium specific
rotations of its trans and cis conformers from 16.73 and $-59.97$ to
23.04 and $-62.61$~deg\,dm$^{-1}$\,(g/mL)$^{-1}$, respectively.
These published values illustrate the magnitude and conformer dependence
of the vibrational correction to an expectation value.

The methyl torsions, at 209.21 and 225.52~\wavenum{} in the
two conformers, supply a direct application of the integrated
periodic DVR of Sec.~\ref{sec:adiabatic-dvr}: the optical rotation is
evaluated along the torsional
path and averaged with the probability density of each torsional
state, followed by a thermal sum. The remaining modes retain their
perturbative property corrections. Each conformer has its own potential,
metric, and property curve; their final averages are combined with
consistent conformer populations. The SI specify this construction
and the exclusion of the torsion from the perturbative bath. The local
one-dimensional perturbative series and the periodic DVR average are
evaluated from the same torsional potential and property surface, making
the latter a numerical reference for the successive diagonal corrections.
The relaxed electronic barriers are 830.59 and 963.59~\wavenum{} for
the trans (C1) and cis (C2) conformers. For the C1 torsional ground state, the fourth-order
property average including $H_5$ (the fitted $V^{(5)}$ and $G'''$ terms)
agrees practically exactly with DVR: the absolute differences are 0.004, 0.071, and 0.064 in specific-rotation units at
355, 589.3, and 633~nm. The corresponding C2 differences are 2.140,
0.447, and 0.377. The 300~K DVR averages of the explicitly sampled
electronic property are $(144.63,25.09,20.66)$ for C1 and
$(-314.19,-74.70,-62.98)$ for C2. The local expansion is therefore an
accurate compact representation when its successive terms and Fourier
fit are stable; the periodic DVR supplies the final LAM result and all
thermally populated torsional states. The SI report the complete
decomposition.

The comparison with the original calculation is performed by replacing
only its 300-K perturbative contribution from the methyl torsion, while
retaining the published contribution of the other 26 modes. At the common
B3LYP/aug-cc-pVTZ level, the independently calculated equilibrium optical
rotations agree within 0.026 specific-rotation units, and the present
barriers of 830.59 and 963.59~\wavenum{} reproduce the published values of
830.6 and 963.8~\wavenum{}. Removing the torsion from both indices of the
published mechanical correction isolates its complete perturbative
contribution before replacement by DVR. The resulting changes in the total
optical rotation are $(-0.32,0.02,0.02)$ for C1 and
$(0.49,0.11,0.09)$ for C2 at 355, 589.3, and 633~nm, respectively.
Adding the DVR torsional increments to the published
B3LYP/d-aug-cc-pVQZ equilibrium properties gives
$(136.88,22.66,18.57)$ for C1 and
$(-293.78,-70.01,-59.04)$ for C2. Including the published correction
from the other 26 modes gives the complete composite values
$(171.72,28.11,23.05)$ and $(-312.59,-74.12,-62.52)$, respectively.
The 300-K perturbative calculation of Ref.~\onlinecite{regan2026optical} and the
thermal perturbative expression implemented here are formally identical:
thermal averaging replaces every $n_j+1/2$ by
$\tfrac12\coth[hc\omega_j/(2k_{\rm B}T)]$.  Direct evaluation with their
published mode-resolved data reproduces their result within the rounding
of the tabulated quantities.  The comparison reported in the SI therefore
contrasts this common second-order perturbation theory (PT2) result
with the independent periodic-DVR
average.  Thermal averaging generally reduces the visible differences
among the local perturbative and periodic descriptions because several
populated torsional levels sample a broader, smoother portion of the same
property surface.  For the present conformers this reduction is clear at
589.3 and 633~nm and particularly for C2; the wavelength-dependent C1 value
at 355~nm shows the wavelength dependence of this trend.  The close 300-K values establish that PT2 is already adequate
for this rotor, while DVR supplies the finite-amplitude reference.  The
local standard, quartic-force-field (QFF), and higher-order entries
in the SI describe the separate torsional ground-state comparison.
The full wavelength-resolved comparison and its algebraic definition are
given in the SI.

\subsection{Rotation--vibration constants and centrifugal distortion}
The full metric in Eq.~\eqref{eq:metric} also supplies rotational
observables, without another electronic-structure stencil.  In the
Eckart body frame let its contravariant blocks be $G^v_{ij}$,
$G^{vr}_{i\tau}$, and $G^r_{\tau\eta}$, where $i,j$ label
dimensionless vibrational normal coordinates and $\tau,\eta$ label
principal rotational axes.  Metric entries and their derivatives
below are converted to wavenumber units, so that
$G^v_{ij}(0)=\omega_i\delta_{ij}$ and
$B^{\rm e}_{\tau}=G^r_{\tau\tau}(0)/2$.
Define $u_{i\tau j}=G^{vr}_{i\tau,j}$ and use the same curvilinear
cubic tensor as in the energy equations,
$\rho_{jii}=f_{jii}+G^v_{ii,j}$.  The latter identity follows
directly from Eq.~\eqref{eq:effective-tensors} for the repeated indices.
The first-order vibration--rotation interaction constant is
\begin{align}
\alpha_{i\tau}={}&-\frac14G^r_{\tau\tau,ii}
-\frac12\sum_{j\ne i}
 \frac{(u_{i\tau j}^{2}+u_{j\tau i}^{2})\omega_j
       -2u_{i\tau j}u_{j\tau i}\omega_i}
      {\omega_i^2-\omega_j^2}
\nonumber\\
&+\frac14\sum_j\frac{\rho_{jii}G^r_{\tau\tau,j}}{\omega_j}.
\label{eq:rovib-alpha}
\end{align}
For exactly degenerate modes, individual $\alpha_{i\tau}$ depend
on the partner basis; only their sum over the degenerate manifold
is reported as an invariant.  Near-degenerate denominators in
Eq.~\eqref{eq:rovib-alpha} are therefore not used to define
individual partner constants.

The ground-state correction is better evaluated directly in the
denominator-free form of Ref.~\onlinecite{mendolicchio2024rovib},
\begin{align}
\Delta B^{\rm vib}_{\tau}
&\equiv B^0_{\tau}-B^{\rm e}_{\tau}
=-\frac12\sum_i\alpha_{i\tau}\nonumber\\
&=\frac18\sum_i\left[
 G^r_{\tau\tau,ii}
 -\sum_j\frac{(u_{i\tau j}+u_{j\tau i})^2}{\omega_i+\omega_j}
 \right]\nonumber\\
&\quad-\frac18\sum_{ij}
 \frac{\rho_{jii}G^r_{\tau\tau,j}}{\omega_j}.
\label{eq:rovib-delta}
\end{align}
The three contributions are respectively rotational-metric
curvature, vibration--rotation coupling, and anharmonic displacement.
Unlike individual near-degenerate interaction constants, their total
ground-state correction has no difference-frequency denominator.
The quartic centrifugal-distortion tensor and the pair-axis
Kivelson--Wilson combination are
\begin{align}
\tau_{\gamma\delta\tau\eta}
 &=-\frac12\sum_i
 \frac{G^r_{\gamma\delta,i}G^r_{\tau\eta,i}}{\omega_i},
 \label{eq:rovib-tau}\\
\tau'_{\tau\tau\eta\eta}
 &=\tau_{\tau\tau\eta\eta}
   +2\delta_{\tau\eta}\tau_{\tau\eta\tau\eta}.
 \label{eq:rovib-tau-prime}
\end{align}
In this convention the effective rotational Hamiltonian contains
\begin{equation}
 H_{\rm eff}^{r}=\sum_\tau B_\tau J_\tau^2
 +\frac14\sum_{\tau\eta}\tau'_{\tau\tau\eta\eta}
 J_\tau^2J_\eta^2+\cdots .
 \label{eq:rovib-effective-rotor}
\end{equation}
We retain the full
four-axis tensor before forming Eq.~\eqref{eq:rovib-tau-prime};
the conversion to a particular reduced spectroscopic Hamiltonian
is separate from its acquisition.

For both $X=C$ and $I$, the calculation procedure samples the same complete
body-fixed metric at $q=0$ and $q_i=\pm h_g$, with
$h_g=0.02$ here.  It transforms each block and derivative with
the frozen GF normal-mode matrix.  The metric points require only
Cartesian geometries and masses, not additional electronic
Hessians or dipoles.  Eckart alignment is applied at every point,
and principal axes within degenerate rotational subspaces are fixed
by a real body-frame gauge.  The implementation accepts three
rotational axes for nonlinear molecules and the two transverse
axes for linear molecules.  This common-metric procedure makes
ESALC-C and ESALC-I subject to the same equations and numerical
checks; their different geometry coordinate representations determine the derivatives.

\subsection{Ammonia rotation--vibration constants}
A two-mode field with nonzero force and rovibrational-metric
couplings verifies the rotational correction against an established
implementation.\cite{mendolicchio2024rovib} The three components
agree within $9.1\times10^{-13}$~MHz.
We applied Eqs.~\eqref{eq:rovib-delta}--\eqref{eq:rovib-tau-prime}
to the saved NH$_3$ ESALC-I and ESALC-C fields.  The complete
rovibrational metric was evaluated on the geometry-only
$q_i=\pm0.02$ stencil; these points require only molecular geometries.
Table~\ref{tab:nh3-rovib} reports the equilibrium constants,
ground-state corrections, and diagonal elements of the unreduced
quartic tensor.  The two transverse principal axes are averaged
for display in this symmetric top; individual transverse values
retain small finite-grid differences.  The maximum unaveraged
I/C difference in $\Delta B^{\rm vib}$ is
$2.61\times10^{-4}$~\wavenum{}, while the largest difference
in the diagonal distortion elements is
$4.34\times10^{-8}$~\wavenum{}.
Changing the geometry-only metric step from $0.02$ to $0.01$
or $0.04$ changes any $\Delta B^{\rm vib}$ by at most
$5.01\times10^{-6}$~\wavenum{} and any element of the quartic
tensor by at most $1.61\times10^{-7}$~\wavenum{} over both coordinate representations.
These comparisons test the local rotation--vibration treatment and
its numerical stability in both representations.

\begin{table*}[tp]
\caption{NH$_3$ B3LYP/6-31G* rovibrational quantities (\wavenum{}).
$\perp$ denotes the mean over equivalent transverse axes; $\parallel$
is the symmetry axis. The unreduced diagonal distortion tensor is shown;
its Kivelson--Wilson coefficient is three times this value
[Eq.~\eqref{eq:rovib-tau-prime}].}
\label{tab:nh3-rovib}
\begin{ruledtabular}
\begin{tabular}{lrrrrrr}
Axis & $B^{\rm e}_I$ & $B^{\rm e}_C$ &
$\Delta B^{\rm vib}_I$ & $\Delta B^{\rm vib}_C$ &
$\tau_{I,\tau\tau\tau\tau}$ & $\tau_{C,\tau\tau\tau\tau}$\\
$\perp$ & 9.770333 & 9.770333 & -0.088475 & -0.088725 &
-0.002678803 & -0.002678760\\
$\parallel$ & 6.328903 & 6.328903 & -0.126742 & -0.126585 &
-0.000669982 & -0.000669980\\
\end{tabular}
\end{ruledtabular}
\end{table*}

\section{Large-amplitude motion and thermochemistry}
\subsection{Periodic hydrogen-peroxide torsion}
The HOOH torsion illustrates the periodic DVR construction of
Sec.~\ref{sec:adiabatic-dvr}. We scanned the dihedral rigidly at
13 points from $0$ to $\pi$ with Gaussian
B3LYP/6-31G* and fitted a real cosine potential. A 256-point
periodic DVR with the Eckart-projected angle-dependent inertia and
kinetic operator $-\partial_\phi B(\phi)\partial_\phi$, where
$B=\mathcal G/2$, gives a
ground-state tunnelling gap of 25.45~\wavenum{} and the next
transition at 244.18~\wavenum{}. A sixth-order Fourier fit reproduces
the sampled potential within 0.20~\wavenum{}; increasing the DVR
grid from 128 to 256 points changes the first six transitions by
less than $2\times10^{-10}$~\wavenum{}. These levels characterize
the double-well torsion on the sampled rigid-path potential; the
local VP2TD field used in Table~\ref{tab:h2o2} is acquired separately.

\subsection{Regularized perturbation theory and thermochemistry}
Spectroscopic VP2TD treats selected near-degenerate states by explicit
polyad diagonalization. For a thermodynamic sum over many occupation
vectors, the state-dependent selection of polyads is inconvenient.
We therefore provide the complementary hybrid degeneracy-corrected
second-order perturbation theory (HDCPT2) in ESALC-I coordinates,
following the curvilinear formulation of
Ref.~\onlinecite{mendolicchio2025thermo}. It uses the same cubic potential,
metric derivatives, quartic expectation values, and equilibrium
geometric scalar potential as the spectroscopic calculation. The resulting smooth energy expression is suited to thermodynamic
sums over many states.

For a cubic coupling $h_{\mathbf v\mathbf m}=\mel{\mathbf v}{H_1}
{\mathbf m}$ and harmonic separation
$\Delta_{\mathbf v\mathbf m}=E_{\mathbf v}^{(0)}-E_{\mathbf m}^{(0)}$,
the ordinary and degeneracy-corrected contributions are
\begin{align}
 C^{\rm PT2}_{\mathbf v\mathbf m}
 &=\frac{|h_{\mathbf v\mathbf m}|^2}{\Delta_{\mathbf v\mathbf m}},
 \nonumber\\
 C^{\rm DC}_{\mathbf v\mathbf m}
 &=\operatorname{sgn}(\Delta_{\mathbf v\mathbf m})
 \left[\sqrt{\epsilon^2+|h_{\mathbf v\mathbf m}|^2}-\epsilon\right],
 \qquad \epsilon=\frac{|\Delta_{\mathbf v\mathbf m}|}{2},
 \label{eq:hdcpt2-pair}
\end{align}
with the symmetric limiting value zero at exact degeneracy. Their
hybrid is $C^{\rm H}=\Lambda C^{\rm PT2}+(1-\Lambda)C^{\rm DC}$, where
\begin{equation}
 \Lambda=\tfrac12\{1+\tanh[\alpha(|h_{\mathbf v\mathbf m}|\epsilon
 -\beta)]\}.
 \label{eq:hdcpt2-switch}
\end{equation}
The implementation uses $\alpha=1$ and
$\beta=5\times10^5$ in its wavenumber-based convention, as in
Ref.~\onlinecite{mendolicchio2025thermo}. Both potential and
curvilinear kinetic cubic couplings enter $h$. The energy evaluated
at the ground state, each fundamental, each first overtone, and
each two-mode combination defines a regularized effective matrix
without fitting a separate force field:
\begin{align}
 \nu_i^{\rm H}&=E_{1_i}^{\rm H}-E_{\mathbf0}^{\rm H},\nonumber\\
 x_{ii}^{\rm H}&=\tfrac12(E_{2_i}^{\rm H}-2E_{1_i}^{\rm H}
 +E_{\mathbf0}^{\rm H}),\nonumber\\
 x_{ij}^{\rm H}&=E_{1_i1_j}^{\rm H}-E_{1_i}^{\rm H}
 -E_{1_j}^{\rm H}+E_{\mathbf0}^{\rm H}\quad(i\ne j).
 \label{eq:hdcpt2-x}
\end{align}
The resulting thermochemical model uses the
explicit quadratic function of integer occupations
\begin{equation}
 \Delta E^{\rm H}(\mathbf v)=\sum_i\nu_i^{\rm H}v_i
 +\sum_i x_{ii}^{\rm H}v_i(v_i-1)
 +\sum_{i<j}x_{ij}^{\rm H}v_iv_j.
 \label{eq:hdcpt2-quadratic}
\end{equation}
In the quadratic VPT2 limit, $x_{ij}^{\rm H}$ has the same convention
as $\chi_{ij}$ of Eq.~\eqref{eq:energy-algebraic}. Because the hybrid
switch depends on the individual state coupling, this quadratic form
reproduces the levels used in Eq.~\eqref{eq:hdcpt2-x} exactly but is
an effective extrapolation for higher occupations.

The simple perturbation theory (SPT) thermodynamic estimate uses $E_{\mathbf0}^{\rm H}$
and the regularized fundamentals as independent oscillator
frequencies:
\begin{equation}
 Q_{\rm SPT}(T)=\exp[-b E_{\mathbf0}^{\rm H}]
 \prod_i[1-\exp(-b\nu_i^{\rm H})]^{-1},
 \label{eq:hdcpt2-spt}
\end{equation}
Here $b=hc/(k_{\rm B}T)$.
Alternatively, TRINITY sums Eq.~\eqref{eq:hdcpt2-quadratic}
explicitly through a finite occupation shell and continues the
remaining tail with positive oscillator transition frequencies
matched at that shell. The same quadratic expression is supplied
to integer-occupation Wang--Landau\cite{wanglandau2001} sampling, whose density of states
is normalized to the number of admissible occupation vectors in a
finite domain. In either route, the domain must terminate before a
softening quadratic model yields nonpositive excitation energies;
the resulting partition function is cutoff dependent and the
highest-shell population or matched-tail weight must be reported.
For a large-amplitude coordinate described by a one-dimensional
adiabatic DVR, we use its bound eigenvalues directly rather than
extrapolating a quadratic $x_{ii}$ to high occupation.  To provide a
compact, smooth representation of a computed ladder, define its
positive gaps $d_n=E_n^{\rm DVR}-E_{n-1}^{\rm DVR}$ and fit
\begin{align}
 d_n^{\rm G}&=d_\infty\exp\!\left[
   \sum_{k=1}^{N_G} a_k
   \exp\!\left(-\frac{(n-c_k)^2}{2s^2}\right)\right],\nonumber\\
 E_n^{\rm G}-E_0^{\rm G}&=\sum_{m=1}^{n}d_m^{\rm G}.
 \label{eq:gaussian-dvr-ladder}
\end{align}
Here $d_\infty>0$ and $0\leq n<N_{\rm bound}$. The parameters
$c_k$, $a_k$, and $s$ are fitted to the DVR gaps and
$d_\infty$ is a positive limiting gap.  The Gaussian correction to
the \emph{logarithm of the gap} vanishes away from its centers, and
the exponential ensures $d_n^{\rm G}>0$; hence the fitted ladder
cannot turn downward as a quadratic occupation model can.  This is
a representation of the computed bound levels: $N_{\rm bound}$ is fixed by the
adiabatic well boundary.  The conformer-local torsional partition is
\begin{equation}
 Q_{\rm tor}(T)=e^{-b E_0^{\rm DVR}}
 \sum_{n=0}^{N_{\rm bound}-1}
 e^{-b(E_n^{\rm G}-E_0^{\rm G})}.
 \label{eq:gaussian-dvr-partition}
\end{equation}
The same finite level table can replace the diagonal quadratic term
in an integer-occupation Wang--Landau calculation, while any
retained intermode $x_{ij}$ terms remain explicit.  The population
of the highest bound level diagnoses the thermal range in which the
conformer-local partition remains adequate.
For a stationary first-order saddle, the corresponding bound-mode
partition is formed on the stable subspace $\mathcal B$ only:
\begin{equation}
 Q_{\rm vib}^{\ddagger}(T)
 =e^{-b E_{0,\mathcal B}^{\ddagger}}
  \sum_{\mathbf v\in\mathcal D_{\mathcal B}}
  e^{-b\Delta E_{\mathcal B}^{\ddagger}(\mathbf v)}.
 \label{eq:ts-bound-partition}
\end{equation}
The unstable reaction coordinate is excluded from the state-count
normalization and requires a separate transmission or tunneling
calculation. Equation~\eqref{eq:ts-bound-partition} defines the use
of the generalized Wang--Landau sampler for bound transition-state
modes; no molecular saddle calculation is included in the results
below.
The equilibrium geometric scalar contributes to the zero-point
energy and absolute partition function, but cancels from excitation
frequencies and $x^{\rm H}$.

\subsection{Adiabatic glycine torsion and thermochemistry}
On the B3LYP-D3BJ/6-31G* glycine-I field, ESALC-I and ESALC-C
have identical harmonic spectra to numerical precision. The selected
C--C torsional Fermi-channel norm falls from 3865.99 to
279.14~\wavenum{} in the internal coordinate representation. This measured reduction
localizes the coupling around a chemically recognizable motion.
The one-dimensional adiabatic DVR gives a first torsional gap of
75.905~\wavenum{} and an IR intensity of
4.336~km\,mol$^{-1}$; the four-moving-bath model with spectator
and dipole-response corrections gives 73.978~\wavenum{} and
5.974~km\,mol$^{-1}$. An independent fourth-order M{\o}ller--Plesset perturbation theory
(MP4)/cc-pVTZ calculation
reported 78 and 86~\wavenum{} with one- and three-dimensional
torsional models, respectively~\cite{senent2005glycine}. The
surfaces differ, and no matched external torsional intensity is
available. The thermochemistry of both conformers is reported in
Table~\ref{tab:glycine-thermo}.

\begin{table}[t]
\caption{Glycine conformer comparison at B3LYP-D3BJ/6-31G* and
298.15~K. $E_0$ is the vibrational zero-point energy in \wavenum{};
$S_{\rm vib}$ is in J\,mol$^{-1}$\,K$^{-1}$; and
$\Delta F=F_{\rm II}-F_{\rm I}$ in kJ\,mol$^{-1}$ includes the
2.108~kJ\,mol$^{-1}$ electronic difference.}
\label{tab:glycine-thermo}
\centering
\resizebox{\columnwidth}{!}{%
\begin{tabular}{lrrrrr}
\toprule
Model & $E_{0,\rm I}$ & $E_{0,\rm II}$ & $S_{\rm I}$ & $S_{\rm II}$ & $\Delta F$\\
Harmonic & 17592.322 & 17731.553 & 44.104 & 35.732 & 5.345\\
SPT & 17296.468 & 17519.986 & 40.423 & 44.766 & 3.566\\
Adiabatic DVR & 17296.468 & 17519.986 & 40.784 & 40.127 & 4.482\\
Wang--Landau & 17296.468 & 17519.986 & 40.933 & 39.973 & 4.538\\
\bottomrule
\end{tabular}
}%
\end{table}

The SPT substitution of conformer II's 100.756~\wavenum{}
harmonic torsion by its 42.380~\wavenum{} first DVR gap adds
7.134~J\,mol$^{-1}$\,K$^{-1}$ to its entropy; changes in the 23
bath frequencies add 1.900~J\,mol$^{-1}$\,K$^{-1}$. For conformer
I the corresponding torsion and bath terms are $-1.375$ and
$-2.306$~J\,mol$^{-1}$\,K$^{-1}$. The opposite signs explain the
SPT entropy trend and its $-0.916$~kJ\,mol$^{-1}$ shift in
$\Delta F$ relative to the finite DVR. The first-gap harmonic
ladder overcounts higher torsional states of the confined conformer
II well. Wang--Landau agrees with direct finite-ladder summation
within 0.15~J\,mol$^{-1}$\,K$^{-1}$; the explicit-$x_{ij}$
algorithmic test is in the SI.

At 410~K, adding the rigid-rotor conformer ratio gives
$\Delta G_{\rm II-I}=4.695$~kJ\,mol$^{-1}$ for the finite-DVR
model. The independent composite electronic-structure and
hindered-rotor treatment reports 4.41~kJ\,mol$^{-1}$, while its
0~K relative energy is 3.73~kJ\,mol$^{-1}$ against
4.782~kJ\,mol$^{-1}$ here~\cite{barone2013glycine}. The 410~K
agreement must therefore be read alongside the different 0~K
reference. Detailed level sums and the rotational correction
are in the SI.

\section{A larger molecular example: naphthalene}
\begin{figure*}[t]
\centering
\includegraphics[width=0.90\textwidth]{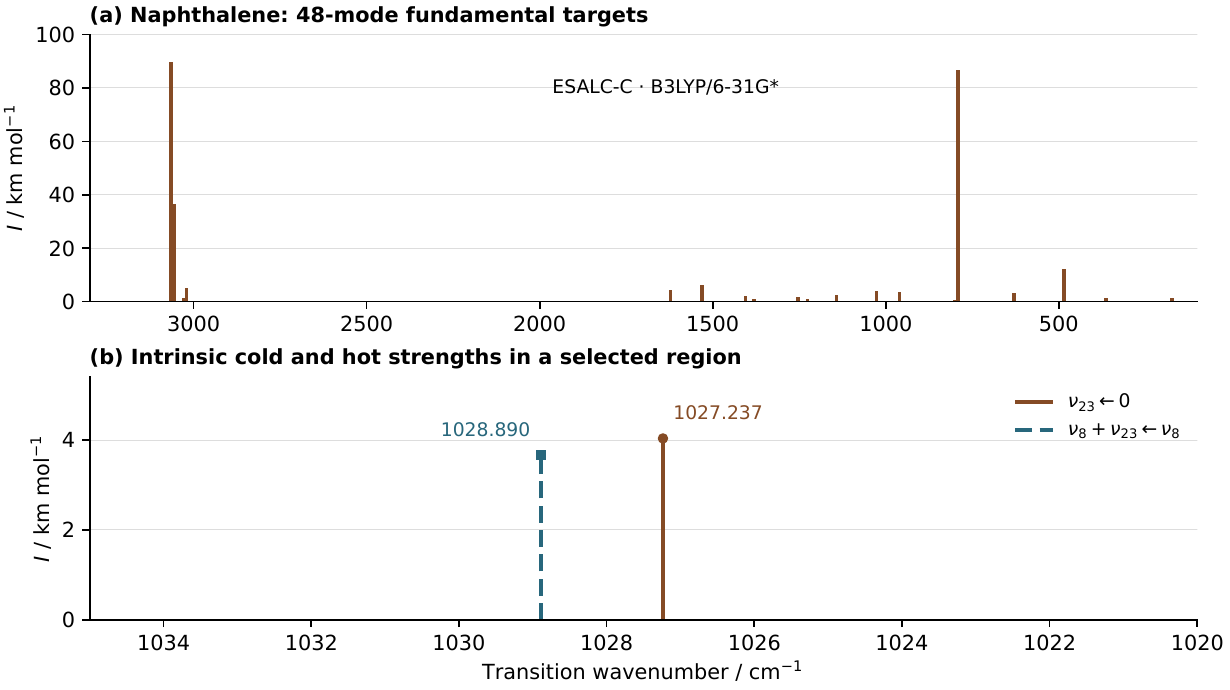}
\caption{Naphthalene ESALC-C VP2TD spectrum at B3LYP/6-31G*.
Top: fundamental-character transitions. Bottom: the mode-23 fundamental
and a selected hot transition from mode~8 into a mixed combination
state. Here modes are ordered by increasing harmonic frequency.
Stick heights are intrinsic integrated IR intensities per occupied
initial level; no thermal population factors are applied.}
\label{fig:naphthalene-spectrum}
\end{figure*}

Naphthalene extends the calculation to 48 vibrational modes on the
same B3LYP/6-31G* electronic level. Its ESALC-C field uses
single-mode displacements of $h=0.16$, the semidiagonal quartic
potential, and second-order transition-property transport. Selection
from all 48 fundamental targets gives 64 states and nine Fermi
polyads, the largest containing six states. The variational domain
allows three quanta in total and per mode, with a target energy
threshold of 0.01~\wavenum{}. Figure~\ref{fig:naphthalene-spectrum}
shows the fundamental-character transitions and a resolved example
of a hot band. The complete spectra from 28 populated fundamental
origins are evaluated in 40.0~s; the total spectroscopic calculation,
including state selection and ground-state intensities, takes
86.9~s on six parallel processes. The SI specify the state space
and numerical results.

Gas-phase measurements provide four fundamental origins and three
hot-band progression spacings.\cite{pirali2009naphthalene}
Table~\ref{tab:naphthalene-hot-spacings} compares the spacings with
the corresponding anharmonic constants. Their mean absolute deviation
is 0.177~\wavenum{}, compared with 6.82~\wavenum{} for the four
absolute fundamentals. Spacings isolate intermode anharmonicity by
cancelling the harmonic carrier frequency. The calculated intensity
example additionally shows how the same property operator acts on
an occupied initial mode and on a mixed final state; these are
intrinsic vibrational strengths, before thermal population weighting.
This 48-mode application combines the algebraic transition formulas,
finite resonance polyads, and parallel property evaluation in a
single molecular calculation.

\begin{table}[t]
\caption{Naphthalene hot-band progression spacings (\wavenum{}).
Mode labels follow Ref.~\onlinecite{pirali2009naphthalene}.
For distinct modes the spacing is $x_{ij}$; the same-mode
progression gives $2x_{ii}$.}
\label{tab:naphthalene-hot-spacings}
\begin{ruledtabular}
\begin{tabular}{lrrr}
Spacing & VP2TD & Experiment & Difference\\
$x_{46,48}$ & $-0.351$ & $-0.300$ & $-0.051$\\
$x_{45,48}$ & $-0.115$ & $-0.290$ & $+0.175$\\
$2x_{47,47}$ & $+0.594$ & $+0.900$ & $-0.306$\\
\end{tabular}
\end{ruledtabular}
\end{table}

\section{A larger flexible example: aspirin}
Aspirin conformer II provides a 21-atom, 57-mode test in which a
high-level harmonic field and a lower-level quartic field are both
available.  The dual-level calculation uses the
revDSD-PBEP86-D4/6-31G* equilibrium Hessian for the normal coordinates,
harmonic frequencies, perturbative denominators, and resonance selection.
The B3LYP/6-31G* cubic and semidiagonal quartic tensors are transported to
that basis with an orthogonal block Duschinsky transformation before the
VP2TD step.  The transformation is nearly diagonal outside a four-mode
block below 110~\wavenum{} and a three-mode block near 1490~\wavenum{};
the complete map is shown in the SI.  Twenty-two Fermi interactions form
13 finite polyads, with maximum dimension five.

\begin{figure*}[t]
\centering
\includegraphics[width=0.90\textwidth]{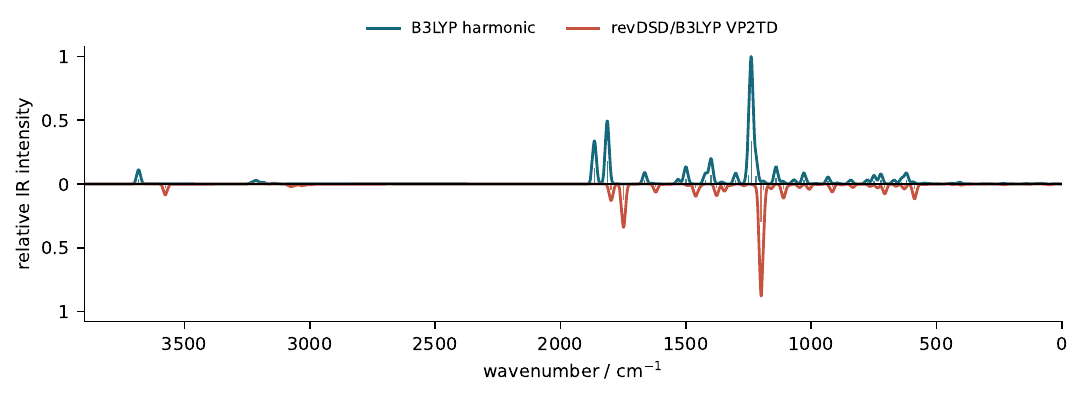}
\caption{Aspirin conformer-II IR spectrum.  The B3LYP/6-31G* harmonic
spectrum is plotted upward and the revDSD/B3LYP dual-level VP2TD spectrum
downward.  Sticks are broadened with Gaussian functions of
18~\wavenum{} full width at half maximum.}
\label{fig:aspirin-dual-spectrum}
\end{figure*}

Figure~\ref{fig:aspirin-dual-spectrum} compares the B3LYP harmonic
spectrum with the dual-level anharmonic band origins.  The latter use the
B3LYP second-order anharmonic intensities carried with the assigned
Duschinsky roots.  The mirrored representation separates shifts and
intensity redistribution throughout the complete range and in the
fingerprint region.  This calculation demonstrates that the same
algebraic and sparse-polyad workflow used for the small validation systems
applies without changing the formalism to a flexible molecule with 57
vibrational modes.

The lowest five normal modes require a separate large-amplitude analysis.
The 26.79~\wavenum{} root is a counter-rotation of the carboxyl and
acetoxy substituents, while the methyl rotor is distributed over the
75.98--107.80~\wavenum{} Duschinsky block.  A two-dimensional adiabatic
DVR was therefore constructed in the collective substituent and methyl
internal-rotation coordinates.  The B3LYP/6-31G* surface uses 14 axial and
eight mixed constrained optimizations, a periodic Fourier representation
for the methyl rotor, and the complete $2\times2$ kinetic matrix.  The
first collective and methyl-like gaps are 27.17 and 84.75~\wavenum{},
respectively.  Their changes relative to independent one-dimensional DVRs
are only $-0.36$ and $-0.11$~\wavenum{}, although the shift reaches
2.35~\wavenum{} among the first ten torsional levels.  Separate
one-dimensional DVRs are consequently sufficient for the two
fundamentals, while the two-dimensional treatment supplies the coupled
upper torsional manifold.  The surface and convergence results are given
in the SI.

\section{Conclusions}
Anharmonic vibrational spectroscopy requires a consistent description of
energy levels, transition amplitudes, and the mixing of nearly resonant
states. The present work brings these elements into a common
perturb-then-diagonalize formulation in rectilinear and curvilinear
coordinates. Its central analytical contribution is the explicit
second-order reduction of transition moments to force constants,
kinetic-metric derivatives, and property derivatives. Specialized
expressions cover fundamentals, overtones, three-quantum transitions,
and hot bands, including the transformation of both the initial and
final states through their resonance polyads. These algebraic forms
complement general matrix and contact-transformation treatments and
make each contribution directly evaluable and independently testable.

The choice of coordinates provides a physical organization of the
anharmonic problem. ESALC-C and ESALC-I share the same harmonic
reference, while the curvilinear formulation retains the nonlinear
geometry map and the associated kinetic terms. The reduction of
intermode couplings observed for methyloxirane and glycine demonstrates
how internal coordinates concentrate anharmonicity around recognizable
chemical motions. In resonant regions, consistent effective
Hamiltonians and property transport connect that interpretation to the
mixed states that appear in a measured spectrum. Independent oscillator
calculations verify the algebraic amplitudes, comparisons with
established tools verify the rectilinear implementation, and measured
water and formaldehyde hot bands provide experimental comparisons
for the resulting vibrational moments and thermal populations.
The 48-mode naphthalene example connects calculated hot-band spacings to
gas-phase spectra.  The 57-mode dual-level aspirin calculation adds a
larger flexible system, identifies the finite Duschinsky blocks required
for tensor transport, and supplies an objective criterion for promoting
two coupled physical coordinates to a two-dimensional DVR.

Large-amplitude motion fits naturally into this framework through an
adiabatic DVR. A periodic or multiwell potential describes the extended
motion, and the transverse vibrations contribute to its effective
potential and properties. Replacing the selected perturbative terms by
DVR averages connects local anharmonic corrections to finite-amplitude
motion without counting the same contribution twice. The optical-rotation
and glycine examples show how this construction extends from individual
spectral transitions to thermal expectation values and conformer
populations. The next nonzero correction to state-specific property
averages, including quintic potential and third metric derivatives,
closely reproduces the C1 methylaziridine torsional DVR result.
Rotation--vibration constants use the same molecular metric,
and regularized perturbative energies and finite DVR ladders supply
thermodynamic state sums, including integer-occupation Wang--Landau
sampling for coupled modes.

The implementation connects these developments in a general molecular
calculation procedure. ORACLE and SMITH construct and organize the symmetry-adapted
coordinates, LINK supplies geometry optimization and access to electronic
calculations, and TRINITY acquires and transforms derivatives, evaluates
the algebraic corrections, and transports properties through the polyads.
Independent displaced geometries can be computed in parallel with
interchangeable electronic-structure programs. This separation makes the
vibrational treatment reusable as electronic methods improve, while
preserving a common description of symmetry, coordinates, and
observables. The resulting framework supports a coherent treatment of
anharmonic spectra and molecular averages across vibrational states,
coordinate representations, and temperatures.

\section*{Supplementary material}
The supplementary material provides the complete reduced equations
for coordinate and momentum properties, cold and hot transitions,
direct-pair resonance selection, numerical validation tables,
derivative-acquisition and convergence controls, molecular metric
checks, and the adiabatic-DVR and thermochemical constructions.

\begin{acknowledgments}
This work was supported by Gaussian, Inc.
\end{acknowledgments}

\section*{Author declarations}
\subsection*{Conflict of interest}
The author has no conflicts to disclose.

\subsection*{Author contributions}
Vincenzo Barone: Conceptualization; Methodology; Software;
Validation; Formal analysis; Investigation; Writing---original
draft; Writing---review and editing.
\appendix
\section{Independent curvilinear oscillator-matrix test}
\label{app:synthetic}
An independent two-mode calculation uses reproducibly generated
symmetric tensors $f_3,f_4,g_1,g_2,D_1,D_2,D_3$ with harmonic
frequencies 970 and 1430~\wavenum{}. The off-diagonal metric
derivatives are nonzero; in particular
$\rho_{112}-\sigma_{112}=34.992$~\wavenum{} in the benchmark's
frequency-scaled convention. We formed
\begin{align}
 H(\lambda)&=H_0+\lambda H_1+\lambda^2H_2,\nonumber\\
 \mu(\lambda)&=D_iq_i+\frac{\lambda}{2}D_{ij}q_iq_j
 +\frac{\lambda^2}{6}D_{ijk}q_iq_jq_k,
 \label{eq:synthetic-scaling}
\end{align}
where $H_1,H_2$ are exactly the ordered operators in
Eq.~\eqref{eq:operators}, including $p_iq_kp_j$ and
$p_iq_kq_lp_j$. Kronecker products of finite one-mode oscillator
matrices, formed independently of the TRINITY contraction routines,
were used to diagonalize $H(\lambda)$. Eigenvectors were identified
by their harmonic-basis overlaps and phase aligned before evaluating
transition moments. Twelve oscillator functions per mode were used;
increasing the basis from ten to twelve changes the largest reported
transition energy by $5.4\times10^{-6}$~\wavenum{} at $\lambda=1$.

The exact eigenvalues at $\lambda=1/16$ and $1/32$ were used to
extract $\chi_{11}$ and $\chi_{12}$ from the $|20\rangle$ and
$|11\rangle$ finite differences, followed by Richardson removal of
the leading $\lambda^2$ remainder after division by $\lambda^2$.
The exact-limit values are $-3.550514154$ and $-1.924003406$~\wavenum{};
Eqs.~\eqref{eq:chi-diag}--\eqref{eq:chi-offdiag} give
$-3.550514174$ and $-1.924003529$~\wavenum{}, respectively.

Table~\ref{tab:synthetic} gives the errors as the perturbation scale is halved.

\begin{table}[tbp]
\caption{Absolute exact-minus-VPT2 errors for the two-mode
curvilinear model. $\Delta\nu_{10}$ is in \wavenum{}; the moment
errors are in atomic units. The frequencies and moments are
compared with the direct-product oscillator diagonalization.}
\label{tab:synthetic}
\begin{ruledtabular}
\begin{tabular}{rrrr}
$\lambda$ & $|\Delta\nu_{10}|$ & $|\Delta M_{10}|$ & $|\Delta M_{20}|$\\
1     & 0.179009 & $7.181\times10^{-5}$ & $9.007\times10^{-4}$\\
1/2   & 0.010928 & $4.071\times10^{-6}$ & $1.115\times10^{-4}$\\
1/4   & 0.000679 & $2.481\times10^{-7}$ & $1.390\times10^{-5}$\\
1/8   & 0.000042 & $1.541\times10^{-8}$ & $1.736\times10^{-6}$\\
\end{tabular}
\end{ruledtabular}
\end{table}

The observed error orders under halving $\lambda$ are 4.015 for
the fundamental frequency, 4.062 for its transition moment, and
3.006 for the overtone moment. Although a generic second-order
truncation has an $O(\lambda^3)$ remainder, parity makes the
energies and fundamental moments even functions of $\lambda$ in
this model: their cubic coefficients vanish. The overtone moment
is odd in $\lambda$ and has the expected cubic remainder. Thus the
test exercises $g_{ij,k}$, $g_{ij,kl}$, $\rho\ne\sigma$, and the
normalization terms independently of derivative acquisition. The
complete calculation data accompany the SI.

\end{document}